\documentclass[fleqn,usenatbib]{mnras}

\usepackage{newtxtext,newtxmath}

\usepackage[T1]{fontenc}

\DeclareRobustCommand{\VAN}[3]{#2}
\let\VANthebibliography\thebibliography
\def\thebibliography{\DeclareRobustCommand{\VAN}[3]{##3}\VANthebibliography}

\usepackage{graphicx}	
\usepackage{amsmath}	

\usepackage{caption}
\usepackage{subcaption}
\usepackage{verbatim}  
\usepackage{float}
\usepackage{xcolor}     
\usepackage{siunitx}
\usepackage{rotating}
\usepackage{amsfonts}
\usepackage{hyperref}
\usepackage{cleveref}
\crefname{section}{§}{§§}
\Crefname{section}{§}{§§}
\usepackage{longtable}
\usepackage{natbib}
\usepackage{verbatim}
\usepackage{lscape}
\usepackage{graphics}
\usepackage{epsfig}
\usepackage{multirow}
\usepackage{setspace}
\usepackage{bigdelim}
\usepackage{bigstrut}
\usepackage{floatflt}	
\usepackage{wrapfig}
\usepackage{multicol}

\usepackage{indentfirst}

\newcommand{\fatm}{$f_{\mathrm{atm}} -q$~}
\newcommand{\jb}{$j_{\mathrm{b}} - M_{\mathrm{b}}$~}

\makeatletter
\newcommand*\mysize{%
   \@setfontsize\mysize{6.8}{10.0}%
}
\makeatother

\sffamily
\def\h1{H\,{\sc i}}

\def\ga{\raisebox{-0.3ex}{\mbox{$\stackrel{>}{_\sim} \,$}}}
\def\la{\raisebox{-0.3ex}{\mbox{$\stackrel{<}{_\sim} \,$}}}

\def\c1{C\,{\sc i}}

\def\NH3{NH$_{3}$}
\def\ch3cn{CH$_{3}$CN}

\def\kms{km s$^{-1}$}
\def\apj{ApJ}
\def\apjs{ApJSS}
\def\apjl{ApJL}

\def\aap{A\&A}

\def\araa{ARA\&A}

\def\pasp{PASP}
\def\apss{Ap\&SS}
\def\aj{AJ}
\def\nat{Nature}
\def\mnras{MNRAS}

\defcitealias{obreschkow14}{OG14}
\defcitealias{obreschkow16}{O16}
\defcitealias{murugeshan2019}{M19}
\defcitealias{Murugeshan20}{M20}

\title[WALLABY Pre-Pilot Survey: AM properties of the Eridanus supergroup]{WALLABY Pre-Pilot Survey: The effects of angular momentum and environment on the \h1 gas and star formation properties of galaxies in the Eridanus supergroup}

\author[C. Murugeshan et al.]{C. Murugeshan,$^{1,2,3}$\thanks{E-mail:chandrashekar.murugeshan@csiro.au} V. A. Kilborn,$^{1,3}$ B.-Q. For,$^{5,3}$ O. I. Wong,$^{2,5,3}$ J. Wang,$^{4}$ \newauthor T. Westmeier,$^{5,3}$ A. R. H. Stevens,$^{5,3}$ K. Spekkens,$^{6}$ P. Kamphuis,$^{7}$ L. Staveley-Smith,$^{5,3}$ \newauthor K. Lee-Waddell,$^{8,3}$ D. Kleiner,$^{9}$ B. S. Koribalski,$^{8,10,3}$ M. E. Cluver,$^{1}$ S.-H. Oh,$^{11}$ \newauthor J. Rhee,$^{5,3}$ B. Catinella,$^{5,3}$ T. N. Reynolds,$^{5,3}$ H. D{\'e}nes,$^{12}$ A. Elagali$^{13}$
\\
$^{1}$Centre for Astrophysics and Supercomputing, Swinburne University of Technology, Hawthorn, Victoria 3122, Australia \\
$^{2}$ATNF, CSIRO, Space and Astronomy, PO Box 1130, Bentley, WA 6102, Australia \\
$^{3}$ARC Centre of Excellence for All Sky Astrophysics in 3 Dimensions (ASTRO 3D), Australia \\
$^{4}$Kavli Institute for Astronomy and Astrophysics, Peking University, Beijing 100871, China \\
$^{5}$International Centre for Radio Astronomy Research, The University of Western Australia, 35 Stirling Highway, Crawley, WA 6009, Australia \\
$^{6}$Royal Military College of Canada, PO Box 17000, Station Forces, Kingston, Ontario, Canada K7K 7B4 \\
$^{7}$Ruhr-Universit\"at Bochum, Faculty of Physics and Astronomy, Astronomical Institute, 44780 Bochum, Germany \\
$^{8}$ATNF, CSIRO, Space and Astronomy, PO Box 76, Epping, NSW 1710, Australia \\
$^{9}$INAF – Osservatorio Astronomico di Cagliari, Via della Scienza 5, 09047 Selargius, CA, Italy \\
$^{10}$Western Sydney University, Locked Bag 1797, Penrith, NSW 2751, Australia \\
$^{11}$Department of Physics and Astronomy, Sejong University, 209 Neungdong-ro, Gwangjin-gu, Seoul 05006, Republic of Korea \\
$^{12}$ASTRON - The Netherlands Institute for Radio Astronomy, NL-7991 PD Dwingeloo, The Netherlands \\
$^{13}$Telethon Kids Institute, Perth Children’s Hospital, Perth, Australia
}

\date{Accepted XXX. Received YYY; in original form ZZZ}

\pubyear{2015}

\begin{document}
\label{firstpage}
\pagerange{\pageref{firstpage}--\pageref{lastpage}}
\maketitle

\begin{abstract}
We use high-resolution ASKAP observations of galaxies in the Eridanus supergroup to study their \h1, angular momentum and star formation properties, as part of the WALLABY pre-pilot survey efforts. The Eridanus supergroup is composed of three sub-groups in the process of merging to form a cluster. The main focus of this study is the Eridanus (or NGC 1395) sub-group. The baryonic specific angular momentum -- baryonic mass ($j_{\mathrm{b}} - M_{\mathrm{b}}$) relation for the Eridanus galaxies is observed to be an unbroken power law of the form $j_{\mathrm{b}} \propto M_{\mathrm{b}}^{0.57 \pm 0.05}$, with a scatter of $\sim 0.10 \pm 0.01$ dex, consistent with previous works. We examine the relation between the atomic gas fraction, $f_{\mathrm{atm}}$, and the integrated atomic disc stability parameter $q$ (the $f_{\mathrm{atm}} - q$ relation), and find that the Eridanus galaxies deviate significantly from the relation owing to environmental processes such as tidal interactions and ram-pressure affecting their \h1 gas. We find that a majority of the Eridanus galaxies are \h1 deficient compared to normal star-forming galaxies in the field. We also find that the star formation among the Eridanus galaxies may be suppressed owing to their environment, thus hinting at significant levels of pre-processing within the Eridanus sub-group, even before the galaxies have entered a cluster-like environment.

\end{abstract}

\begin{keywords}
galaxies: evolution-- galaxies: fundamental parameters-- galaxies: ISM-- galaxies: kinematics and dynamics
\end{keywords}



\section{Introduction}

\noindent The neutral atomic hydrogen (\h1) gas in galaxies is a fundamental constituent, as it forms the primary reservoir from which molecular gas and eventually stars are formed. The \h1 gas content in galaxies is set by both intrinsic properties and processes such as mass, angular momentum (AM), star formation and feedback, and/or a number of environmentally driven processes including tidal interactions and ram-pressure stripping. Therefore, studying the \h1 properties in relation to other observable global properties of galaxies, and the various processes that affect the \h1 gas is critical to our understanding of how galaxies evolve. In addition, the \h1 gas in galaxies is typically more extended than their stellar discs \citep{verheijen01}, making high (spatial and spectral) resolution \h1 observations of galaxies extremely important, as this allows us to trace the rotation curves of galaxies out to large radii. High-resolution \h1 observations of galaxies therefore not only enable us to study the kinematics of galaxies, but also accurately measure their AM.

It is well established that the \h1 gas fraction ($M_{\textrm{\h1}}/M_{\star}$) of late-type galaxies is correlated with their stellar mass ($M_{\star}$), which is a commonly used scaling relation in \h1 astronomy (\citealt{haynes84};~\citealt{chamaraux86};~\citealt{solanes96};~\citealt{denes14}). In addition, the \h1 content in galaxies has been observed to correlate with a number of other observables such as stellar surface mass density, NUV - \textit{r} colour and the $R$-band magnitude (\citealt{catinella10},~\citeyear{catinella12},~\citeyear{catinella18};~\citealt{denes14};~\citealt{brown15},~\citeyear{brown17}). By studying these scaling relations for large samples of galaxies, we are able to understand how the \h1 gas content in galaxies is associated with other global properties. Galaxies that deviate from the mean trend of the various scaling relations are interesting, as they present to us an opportunity to probe the underlying physics driving their \h1 gas properties. 

Among the various intrinsic properties of galaxies, mass and AM are believed to play important roles in regulating their \h1 content (see for example \citealt{zasov89};~\citealt{huang12};~\citealt{denes14};~\citealt{maddox15}). \citet{obreschkow16} [hereafter \citetalias{obreschkow16}] showed that the atomic gas fraction, $f_{\mathrm{atm}} = \frac{1.35M_{\textrm{\h1}}}{M_{\mathrm{b}}}$, of disc galaxies is correlated with their integrated disc stability parameter $q = \frac{j_{\mathrm{b}}\sigma}{G M_{\mathrm{b}}}$, in the following manner,
\begin{equation}
    f_{\textrm{atm }}=\min \left\{1,2.5 q^{1.12}\right\},
\end{equation}
where $M_{\textrm{\h1}}$ is the \h1 mass of the galaxy, $M_{\mathrm{b}}$ is the total baryonic mass, $j_{\mathrm{b}}$ is the baryonic specific angular momentum (sAM), $\sigma$ is the velocity dispersion of the Warm Neutral Medium (WNM) and $G$ is the universal gravitational constant. Galaxies with a lower $q$ will promote star formation processes driven by gas fragmentation and collapse, while larger values of $q$ imply a higher stability against such mechanisms. The $q$ parameter in this sense can be thought of as a global analog to the local Toomre parameter $Q$ \citep{toomre64}. A large sample of late-type galaxies from isolated environments (where the influence of environmental processes are minimal) are observed to follow the \fatm relation consistently (\citetalias{obreschkow16};~\citealt{lutz17};~\citealt{murugeshan2019},\citeyear{Murugeshan20}).

However, as mentioned earlier, the \h1 gas in galaxies can be easily influenced by environmental processes, which is particularly true in high-density environments, where processes such as tidal interactions (\citealt{Merritt83};~\citealt{Byrd90};~\citealt{Rots90}), ram-pressure (\citealt{gunn72};~\citealt{vollmer01};~\citealt{cayatte94};~\citealt{Chung09}), strangulation~\citep{larson80}, harassment \citep{moore96} and thermal evaporation \citep{cowie77} are prevalent. Such environmental processes have been observed to lower the \h1 gas fractions in late-type galaxies residing in high-density environments (see for example \citealt{davies73};~\citealt{giovanelli85};~\citealt{solanes01};~\citealt{denes14}). 

The \citetalias{obreschkow16} model is an idealized analytic model for equilibrium discs, and does not account for evolutionary or environmental processes that may be affecting the \h1 gas in galaxies. In order to understand how environmental effects affect galaxies in this parameter space, both simulation and observational studies are required. \citet{Stevens18} used semi-analytic models (SAMs) for the first time to study the \fatm parameter space, and found that two main environmental mechanisms -- mergers and ram-pressure stripping -- were responsible for breaking the link between $f_{\mathrm{atm}}$ and $q$. However, empirical  studies are necessary to test and verify this connection. The \fatm parameter space thus provides the opportunity to identify galaxies whose \h1 gas fractions or AM properties have been affected by environmental processes. 

While several theoretical studies have investigated the role of sAM and environment on the atomic gas fraction of galaxies (\citealt{lagos17};
\citealt{Stevens18},~\citeyear{Stevens19};~\citealt{Wang18};~\citealt{Zoldan18}), only two recent empirical investigations have studied the behaviour of galaxies on the \fatm relation as a function of the galaxies' local environments. The first work was conducted by \citet{Li20}, who examined the behaviour of late-type galaxies in the Virgo cluster, using the VIVA sample \citep{Chung09}. They observed that Virgo galaxies are not only conventionally \h1-deficient relative to field galaxies of the same type, but they are also systematically offset below the \fatm relation, indicating that environmental processes have significantly affected their \h1 gas content. By comparing their results with simulated data from the EAGLE simulations \citep{Schaye15}, they attribute the \h1-deficiency in the Virgo cluster galaxies to be driven primarily by ram-pressure stripping. Under this scenario, as galaxies are entering the cluster potential, ram-pressure effects take over and the galaxies lose a large fraction of their \h1 gas, which results in the overall reduction of their $f_{\mathrm{atm}}$. However, the $j_{\mathrm{b}}$ values of these galaxies are not severely affected, which implies that the $q$ values remain fairly unchanged. The combined result then is that galaxies will systematically drop off the \fatm relation, supporting the earlier predictions made by \citet{Stevens18,Stevens19}. 

The second empirical work to study the effects of the environment on galaxies in the \fatm parameter space was undertaken by \citet{Murugeshan20} [hereafter \citetalias{Murugeshan20}]. They used 114 late-type galaxies from the WHISP survey \citep{whisp02}, in low- and intermediate density environments, and find that galaxies that are currently interacting (or have recently interacted, thereby accreting \h1 gas from their smaller companions) showed enhanced $f_{\mathrm{atm}}$ values, lower $q$ values, and increased star formation rates (SFR) for their stellar mass. Interactions and mergers can have a significant influence on the way galaxies evolve. \citet{Lagos18} used hydrodynamical simulations to study the effects of mergers on the stellar sAM ($j_{\star}$) of galaxies, and found that mergers tend to reduce $j_{\star}$, with dry major mergers showing the largest effect. However, they find that wet mergers on the other hand may marginally increase (by $\sim 10\%$) $j_{\star}$. Clearly, the location of galaxies in the \fatm parameter space is governed by both $M_{\mathrm{b}}$ and $j_{\mathrm{b}}$. In order to properly understand the behaviour of galaxies in this parameter space, it is imperative to populate this parameter space with galaxies from various environments and at various stages of their evolution. 

The Widefield ASKAP L-Band Legacy All-sky Blind SurveY (WALLABY; \citealt{Koribalski20}) which aims to deliver high-resolution \h1 images for over 5000 galaxies uniformly across various environments, will be unprecedented in terms of its scope for enabling us to understand the different processes affecting the \h1 content of galaxies. WALLABY uses the Australian SKA Pathfinder (ASKAP;~\citealt{Hotan21}) as its survey instrument (see Section~\ref{subsec:Obs} for more details). With such a large sample, we will be able to populate the \fatm parameter space systematically for the first time with galaxies from different environments, thus giving us the opportunity to study the degree of influence of both AM and environmental processes on the gas evolution of galaxies. In light of this, we make use of the WALLABY pre-pilot ASKAP observations of the Eridanus supergroup (see also \citealt{For21};~\citealt{Wong21}), to study the influence of both AM and the environment on the \h1 gas content of the galaxies.

\subsection{The Eridanus supergroup}

\noindent As part of the Southern Sky Redshift Survey (SSRS;~\citealt{dacosta88}), the Eridanus group of galaxies was found to be part of a filamentary structure that encompassed the Fornax cluster and the Dorado group. \citet{Willmer89} classified the Eridanus group of galaxies as a cluster composed of three to four subgroups. On the other hand, \citet{Garcia93} and \citet{Barton96} identified only two optically classified sub-groups, centered on NGC 1407 and NGC 1332 respectively. \citet{Omar05a} conclude that the Eridanus group of galaxies is a loose group.

A detailed investigation of the Eridanus group of galaxies was later made by \citet{Brough06}, who define this to be a supergroup, where three sub-groups of galaxies -- the Eridanus (or NGC 1395), NGC 1407 and NGC 1332 sub-groups, are in the process of merging to form a cluster. They find that the NGC 1407 sub-group is the most evolved among the three sub-groups, and is composed of a large fraction of early-type galaxies, while the NGC 1395 and NGC 1332 sub-groups are observed to contain larger fractions of late-type galaxies. For consistency, here after, we will refer to the NCG 1395 sub-group as the Eridanus sub-group, which is the main focus of this study.

The NGC 1407 and NGC 1332 sub-groups have been studied as part of the Group Evolution Multiwavelength Study (GEMS) by \citet{osmond04}.
Using ROSAT X-ray data, both \citet{Trinchieri2000} and \citet{Brough06} showed that there is extended X-ray emission associated with the NGC 1407 intragroup gas, while the X-ray emission from the NGC 1332 sub-group is localised to the NGC 1332 galaxy itself. \citet{Omar05a} also find intragroup X-ray emission associated with the Eridanus sub-group. 

In addition, as part of the southern GEMS project, \citet{Kilborn09} examined the \h1 properties of the NGC 1407 and NGC 1332 sub-groups using the 64-m Parkes single-dish telescope, and found that the \h1 gas distribution in the two groups differ, in the sense that for the NGC 1407 group there is a lack of \h1 in the central region of the group, while the NGC 1332 group is observed to have significantly more gas in the central region. Using high-resolution GMRT observations of the \h1 gas for a sub-sample of Eridanus galaxies, \citet{Omar05b} find that the Eridanus galaxies are about 2 -- 3 times more \h1-deficient compared to galaxies of similar type in the field. They suggest that the \h1 deficiency in the Eridanus galaxies is driven by tidal interactions rather than ram-pressure, as they observe that the level of ram-pressure is about two orders of magnitude lower compared to that in typical clusters.  

A number of previous works have shown that even in the group regime, fast gas-removing processes such as ram-pressure cannot be fully ruled out, and are likely playing an important role in affecting the \h1 gas properties of the galaxies (\citealt{Chamaraux04};~\citealt{kilborn05},\citeyear{Kilborn09};~\citealt{sengupta06};~\citealt{westmeier11};~\citealt{hess13};~\citealt{denes16};~\citealt{brown17};~\citealt{stevens17}). These studies hint at a scenario where a significant amount of pre-processing may be prevalent in group-like environments. The Eridanus supergroup is therefore an interesting field, as it is a supergroup composed of multiple sub-groups, which will eventually merge to form a cluster \citep{Brough06}. Studying the \h1 and other global properties of the Eridanus galaxies therefore provides us the unique opportunity to understand both the level of pre-processing within the different sub-groups as well as their in-fall boundaries. 

\section{Data}
\subsection{Observations}
\label{subsec:Obs}
\begin{figure}
\centering
\hspace*{-0.4cm}
\includegraphics[width=9cm,height=7.8cm]{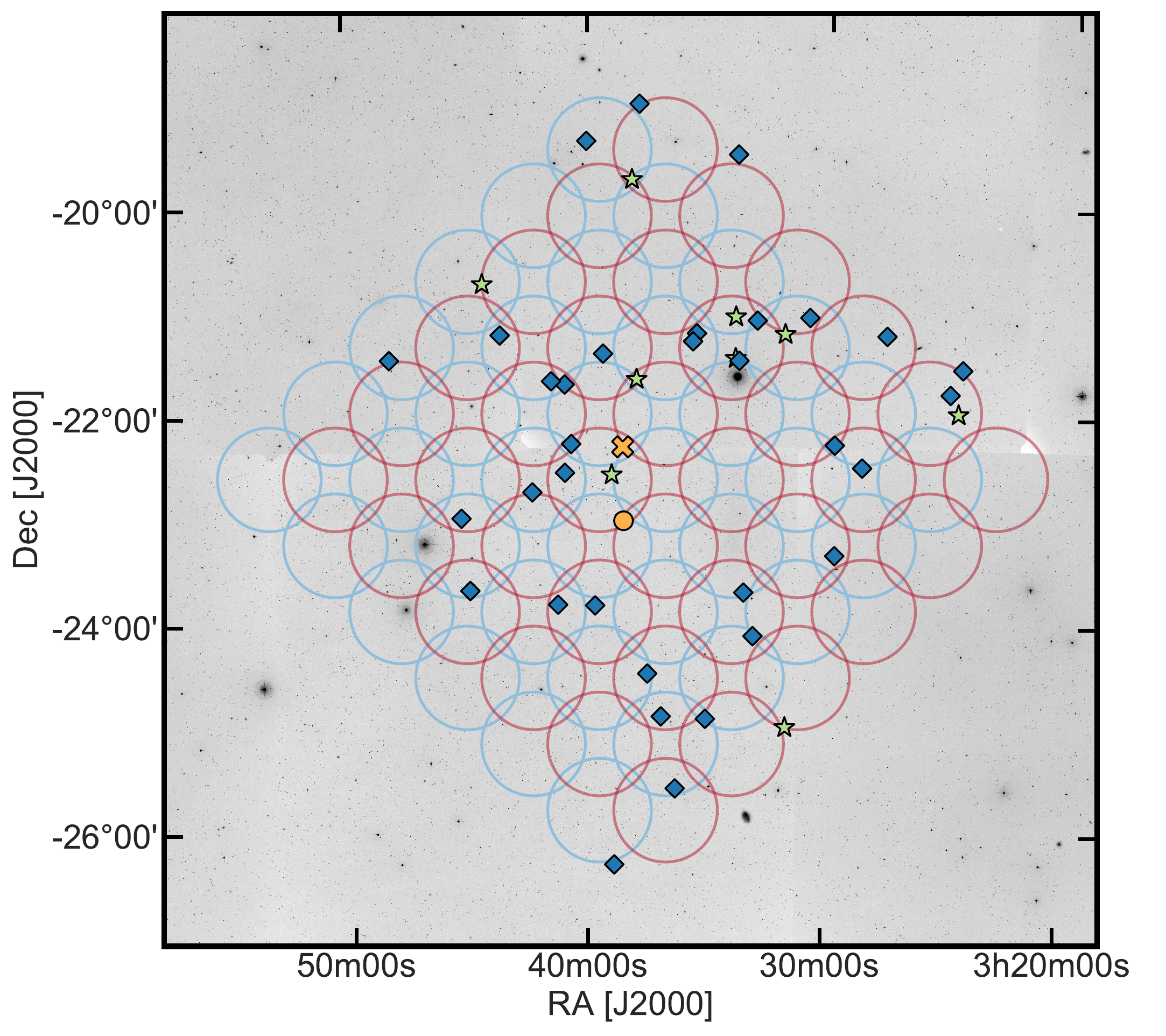}
\caption{Footprints A (light blue) and B (red), overlaid on DSS image of the Eridanus field. The blue diamond points denote the Eridanus galaxies, while the background galaxies are represented by the green stars. The yellow circle represents the central large elliptical NGC 1395 of the Eridanus sub-group, while the yellow cross marks the centroid of the Eridanus sub-group derived by \citet{Brough06}, using a friends-of-friends group-finding algorithm.}
\label{Fig.1}
\end{figure}

\noindent The \h1 data for the current work has been acquired via ASKAP observations of the Eridanus supergroup, as part of the WALLABY pre-pilot survey efforts. ASKAP \citep{Hotan21} is a radio interferometer located at the Murchison Radio Observatory (MRO), in an isolated area in Western Australia. ASKAP consists of 36 antennas, each 12 m in diameter and equipped with the Mk II phased array feeds (PAFs; \citealt{DeBoer09};~\citealt{Chippendale10};~\citealt{hotan14}). Using the PAF technology, each ASKAP antenna is able to form 36 beams on the sky simultaneously. Typically, the 36 beams are arranged in a $6 \times 6$ square or diamond like pattern (also called footprints). Each footprint has a field of view (FOV) $\sim 30$ deg$^{2}$ at 1.4 GHz. By interleaving two footprints (footprint A and B), we achieve close to uniform rms across the field. The observations of the Eridanus field were carried out in March 2019, for both footprints A and B, with a combined integration time of 10.8 h. The observing strategy involved interleaving two diamond-shaped ASKAP footprints to mostly observe galaxies belonging to the Eridanus sub-group (see Fig.~\ref{Fig.1}). The Eridanus sub-group is the first pre-pilot field for which observations were made with all 36 antennas and using a full bandwidth of 288 MHz. For a full description of the observations and quality of data, we refer the reader to \citet{For21}. 

\subsection{Data reduction}
\label{sec:data_reduction}
\noindent In this work, only 144 MHz of the data corresponding to the higher frequency end (1295.5 to 1439.5 MHz) of the total 288 MHz band was processed, as the Eridanus sub-group and the background galaxies are contained within this velocity range (1000 to 8100 \kms). The ASKAP observations have been reduced using the \textsc{AskapSoft} pipeline [version 0.24.7] \citep{Whiting20}, which is the default software used to process and image ASKAP observations. The processing strategy involved first reducing the data for the two footprints independently, and then finally mosaicking the individual beams from each footprint to make the final image cubes. Standard processing steps were involved, such as flagging, bandpass calibration, continuum imaging, continuum subtraction, spectral imaging and image-based continuum subtraction. A Robust weighting of 0.5 was used for the spectral imaging step, and the final mosaicked cube was convolved to a resolution of 30 arcsec, and having a channel width of $\sim 4$~\kms, corresponding to the original channel resolution of $18.5$~kHz. For more details on each processing step, we refer the reader to previous early science papers (\citealt{Kleiner19}; \citealt{Reynolds19}; \citealt{Lee-Waddell19}; \citealt{Elagali19}; \citealt{For19}), which have described both \textsc{AskapSoft} and the ASKAP pipeline in great length. The rms noise in the final mosaicked image cube is $\sim 2.6$ mJy per beam per channel, corresponding to a $3\sigma$ \h1 column density of $N_{\textrm{\h1}} \sim 4 \times 10^{19}$ cm$^{-2}$. Following the image processing, \texttt{SoFiA} \citep{Serra15}, an efficient source finding software was run on the final spectral cube to automatically detect the \h1 sources.

From the final mosaicked image cube, we split out smaller ($0.2 \times 0.2$~deg$^{2}$) cubelets for the individual galaxy detections. However, as the signal-to-noise ratio (SNR) for many sources in the sample was low, we performed a Hanning smoothing along the velocity axis and smoothed the cubes to a 12 \kms ~velocity resolution, in order to improve the SNR. This reduces the rms in the smoothed cubes to $\sim 1.8$ mJy per beam per channel, resulting in a $3\sigma$ \h1 column density limit of $N_{\textrm{\h1}} \sim 8 \times 10^{19}$ cm$^{-2}$. We use these cubelets and the resulting moment-0 (intensity) and moment-1 (velocity) maps for the rest of the analysis in this work. For the full catalogue and description of the source properties, we refer the reader to \citet{For21}. 

\subsection{The sample}
\label{Subsec:HI-properties}

\noindent The final source catalogue from \texttt{SoFiA} consists of 55 \h1 detections -- 43 detections in the Eridanus field and 12 detections corresponding to background field galaxies. Of the total 55 galaxies, we select galaxies that have an \h1 diameter, $D_{\textrm{\h1, max}} \geqslant 120$ arcsec, so that we have at least four resolution elements across the major axis, to be able to model the galaxies accurately using 3D fitting techniques (for more details see Section \ref{subsec:measuring_AM}). $D_{\textrm{\h1, max}}$ is the maximal \h1 diameter out to which the tilted-ring modelling is performed, and corresponds to the radius where the \h1 column density drops to the $3\sigma$ \h1 column density limit of $N_{\textrm{\h1}} \sim 8 \times 10^{19}$ cm$^{-2}$. This is to ensure that we do not miss emission in the very outskirts of the galaxies, which also enables us to derive more accurate AM values. This selection criteria leaves us with the fiducial sample consisting of a total of 42 galaxies -- 33 Eridanus galaxies and 9 background galaxies.  The sample and all the relevant derived properties of the galaxies are listed in Table~\ref{tab:sample}.

\subsection{Measuring the total baryonic mass ($M_{\mathrm{b}}$) and baryonic specific angular momentum ($j_{\mathrm{b}}$)}
\label{subsec:measuring_AM}

\noindent We now describe the methods employed in this work to measure the total baryonic mass ($M_{\mathrm{b}}$) and baryonic specific AM ($j_{\mathrm{b}}$) for the sample galaxies. We first perform 3D tilted-ring fitting \citep{rogstad74} using \texttt{3DBarolo}~\citep{teodoro15} for 42 of the 55 \h1 detections in the field. The remaining 13 detections were not well resolved for robust kinematic modelling, and for this reason they were omitted from the analysis. For a step-by-step description of the 3D fitting method, we refer the reader to \citet{murugeshan2019}. The 3D modelling involves fitting $n$ independent tilted rings (where $n$ is the total number of resolution elements across the major axis) of width 15 arcsec (half the FWHM of the resolution element). In the fitting procedure we fix the kinematic centre to the optical/NIR centre (obtained from NED\footnote{\url{http://ned.ipac.caltech.edu/}}), and the systemic velocity and position angle (PA) to the values determined by SoFiA. For the inclination angle ($i$), we use the $B$-band optical surface photometry based estimate by \citet{Lauberts89}, as the initial guess. We allow PA, $i$ and the rotation velocity ($v_\mathrm{rot}$) to vary as free parameters.

The fitting is performed in two stages -- in the first stage the user-defined initial guesses for the galaxy centre, systemic velocity, PA, $i$ and $v_\mathrm{rot}$ are assumed for the fitting, and in the second stage the code uses the best fit values from the first stage to fit the final model. We allow both the PA and $i$ to vary within a range of $\pm 10 \degr$ ~about the inputted value for the fitting, thus enabling the code to reasonably model any kinematic warps. Once the best fitting model is derived, the code outputs the best fitting values for PA, $i$ and $v_\mathrm{rot}$ for every ring. In order to verify the goodness of the fit, we inspect if the derived line-of-sight velocities $v_{\textrm{\tiny LOS}} = v_{\mathrm{rot}}\sin(i)$ are in agreement with the position--velocity (PV) diagrams of the galaxies. In Appendix~\ref{appendix:fitting_models}, we present the fitted model parameters and PV diagrams for two specific galaxies to demonstrate the robustness of the modelling. The tilted-rings are then projected onto the moment-0 \h1 integrated flux map to compute the \h1 mass within each ring, using the following relation,
\begin{equation}
     M_{\textrm{HI}}\left[\mathrm{M}_{\odot}\right]= 2.356 \times 10^{5} D^{2} S_{\mathrm{int}},
\end{equation}
where, $D$ is the distance to the source in Mpc and $S_{\mathrm{int}}$ is the total integrated flux from all pixels enclosed within the ring in Jy~\kms. We compute the uncertainties on $M_{\textrm{HI}}$ by propagating the errors associated with $D$ and $S_{\mathrm{int}}$. We assume a nominal error of 10\% on the distance ($D$) and compute the error on $S_{\mathrm{int}}$ using the relation introduced in \citet{Koribalski04}.

The stellar mass within each ring is calculated by projecting the tilted-rings onto the 2MASS \citep{2mass06} $K_{s}$-band background subtracted mosaic of the galaxies. The sum of the $K_s$ magnitudes within each ring is converted to a stellar mass following the relation described by Eq.3 in \citet{wen13} as follows 
\begin{equation}
    \begin{aligned}
\log _{10}\left(\frac{M_{\star}}{\mathrm{M}_{\odot}}\right)=&(-0.498 \pm 0.002)+(1.105 \pm 0.001) \\
& \times \log _{10}\left(\frac{v L_{v}\left(K_{s}\right)}{\mathrm{L}_{\odot}}\right).
\end{aligned}
\end{equation}
Where $L_{v}$ is the luminosity, derived using the extinction-corrected $K_s$-band magnitude.

The total \h1 and stellar masses are computed by summing their respective mass within each ring. The total baryonic mass is calculated using the relation $M_{\mathrm{b}} = M_{\star} + 1.35 M_{\small \mathrm{\textrm{\h1}}}$, where $M_{\small \textrm{\h1}}$ is the total \h1 mass and $M_{\star}$ is the total stellar mass. The factor 1.35 accounts for the local universal 26\% He fraction. The total baryonic specific AM is computed using the following relation
\begin{equation}
    j_{\mathrm{b}} =\frac{\sum_{i} (1.35 M_{\mathrm{\tiny \textrm{\h1}},i} + M_{\star,i}) v_{\mathrm{rot},i}r_{i}}{\sum_{i} (1.35 M_{\mathrm{\tiny \textrm{\h1}},i} + M_{\star,i})},
\end{equation}
Where, $r_{i}$ is the radius of the $i^{th}$ ring in kpc and $v_{\mathrm{rot},i}$ is the rotation velocity corresponding to that ring in \kms. 

We note the following approximations have been made in our analysis without significantly affecting our $M_{\mathrm{b}}$ and $j_{\mathrm{b}}$ measurements -- a) As a large fraction ($\sim 76\%$) of the sample galaxies are low-mass spirals ($\log_{10}(M_{\star})$[M$_{\odot}] < 9.5$) with no prominent bulge component, the potential contribution from the bulge to the $M_{\mathrm{b}}$ and $j_{\mathrm{b}}$ measurements will be negligible. We have therefore not implemented any corrections to account for this effect in our analysis. b) We have not accounted for the molecular gas (H$_2$) contribution in our measurements, as CO data are unavailable for the sample. However, the H$_2$ gas mass in low-mass spirals typically constitutes $\leq 6\%$ of the total baryonic mass (see for e.g.,~\citealt{Saintonge17};~\citealt{Isbell18}). Therefore, the lack of H$_2$ data should not affect our $M_{\mathrm{b}}$ and $j_{\mathrm{b}}$ measurements by more than a few percent. c) Lastly, $j_{\mathrm{b}}$ is computed assuming that all baryonic matter (stars + \h1 gas) in the galaxies are co-moving with the same rotation velocity. This may not always be the case, however, the \h1 gas forms a dominant part of the total baryonic mass budget of the Eridanus galaxies, and is likely to be the largest contributor to their $j_{\mathrm{b}}$ values. We verify this by computing the ratio of the specific angular momentum of the \h1 gas to that of the stars, $j_{\textrm{\h1}}/j_{\star}$, and plotting it against their total baryonic mass, $M_{\mathrm{b}}$ (Fig.~\ref{Fig.B1} in Appendix~\ref{appendix:j_fracVsMb}). We observe no correlation between $j_{\textrm{\h1}}/j_{\star}$ and $M_{\mathrm{b}}$ and find that the mean value of $j_{\textrm{\h1}}/j_{\star} \sim 2$ for the Eridanus sample (in agreement with the observations made by \citet{ManceraPina21b} for their sample of disc galaxies).

Finally, we compute the atomic gas fraction and stability parameter following \citetalias{obreschkow16} as
\begin{equation}
    f_{\mathrm{atm}} = \frac{1.35 M_{\textrm{\h1}}}{M_{\mathrm{b}}}
\end{equation}

\begin{equation}
    q = \frac{j_{\mathrm{b}}\sigma_{\textrm{\h1}}}{G M_{\mathrm{b}}}.
\end{equation}
In this work, we have fixed the \h1 velocity dispersion ($\sigma_{\textrm{\h1}}$) value to 10~\kms. This is mainly due to the following reasons --  a) The \h1 cubes used for the 3D tilted-ring modelling were smoothed to a velocity resolution of 12 \kms ~(for details see Section~\ref{sec:data_reduction}), implying that fitting for $\sigma_{\textrm{\h1}}$ which typically has values in the range 8 -- 13 \kms ~\citep{leroy08} will no longer be reliable, making it hard to constrain the fitted $\sigma_{\textrm{\h1}}$ value. b) As the SNR in the data cubes are low, having an additional free parameter to fit for will possibly make the overall fits less physically accurate. c) A number of previous works have shown that the mean $\sigma_{\textrm{\h1}}$ value for large samples of late-type galaxies is $\sim 10$ \kms ~(\citetalias{Murugeshan20}; \citealt{Mogotsi16}; \citealt{Tamburro09}) and corresponds to the expected velocity dispersion driven by turbulence in the gas disc (\citealt{Tamburro09};~\citealt{Krumholz18}). 

\noindent The uncertainties on $M_{\mathrm{b}}$, $j_{\mathrm{b}}$, $f_{\mathrm{atm}}$ and $q$ were calculated by propagating the errors associated with $M_{\star}$, $M_{\textrm{\h1}}$ and $v_\textrm{rot}$ using standard error propagation.

\section{Results}
\label{sec:results}
\noindent The majority of the detected galaxies in the Eridanus group are disc-dominated low-mass spirals and dwarfs (with a median stellar mass of $\log_{10}(M_{\star})$[M$_{\odot}] \sim 8.8$). The fact that the Eridanus galaxies are low-mass systems and residing in high-density environments presents us with the opportunity to study the effects of environmental processes on their \h1 and star formation properties. In the following sections, we investigate the effects of mass and sAM, as well as the environment on the global properties of the Eridanus galaxies. 

\subsection{The \jb relation}
\label{Subsec:jb-Mb relation}

\begin{figure}
\hspace*{-0.5cm}
\includegraphics[width=9cm,height=7cm]{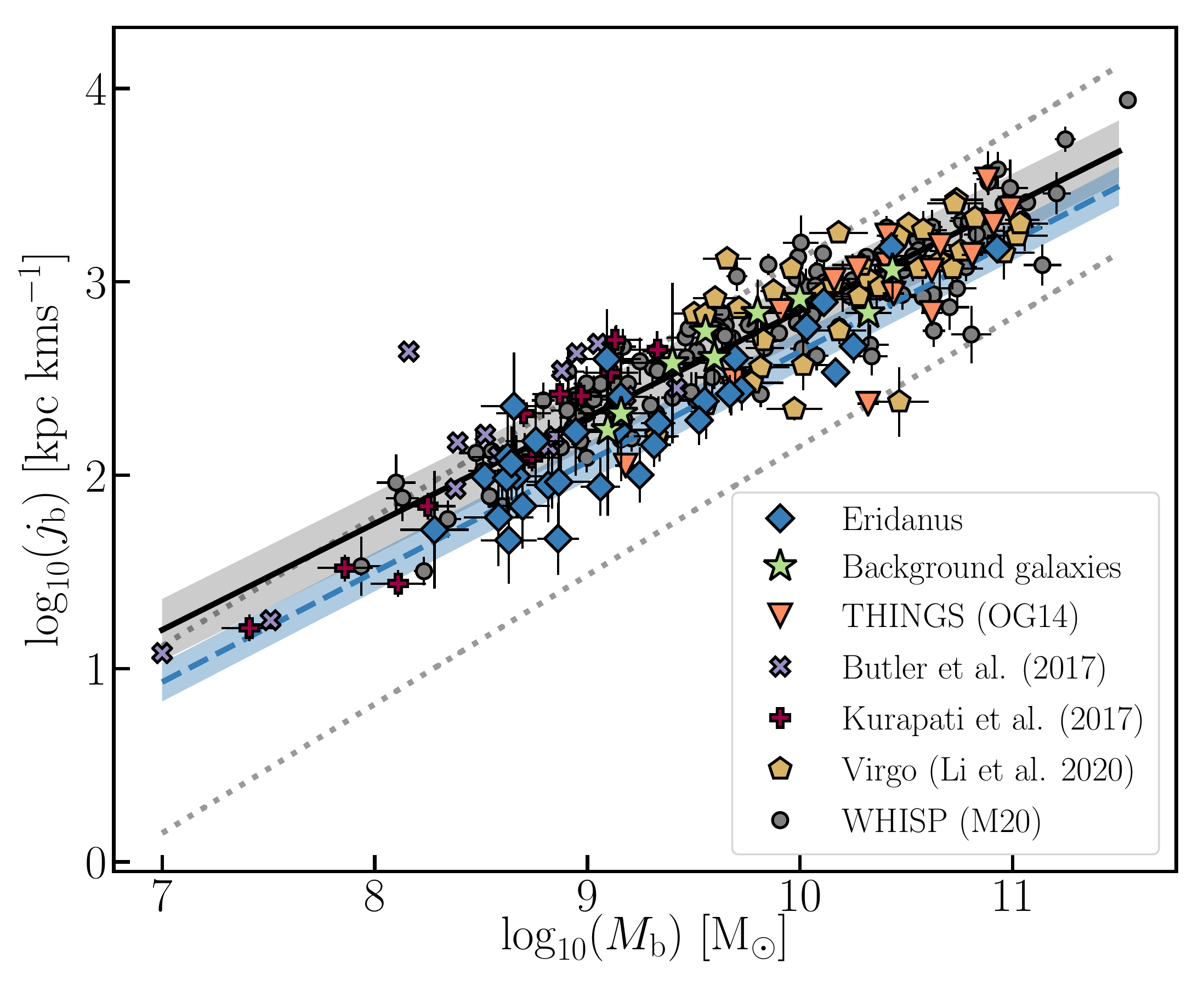}
\caption{The \jb relation for 235 late-type galaxies. The best fit line to all galaxies (solid black line) has a slope $0.55 \pm 0.02$ and $1\sigma$ scatter of $0.16$ dex (light grey shaded region). The best fit line to the Eridanus galaxies alone (dashed blue line) has a slope $0.57 \pm 0.05$ and a scatter of $0.10$ dex (light blue shaded region). The dotted lines show the expected range of $j_{\mathrm{b}}$ values based on $\Lambda$ Cold Dark matter ($\Lambda$CDM) predictions (see \citetalias{obreschkow14} for more details).}
\label{Fig.2}
\end{figure}

\noindent Fig.~\ref{Fig.2} shows the \jb relation for 235 galaxies (including the Eridanus and background galaxies from the current work), gathered from studies which have accurately measured their $j_{\mathrm{b}}$ and $M_{\mathrm{b}}$ values. The sample includes 16 THINGS galaxies \citep{things08}, for which the $j_{\mathrm{b}}$ and $M_{\mathrm{b}}$ measurements where made by \citep{obreschkow14} [hereafter \citetalias{obreschkow14}], 14 dwarf galaxies from the LITTLE THINGS survey \citep{littlethings12} for which the measurements were made by \citet{butler17}, 11 void dwarf galaxies from \citet{Kurapati18}, 38 late-type galaxies from the Virgo cluster from \citet{Li20}, 114 late-type galaxies from the WHISP sample \citep{whisp02}, for which the measurements where made by \citetalias{Murugeshan20}, 33 galaxies in the Eridanus supergroup and 9 background galaxies, both from this study. Collectively, this forms the largest sample of late-type galaxies to be studied on the \jb parameter space. We fit\footnote{Fitting was performed using Hyper-Fit, an R package for fitting multi-dimensional data. See \citet{Robotham15} for more details.} a linear regression to the relation (accounting for the errors on $j_{\mathrm{b}}$ and $M_{\mathrm{b}}$) for all galaxies (excluding the Eridanus sample) and find a best fitting line with slope $0.55 \pm 0.02$ and intercept $-2.65 \pm 0.17$ (solid black line in Fig.~\ref{Fig.2}) with an intrinsic scatter of 0.16 dex. We also fit a line to the Eridanus galaxies alone and find a best fitting line with slope $0.57 \pm 0.05$, intercept $-3.06 \pm 0.45$ and an intrinsic scatter of 0.10 dex. 

We perform an analysis of covariance (ANCOVA) hypothesis test to check if the slopes and intercepts of the \jb relation for all galaxies and the Eridanus sample are statistically different. We find that the slopes of the two samples are not significantly different ($p$-value = 0.85) at an $\alpha = 0.05$ significance level. We then examined if the intercepts of the Eridanus sample and all other galaxies are statistically different and find that they are ($p$-value = 0). In Section~\ref{sec:Discussion} we discuss in more detail if this effect is simply due to selection biases associated with the different samples or driven by the environment.

\subsection{The \fatm relation for the Eridanus galaxies}
\label{Subsec:fatm-q relation}

\begin{figure}
\hspace{-0.5cm}
    \includegraphics[width=9cm,height=7cm]{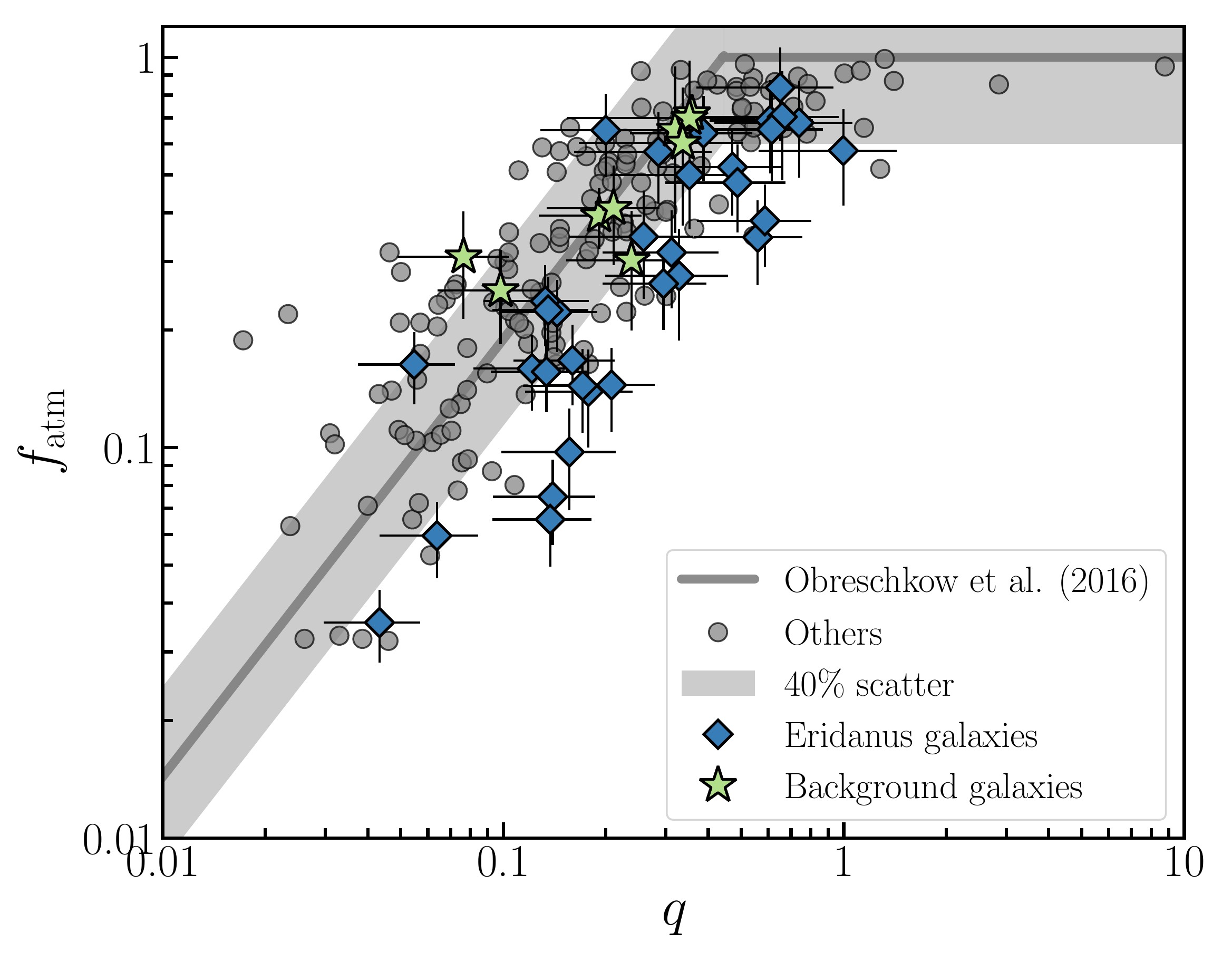}
    \caption{The relationship between the global disc stability parameter ($q$) and the atomic gas fraction ($f_{\mathrm{atm}}$) of galaxies. Blue diamonds are the 33 Eridanus galaxies. The green stars are the background galaxies. The grey circles represent galaxies from previous studies exploring the \fatm parameter space. The light grey region is defined in \citetalias{obreschkow16} as the 40\% empirical intrinsic scatter in the \h1 velocity dispersion ($\sigma_{\textrm{\h1}}$) of galaxies.} 
    \label{Fig.3}
\end{figure}

\noindent Fig.~\ref{Fig.3} shows the distribution of the Eridanus galaxies (blue diamonds) in the \fatm parameter space. We also plot the background galaxies (green stars) in order to compare their properties with the Eridanus sample. In addition, we also plot galaxies from relevant previous studies that have measured the $f_{\mathrm{atm}}$ and $q$ values for galaxies residing in low-density environments (grey circles in Fig.~\ref{Fig.3}). This includes the 16 late-type galaxies from the THINGS sample (\citetalias{obreschkow16}), 14 dwarf galaxies from LITTLE THINGS \citep{butler17}, 13 \h1-excess galaxies from the HIX survey \citep{lutz17}, 6 \h1-deficient galaxies from \citet{murugeshan2019} and 114 late-type galaxies from the WHISP sample from \citetalias{Murugeshan20}. The vast majority of the galaxies in this large sample follow the \fatm relation consistently, including those that are conventionally categorised as \h1-excess and \h1-deficient. However, we find that most of the galaxies in the Eridanus sub-group deviate from the relation, where their observed $f_{\mathrm{atm}}$ values are significantly lower for their $q$ values. This may be an indication that the Eridanus galaxies are subject to environmental effects such as tidal and ram-pressure stripping of the \h1 gas. A detailed discussion pertaining to the likely environmental processes that may be affecting the \h1 gas of galaxies in the Eridanus sub-group is made in Section~\ref{sec:Discussion}.

In contrast, the background galaxies which are in the field behind the Eridanus supergroup are observed to follow the \fatm relation consistently (see green stars in Fig.~\ref{Fig.3}), as they are not likely to be affected severely by environmental processes. This result shows the stark contrast seen between the behaviour of galaxies residing in high- and low-density environments on the \fatm parameter space. 

The deviation of galaxies from the \fatm relation driven by environmental effects can be used to quantify the level of \h1 deficiency in the galaxies (see \citealt{Stevens18,Stevens19};~\citealt{Li20};~\citealt{Cortese21}). Galaxies that have significantly lower $f_{\mathrm{atm}}$ for their $q$ are those systems whose \h1 gas has likely been stripped by environmental processes (e.g., \citealt{Li20}). On the other hand, galaxies that have significantly higher $f_{\mathrm{atm}}$ values for their $q$ are those systems that have possibly accreted gas from an interaction, and the \h1 gas has not yet settled into an equilibrium disc (see e.g.,~\citealt{Wang18};~\citetalias{Murugeshan20}). In Section~\ref{Subsec:HI-deficiency}, we will use the \fatm relation as a diagnostic plot to infer the \h1 deficiency among the Eridanus galaxies, by comparing their theoretically expected $f_{\mathrm{atm}}$ values with their observed $f_{\mathrm{atm}}$ values. This is a more physically motivated way to define \h1 deficiency driven by environmental processes, based on the disc stability of galaxies.

\subsection{Effects of the local environment}
\label{Subsec:Env. effect}
\noindent In this section we present results and discuss the effects of the local environment on the $f_{\mathrm{atm}}$ and $q$ values of the Eridanus galaxies.
We probe the local density of the environment using the projected nearest-neighbour density metric ($\Sigma_2$ [Mpc$^{-2}$]), following the justifications made by \citetalias{Murugeshan20}. The $\Sigma_2$ metric is defined as $\Sigma_2 = 2/\pi D^2$, where $D$ is the projected distance to the 2nd nearest-neighbour within $\pm ~500$~\kms. We compute the  $\Sigma_2$ values using the 2MASS Redshift Survey (2MRS) catalogue \citep{huchra12}, which contains spectroscopic redshifts for over 43,500 galaxies with $K_s \leq 11.75$ mag and |b| > 5 $\degr$. The 2MRS survey is complete to 97.6\% within these limits. As the 2MRS magnitude limits are based on $K_s$-band magnitudes, it is likely to be more sensitive to galaxies that are redder. To avoid any potential biases, we make the 2MRS catalogue volume-limited with the following two steps:
\begin{enumerate}
    \item We first make a velocity cut to the original 2MRS catalogue, by selecting galaxies within the velocity range 200 -- 12000 \kms. 
    \item Following this, we add an absolute $K_{s}$-band magnitude cut to sources with $M_K < -23.45$, corresponding to the survey's limiting apparent magnitude of 11.75 mag at the highest velocity/redshift edge (12000 \kms). This makes the 2MRS sample volume-limited.
\end{enumerate}

\begin{figure}
\hspace{-0.5cm}
    \includegraphics[width=9.5cm,height=7.2cm]{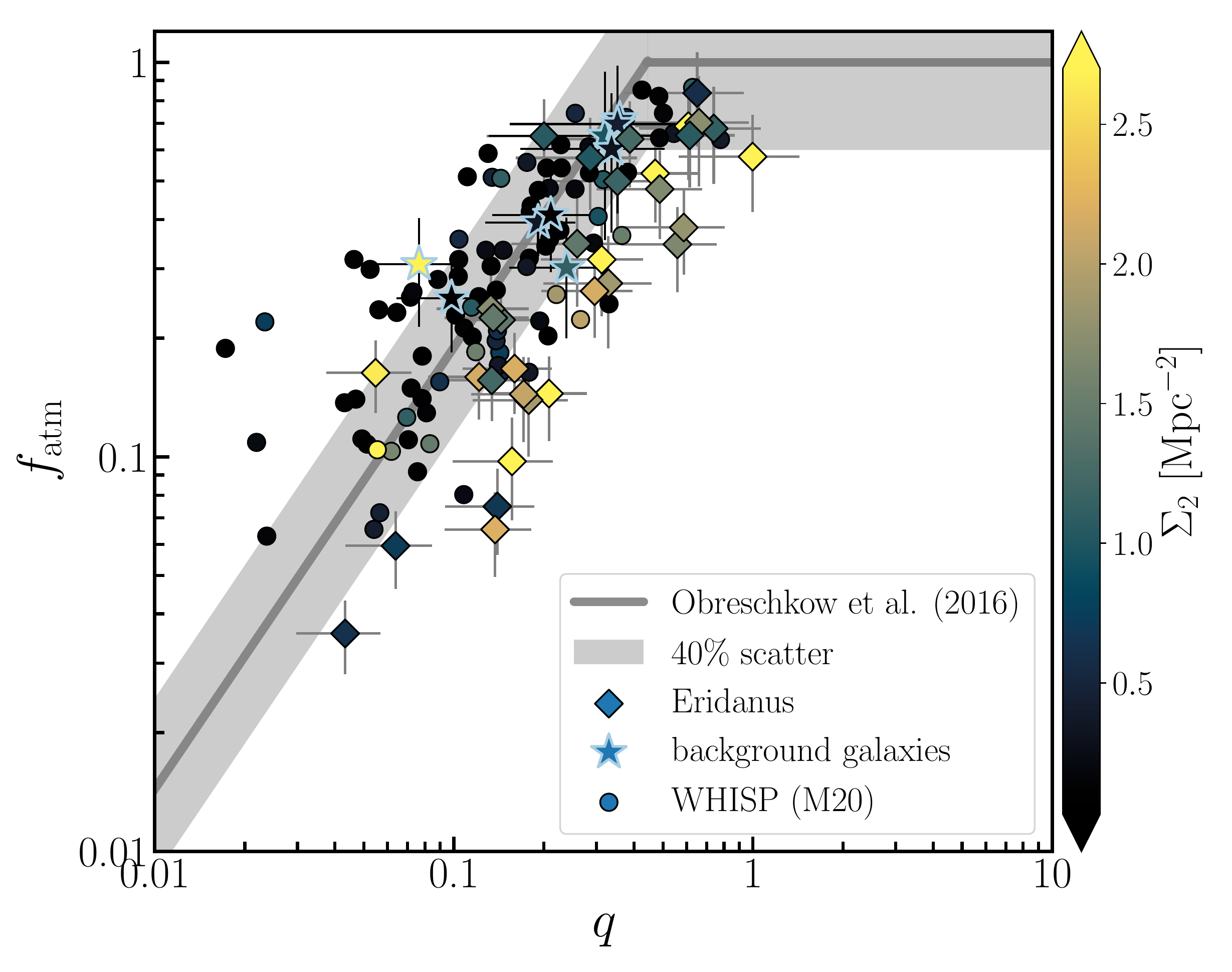}
    \caption{The $f_{\mathrm{atm}} - q $ relation for the 33 Eridanus galaxies (diamonds), color-coded by their environment density, $\Sigma_2$. For comparison, also plotted are WHISP galaxies (circles) from \citetalias{Murugeshan20}. We observe that the Eridanus galaxies are at comparatively higher environment densities compared to the WHISP galaxies. Also shown are the 9 background galaxies (stars), most of which are from low-density environments and follow the relation consistently.}
    \label{Fig.4}
\end{figure}

Fig.~\ref{Fig.4} shows the \fatm relation for the Eridanus and background galaxies, with their $\Sigma_2$ values shown in the colour bar. Also plotted are the WHISP galaxies from \citetalias{Murugeshan20}, most of which are from relatively isolated environments or from loose groups. We see that the $\Sigma_2$ values for the Eridanus galaxies are significantly higher than those of the WHISP galaxies. Furthermore, we find that galaxies residing in higher densities show a larger deviation from the \fatm relation compared to those in lower densities within the Eridanus supergroup. Additionally, in Section~\ref{Subsec:HI-deficiency}, where we explore the disc stability-based \h1 deficiency parameter, we show that galaxies residing in higher local densities are observed to have a larger \h1-deficiency (see Fig.~\ref{Fig.E1} in Appendix~\ref{appendix:Delta_f_qVsSigma2}). This result comes in support of the fact that galaxies with close neighbours in the Eridanus supergroup have possibly interacted previously and have lost a significant amount of their \h1 gas. In addition to the Eridanus galaxies, we also examine the behaviour of the background field galaxies in the \fatm parameter space (indicated by the stars in Fig.~\ref{Fig.4}). We see that all background galaxies (with the exception of one) are from relatively low-density environments compared to the Eridanus galaxies and follow the \fatm relation consistently.

We also visually examined both the \h1 and optical morphology of the Eridanus galaxies in order to look for any signs of environmental effects. The \h1 gas is easily perturbed due to environmental processes such as tidal interactions and ram-pressure (\citealt{hibbard96}). Each of these processes leave different signatures on the \h1 and optical morphology of galaxies. Tidal interactions typically tend to affect both the \h1 gas and the stellar discs, where distortions and asymmetries in both the components can be observed (see for example~\citealt{Putman98};~\citealt{MartinezDelgado10}). Ram-pressure on the other hand usually only affects the \h1 gas, while the stellar component remains fairly unperturbed (\citealt{cayatte94};~\citealt{vollmer01}). 

Upon examination, we find that most galaxies in the Eridanus sample show a disturbed \h1 morphology. Some galaxies are observed to have an asymmetric \h1 distribution without any sign of perturbation to their stellar discs, while others in the sample show both disturbed \h1 and stellar discs, hinting that a number of processes observed in high-density environments such as tidal interactions, minor mergers and ram-pressure effects may be prevalent in the Eridanus supergroup. This observation is consistent with a previous work that quantified the degree of lopsidedness of the \h1 gas among the Eridanus galaxies compared to field galaxies using Fourier component analysis of their \h1 maps (see \citealt{Angiras06}). For a detailed discussion on the \h1, optical and kinematic morphologies of individual galaxies in the Eridanus sample, we refer the reader to \citet{For21}. However, in the context of this work, we discuss the morphologies of a few select galaxies that show signatures that are indicative of tidal and/or ram-pressure effects.

We first discuss candidates in our sample that show clear signs of ongoing tidal interactions, which is evidenced in their \h1 and optical morphology, as in the case of NGC 1359 and NGC 1385. NGC 1359 is in the process of a minor merger (\citealt{Omar05a}). The \h1 morphology of NGC 1359 is observed to be disturbed, with tidal plumes and tails associated with the galaxy. The optical image of the galaxy also shows a distorted and irregular stellar disc, indicative of an ongoing merger. NGC 1385 also shows signs of interaction. We observe an extended tidal tail to the south-west of the galaxy. The SFRs of both of these galaxies is significantly enhanced compared to the rest of the Eridanus galaxies. Tidal interactions and mergers among galaxies have been observed to enhance the SFR of the interacting galaxies (\citealt{Kennicutt87};~\citealt{Nikolic04};~\citealt{ellison11}). Additionally, such gas-rich interactions and minor mergers may lower the $q$ values of the galaxies and increase their $f_{\mathrm{atm}}$ values (see~\citealt{ellison18};~\citetalias{Murugeshan20}). We discuss this in more detail in Section~\ref{subsec:star formation properties} for these two galaxies, as a case study that demonstrates this behaviour.

\begin{figure*}
\hspace{-0.4cm}
    \includegraphics[width=17cm,height=7cm]{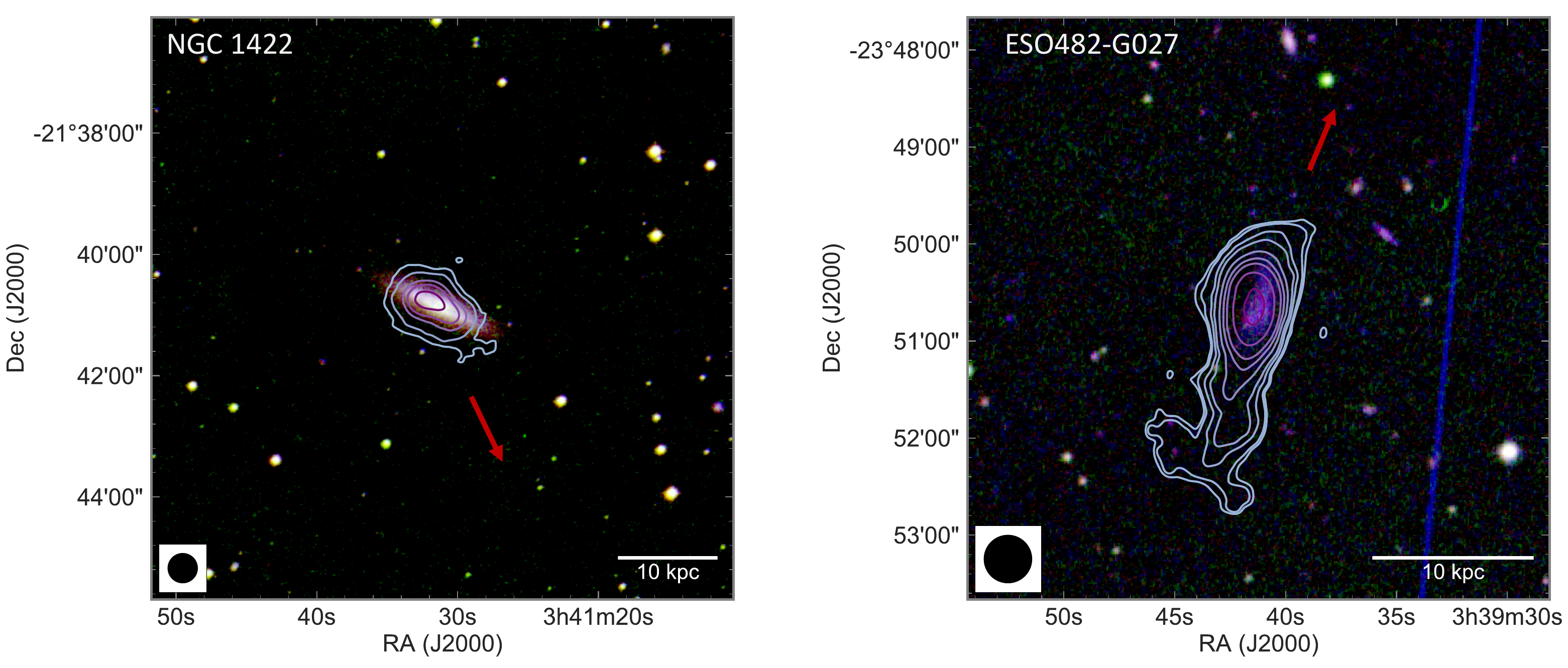}
    \caption{Two galaxies -- NGC 1422 and ESO482-G027 -- in the Eridanus sub-group exhibiting a truncated and lopsided \h1 distribution, respectively. The \h1 contours of the galaxies have been overlaid on top of their composite DSS optical images. The red arrows point in the direction of the large group central elliptical galaxy NGC 1395. In the case of NGC 1422, we find that the stellar disc does not show any signs of perturbation, while the \h1 gas distribution is severly truncated. This is a typical signature of galaxies affected by ram-pressure. ESO482-G027 on the other hand is observed to have an extended tail pointing away from the group central galaxy NGC 1395. In addition, we see that the \h1 contours are compressed to the north-west side of the galaxy. Both these features allude to ram-pressure effects. The contours for NGC 1422 correspond to column densities of 0.7, 2, 5, 7 and $10 \times 10^{20}$ cm$^{-2}$. The contours levels for ESO482-G027 are at 0.7, 0.9, 1.5, 2, 3, 4, 5, 7 and $10 \times 10^{20}$ cm$^{-2}$. For detailed maps of all the Eridanus and background galaxies, we refer the reader to the main catalogue paper by \citet{For21}}
    \label{Fig.5}
\end{figure*}

A few galaxies in the Eridanus group show asymmetric \h1 morphologies (for example ESO482-G027, IC 1952, ESO482-G013) and/or exhibit truncated \h1 discs as in the case of NGC 1422, ESO548-G070 and IC 1953, without exhibiting any visible perturbations in their stellar discs. In Fig.~\ref{Fig.5} we show the \h1 contours of NGC 1422 and ESO482-G027 overlaid on top of their composite DSS optical images. It is clearly seen that NGC 1422 exhibits a truncated \h1 distribution compared to its stellar disc. ESO482-G027 is observed to have an extended tail-like feature towards the south-east end, with the \h1 contours compressed towards the north-west end, and the tail pointing away from the large group central galaxy NGC 1395.

This suggests that processes other than tidal interactions such as ram-pressure stripping may be prevalent in the Eridanus sub-group. In order to study the extent of influence of ram-pressure on the Eridanus galaxies, we compute the degree of ram-pressure exerted by the intra-group medium (IGM) of the Eridanus sub-group using the traditional formulation prescribed by \citet{gunn72}. The condition for ram-pressure stripping can be given as 
\begin{equation}
    \rho_{\mathrm{IGM}} v_{\mathrm{gal}}^{2} \geq 2 \pi G \Sigma_{\star} \Sigma_{\textrm{\h1}},
\end{equation}
where $\rho_{\mathrm{IGM}}$  is the IGM density at the projected distance of the individual galaxies from the group central galaxy NGC 1395. $v_{\mathrm{gal}}$ is the relative velocity of the galaxies in the Eridanus sub-group with respect to the IGM, which can be approximated by the dispersion velocity of the Eridanus sub-group. $\Sigma_{\star}$ and $\Sigma_{\textrm{\h1}}$ denote the stellar and gas surface densities respectively, and $G$ is the gravitational constant. As mentioned previously, diffuse X-ray emission surrounding NGC 1395 has been observed by previous studies (\citealt{Trinchieri86};~\citealt{Omar05b};~\citealt{Fukazawa04}). \citet{Fukazawa04} make use of the archival ASCA X-ray data to model the diffuse X-ray surrounding the large central elliptical NGC 1395 in the Eridanus sub-group. They fit a single-$\beta$ model by assuming a Raymond-Smith plasma \citep{Raymond77} corresponding to an energy of $\sim 0.76$ keV and a solar abundance of 0.25. We use the following relation to estimate the IGM density at the projected distance to each galaxy in the Eridanus sub-group,
\begin{equation}
    \rho_{\mathrm{IGM}}(r)=\rho_{\mathrm{IGM},0}\left[ 1+\left(\frac{r}{r_{c}}\right)^{2}\right]^{-3 \beta / 2},
\end{equation}
here, $\rho_{\mathrm{IGM},0} = 10.2 \times 10^{-3}$ cm$^{-3}$ is the central IGM density, $r_{c} = 3.7$ kpc is the core radius and $\beta = 0.38$ is the exponent from the single-$\beta$ model fit to the X-ray data by \citet{Fukazawa04}. The dispersion velocity of the Eridanus sub-group differs significantly between previous works. \citet{Brough06} use a friends-of-friends algorithm and estimate the value to be $\sim 156$~\kms, while \citet{Tully15} find the group velocity to be $\sim 228$ \kms. We use both these values to approximate our uncertainty on the level of ram-pressure experienced by galaxies in the Eridanus sub-group. We find that assuming $v_{\mathrm{gal}} = 156$~\kms ~for the group, only 7 of the 27 galaxies in the Eridanus sub-group may be affected by ram-pressure. However, when assuming $v_{\mathrm{gal}} = 228$~\kms, we find that as many as 13 galaxies ($\sim$ 48\%) may be affected by ram-pressure to column densities N$_{\textrm{\h1}} \leq 1.8 \times 10^{20}$ cm$^{-2}$. This may explain the curious case of some galaxies in the Eridanus supergroup exhibiting asymmetric \h1 morphologies and truncated \h1 discs (see Fig.~\ref{Fig.5}).

\subsection{\h1 deficiency in the Eridanus galaxies based on their disc stability}
\label{Subsec:HI-deficiency}

\noindent In the previous sections, we showed that the Eridanus galaxies deviate from the \fatm relation, with galaxies having consistently lower $f_{\mathrm{atm}}$ values for their $q$ values. We attribute this to environmental effects on the \h1 content of the Eridanus galaxies. In order to quantify the degree of this offset, we make use of the disc stability-based \h1 deficiency parameter ($\Delta f_q$) which was first briefly explored by \citet{Stevens18}, and later tested with empirical data by \citet{Li20}. $\Delta f_q$ can be defined as follows (Eq. 4 in \citealt{Li20}),
\begin{equation}
    \Delta f_{q}=\log _{10}\left(\min \left\{1,2.5 q^{1.12}\right\}\right)-\log _{10}\left(f_{\mathrm{atm}}\right).
\end{equation}
The more deficient in \h1 a galaxy is, the higher the value of $\Delta f_q$. According to the model in \citetalias{obreschkow16}, the atomic gas fraction ($f_{\mathrm{atm}}$) in rotationally supported exponential disc galaxies depends on their $q$ values. Unperturbed late-type galaxies will follow the relation consistently as shown by previous studies (\citealt{lutz17};~\citealt{murugeshan2019}), however, galaxies that are perturbed by environmental processes will deviate from the \fatm relation (\citealt{Stevens18};~\citealt{Li20};~\citetalias{Murugeshan20}). The $\Delta f_{q}$ parameter simply measures the difference in the expected $f_{\mathrm{atm}}$ value (for a given $q$) and the observed $f_{\mathrm{atm}}$ value of the galaxies. 

\begin{figure}
    \includegraphics[width=8cm,height=6.5cm]{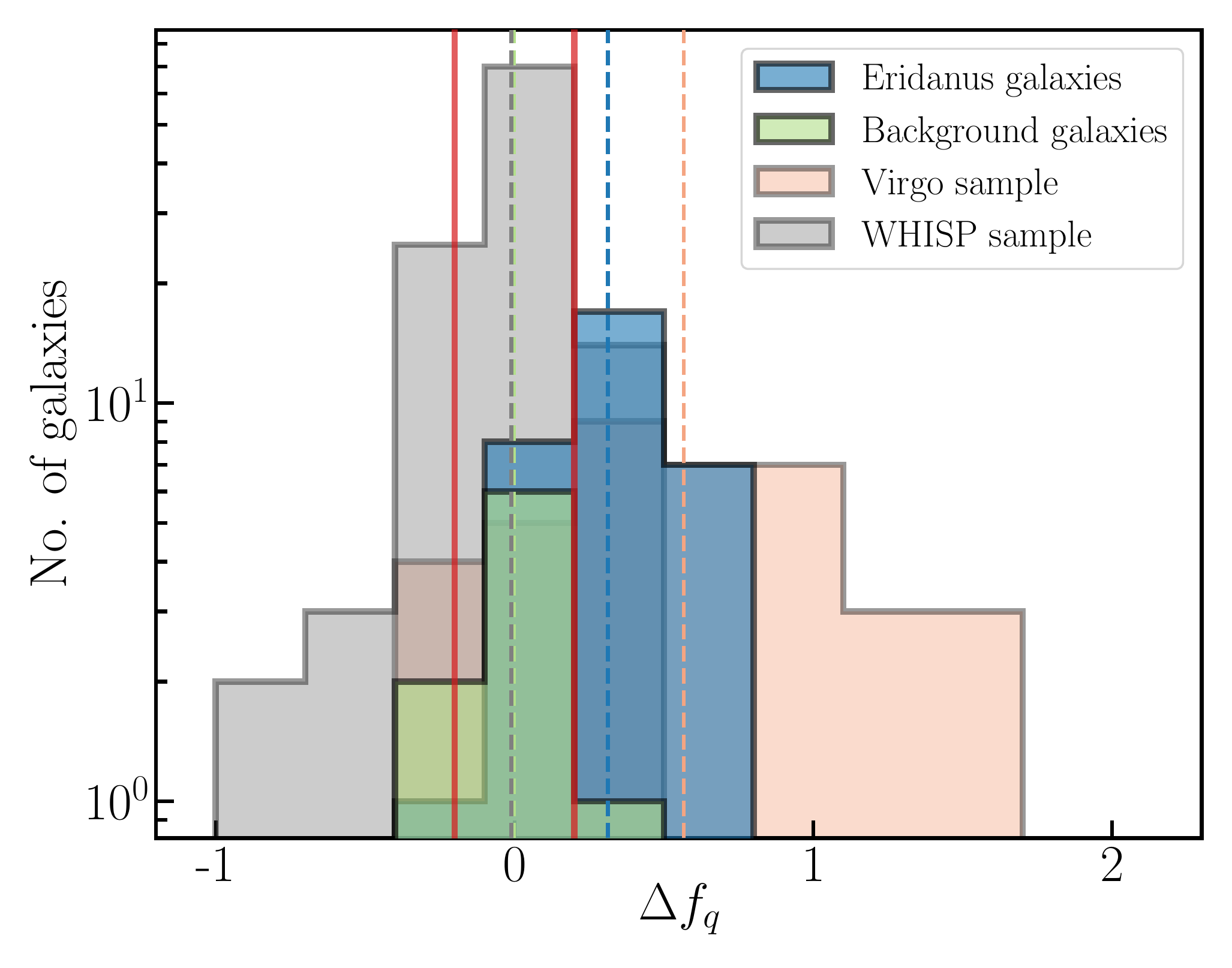}
    \caption{Distribution of the \h1 deficiency parameter, $\Delta f_q$, for the different samples. The red solid lines demarcate the \h1 -deficient, -normal and -excess regions and is equal to 0.2 dex. The blue, green, orange and grey dashed lines show the mean (with standard deviation included within brackets) $\Delta f_q$ value for the Eridanus 0.31 (0.19), background $-0.0025 (0.1476)$, WHISP $-0.011 (0.225)$ from \citetalias{Murugeshan20} and the Virgo samples 0.57 (0.53) from \citet{Li20}, respectively.} 
    \label{Fig.6}
\end{figure}

Fig.~\ref{Fig.6} shows the distribution of the $\Delta f_{q}$ values for the Eridanus and background galaxies. In addition, we also plot the WHISP galaxies from \citetalias{Murugeshan20}, which are all late-type galaxies from low-density environments, for comparison. In this work, we define galaxies as \h1-deficient if their $\Delta f_{q} > 0.2$ and \h1-excess if their $\Delta f_{q} < -0.2$ (represented by the red vertical lines in Fig.~\ref{Fig.6}). This is based on the observation that the intrinsic dispersion of galaxies following the \fatm relation is $\sim 0.2$ dex for a large sample of normal star-forming late-type galaxies (\citetalias{obreschkow16};~\citetalias{Murugeshan20}).

The mean $\Delta f_{q}$ value (with standard deviation included in brackets) for the Eridanus galaxies is found to be 0.31 (0.19). In comparison, the mean $\Delta f_{q}$ value for the background and the WHISP galaxies is found to be $-0.0025 (0.1476)$ and $-0.011 (0.225)$. On the other extreme, we use the $\Delta f_{q}$ values published by \citet{Li20} for the Virgo cluster galaxies and find a mean $\Delta f_{q}$ value of 0.57 (0.53), which is unsurprisingly higher than the Eridanus galaxies. Clearly, we see that the Eridanus galaxies are more \h1 deficient compared to galaxies from low-density environments, however they are not as deficient as galaxies in the Virgo cluster on average. 

We perform several pairwise two-sample KS tests to examine if the distribution of the $\Delta f_{q}$ values for the Eridanus, background and the Virgo galaxies are significantly different from the WHISP control sample. The KS test statistics are listed in Table.~\ref{tab:KS-test-Delta_fq} for the different samples. We find that both the Eridanus and Virgo samples are significantly different (at a significance level of 5\% or $\alpha = 0.05$) compared to the WHISP sample, while the background galaxies are not. This shows that the Eridanus galaxies are indeed significantly more \h1-deficient compared to the normal star-forming galaxies in the WHISP sample. Additionally, in Fig.~\ref{Fig.E1}
in Appendix~\ref{appendix:Delta_f_qVsSigma2}, we show that the $\Delta f_{q}$ values of both the Eridanus and the WHISP galaxies are observed to increase with increasing local environment density (represented by $\Sigma_2$), consistent with a number of previous studies that have highlighted the effects of the dense environment on the \h1 gas fractions of galaxies (e.g \citealt{giovanelli85};~\citealt{solanes01};~\citealt{denes14}).

While traditional \h1 deficiency indicators are calibrated based on empirical samples of isolated late-type galaxies, the $\Delta f_{q}$ parameter does not require any empirical sample for its calibration. It is purely based on the level of expected \h1 saturation in late-type disc galaxies for a given disc stability. It is therefore very straightforward to incorporate this parameter in both observational and simulation studies. In order to examine how the $\Delta f_{q}$ parameter compares with more traditional \h1 deficiency parameters widely used in the literature, we plot the $\Delta f_{q}$ values of the Eridanus galaxies against other \h1 deficiency indicators in Appendix~\ref{appendix:HI-deficiency}, and find a good positive correlation in general, albeit with large scatter.

In addition to its use as an \h1-deficiency indicator, the $\Delta f_{q}$ parameter has been shown to be useful in other ways. For instance, \citet{Stevens18} in their semi-analytic study of the \fatm parameter space introduced a parameter similar to $\Delta f_{q}$ and used it to predict the absolute \h1 mass of galaxies that deviate from the \fatm relation due to environmental effects (see Fig. 8 in their paper). Such predictions are very useful to design targeted \h1 surveys to detect galaxies that have a low \h1 gas fraction ($\log_{10}(M_{\textrm{\h1}})$[M$_{\odot}] \leq 8$).

\begin{table}
    \caption{Two-sample KS test statistics for the different samples comparing their $\Delta f_{q}$ distributions with that of the WHISP control sample. $\Delta f_{q}$ values listed are the means for each sample, with the standard deviation given within brackets.}
    \label{tab:KS-test-Delta_fq}
    \begin{tabular}{lcccc}       
    \hline \hline
        Sample & Size & $\Delta f_{q}$  & $D$ statistic & $p$-value  \\
            &  &  &  &  \\
    \hline         
        Eridanus & 33    &  0.31 (0.19)     & 0.65  & $2.13 \times 10^{-10}$  \\
        Background  & 9  &  $-0.0025 (0.1476)$     & 0.25  & 0.6  \\
        Virgo   & 38     &   0.57 (0.53)       & 0.66  & $1.03 \times 10^{-11}$ \\
    \hline
    WHISP   &   114    &    $-0.011 (0.225)$   &   --    & --  \\
    \hline
    \end{tabular}
\end{table}

\subsection{Star formation properties}
\label{subsec:star formation properties}
\noindent In this section we present results pertaining to the star formation (SF) properties of the Eridanus and background galaxies, such as their specific star formation rate (sSFR = SFR/$M_{\star}$) and \h1 based star formation efficiency (SFE = SFR/$M_{\textrm{\h1}}$), and compare them with the SF properties of a sample of late-type galaxies in low-density environments from the WHISP survey (\citetalias{Murugeshan20}). The objective of this analysis is to understand if and how the environment and internal properties such as mass and AM are playing a role in regulating the star formation in the sample galaxies.

We make use of both \textit{GALEX} \citep{Martin05} and \textit{WISE} \citep{Wright10} FUV and MIR photometries respectively, to estimate the SFR of the galaxies. The dust un-attenuated part of the total SFR is measured using \textit{GALEX} FUV luminosities, while the dust attenuated part is measured using \textit{WISE} W4 luminosities. For more details on the methods employed to measure the SFR for the Eridanus and background galaxies, we refer the reader to \citet{Wang17} and \citet{For21}. We were able to derive robust SFR values for 26 of the 33 Eridanus galaxies and 5 of the 9 background galaxies. \citetalias{Murugeshan20} were only able to measure the SFR for 86 of the 114 WHISP galaxies in their sample. The sample size for each of the three samples is therefore reduced to the above mentioned numbers for the rest of the analysis concerning their star formation properties. 

\begin{figure*}
\hspace{-0.8cm}
\begin{subfigure}{.5\textwidth}
  \centering
  \includegraphics[width=8.5cm,height=7.2cm]{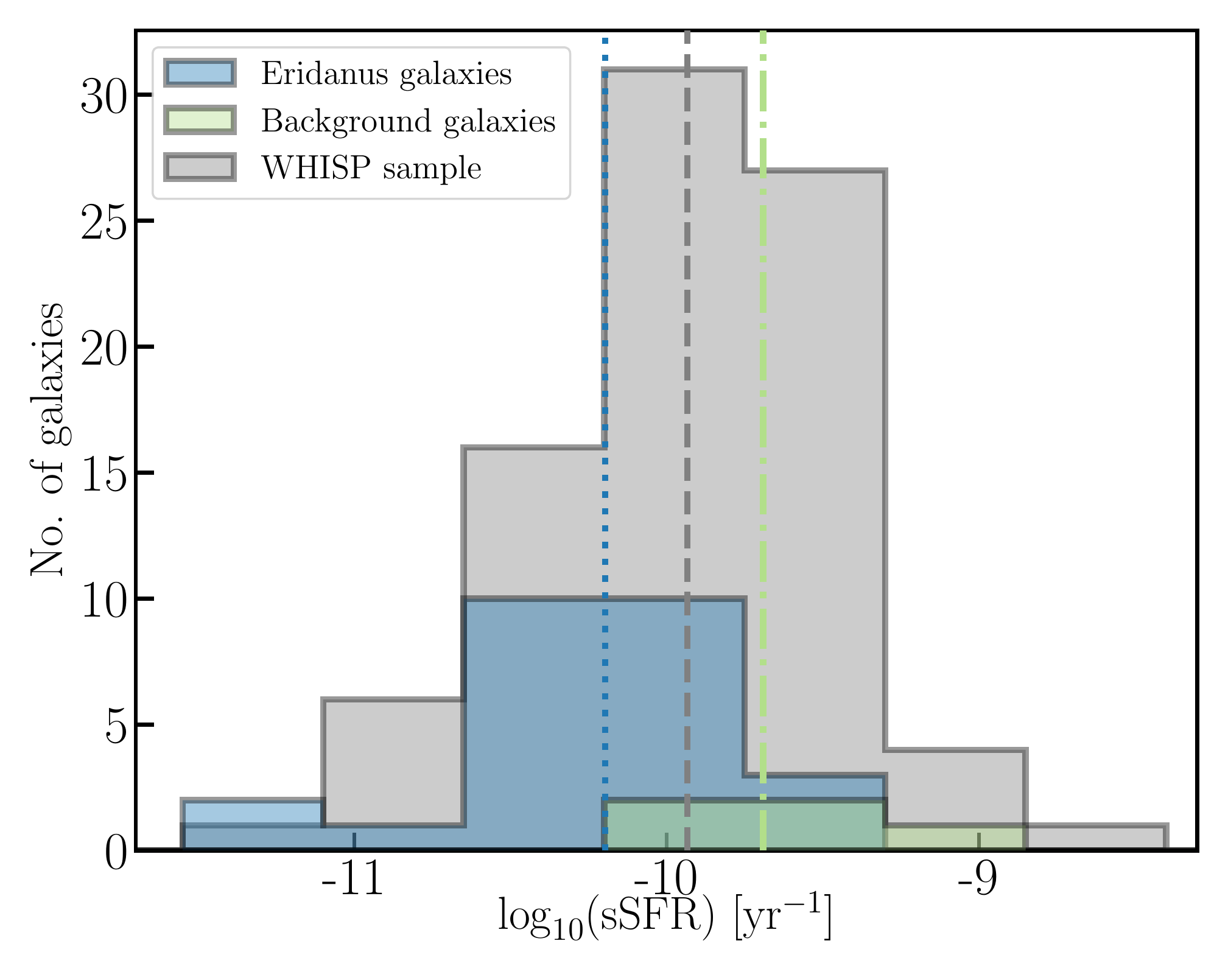}
  \caption{}
  \label{Fig.7.a}
\end{subfigure}%
\begin{subfigure}{.5\textwidth}
  \hspace{-0.3cm}
  \includegraphics[width=9.5cm,height=7.2cm]{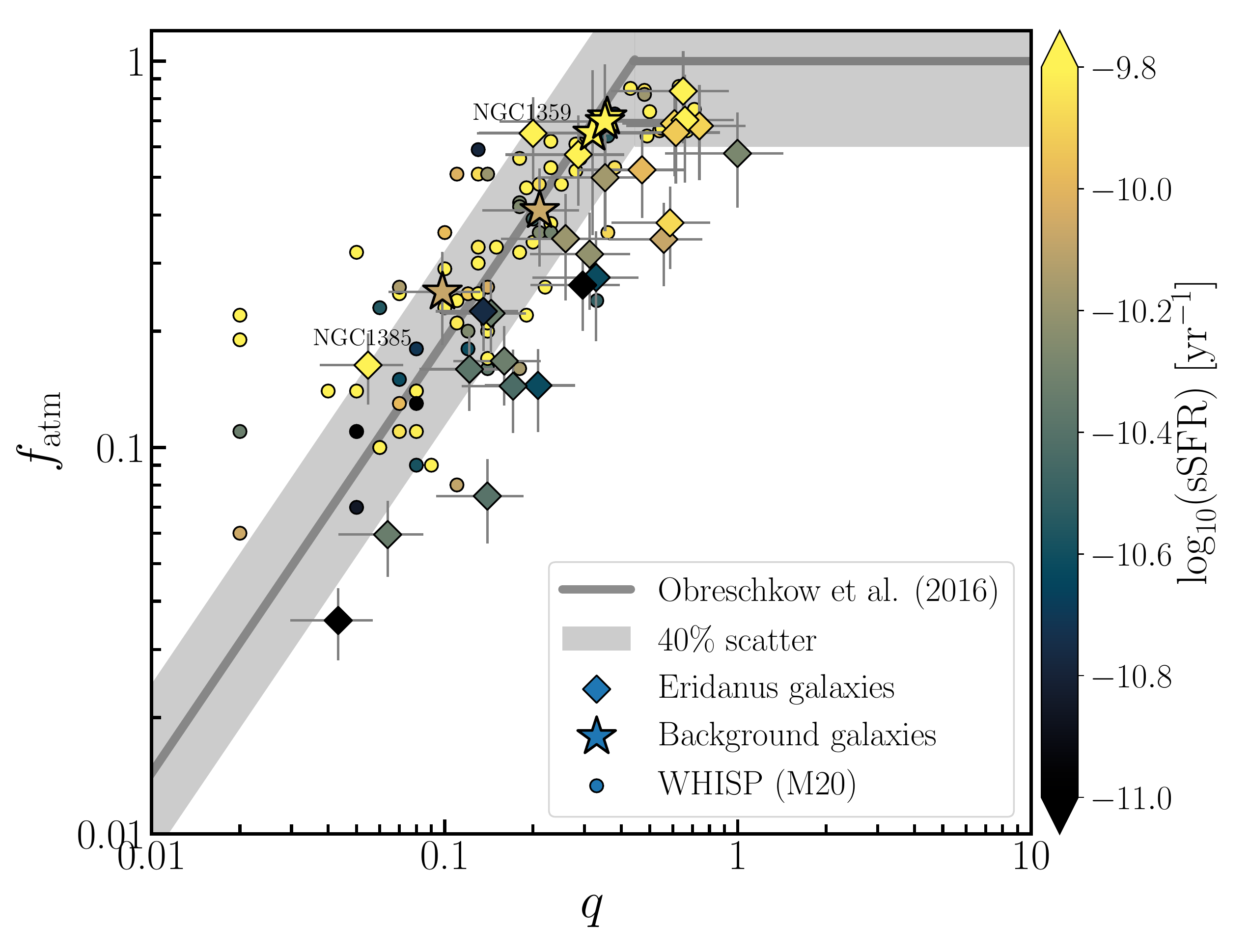}
  \caption{}
  \label{Fig.7.b}
\end{subfigure}
\caption{(a) Distribution of the sSFR of the Eridanus, background and WHISP samples. The mean value of the sSFR for the Eridanus, background and WHISP samples are represented by the blue dotted, green dash-dot and grey dashed lines respectively. 
(b) The $f_{\mathrm{atm}} - q -$ sSFR parameter space for the 33 Eridanus galaxies (diamonds) and background galaxies (stars) color coded by their $\log_{10}$(sSFR) values. A general trend of increasing sSFR with increasing $q$ is observed for the Eridanus galaxies. For comparison, also plotted are the 86 WHISP galaxies (circles) from \citetalias{Murugeshan20}. Also highlighted are two galaxies, NGC 1385 and NGC 1359 in the Eridanus sub-group, each of which are currently interacting with their companions. Such interacting systems have been observed to have higher $f_{\mathrm{atm}}$ values for their $q$, as they are actively accreting \h1 gas from their companions. This gas accretion results in the funneling of fresh gas to their centers to form stars hence enhancing their SFR (e.g. see~\citealt{ellison10};~\citealt{ellison18}).}
\label{Fig.7}
\end{figure*}

Fig.~\ref{Fig.7.a} shows the distribution of the sSFR of the Eridanus, background and the WHISP galaxies.
We find that the mean sSFR (with standard deviation in brackets) of the Eridanus galaxies is $\log_{10}$(sSFR) [yr$^{-1}$] $\approx -10.20 (0.45)$, that for the background galaxies is $\log_{10}$(sSFR) [yr$^{-1}$] $\approx -9.69 (0.34)$, while the mean sSFR of the WHISP sample is $\log_{10}$(sSFR) [yr$^{-1}$] $\approx -9.93 (0.49)$. In addition, we also perform a two sample KS test to compare the sSFR distributions of the Eridanus and background galaxies with the WHISP control sample, and check if the two sample distributions are different from the control sample. Table.~\ref{tab:KS-test-sSFR} shows the KS test results, and we find that the distribution of the sSFR values of the Eridanus galaxies differs significantly from the WHISP sample. In contrast, the background galaxies are not significantly different from the WHISP sample.

 \begin{table}
    \caption{KS test results for the Eridanus and background galaxy samples when comparing their sSFR distributions to the WHISP control sample (N=86).}
    \label{tab:KS-test-sSFR}
    \begin{tabular}{lccc}       
    \hline \hline
        Sample & Size & $D$ statistic & $p$-value  \\
             &  &  &  \\
    \hline         
        Eridanus    & 26    & 0.38  & 0.0046  \\
        Background  & 5     & 0.41  & 0.3  \\
    \hline
    \end{tabular}
\end{table}

The sSFR value of the Eridanus galaxies on average being lower compared to both the WHISP and the background galaxies possibly hints at environmental effects influencing the star formation of the Eridanus galaxies. It has been observed that galaxies in low-density environments on average have higher sSFR compared to galaxies in high-density environments (see for example~\citealt{Moorman16}). Additionally, it has been shown that the sSFR of galaxies positively correlates with their \h1 gas fraction (\citealt{Fumagalli08};~\citealt{Zhou18}). As shown in the previous sections, the \h1 gas fraction (or $f_{\mathrm{atm}}$) among the Eridanus galaxies is significantly lower for their $q$ values. The \h1 deficiency is possibly driven by environmental processes such as tidal interactions and ram-pressure. As such, we can associate the observed low sSFR among the Eridanus galaxies to environmental processes affecting their \h1 gas fractions. Within the Eridanus sample, we observe that the low-mass galaxies have a higher sSFR compared to the more massive (and low AM) systems. This is consistent with previous studies, and indicates that the star formation in low-mass systems contribute more to their growth compared to high-mass (and low AM) systems (see for example~\citealt{Vulcani10};~\citealt{Zhou18}), even in high-density environments such as the Eridanus sub-group.

\citetalias{Murugeshan20} found that galaxies that are currently interacting and actively accreting \h1 gas from their smaller dwarf companions showed enhanced SFR. In addition to this, they find that the $f_{\mathrm{atm}}$ values for such systems are higher, while their $q$ values are reduced due to an overall reduction in $j_{\mathrm{b}}$. In the Eridanus sample, NGC 1359 and NGC 1385 are currently undergoing a merger with their companions, and exhibiting clear signs in the form of tidal tails and debris, which are associated with the two galaxies. It is also observed that the sSFR of these two galaxies is enhanced, while the $f_{\mathrm{atm}}$ values are marginally higher than what is expected for their $q$ (see Fig.~\ref{Fig.7.b}). These are the only two galaxies in the Eridanus sample that have $\Delta f_q < 0$, indicating that they are \h1-normal (or even marginally \h1-excess) for their integrated disc stability. The enhanced $f_{\mathrm{atm}}$ and sSFR may be associated with the accretion of fresh gas during the process of merging leading to increased star formation, consistent with the observations made by \citet{ellison18} and \citetalias{Murugeshan20}.  

\begin{figure*}
\hspace{-0.8cm}
\begin{subfigure}{.5\textwidth}
  \centering
  \includegraphics[width=8.5cm,height=7.2cm]{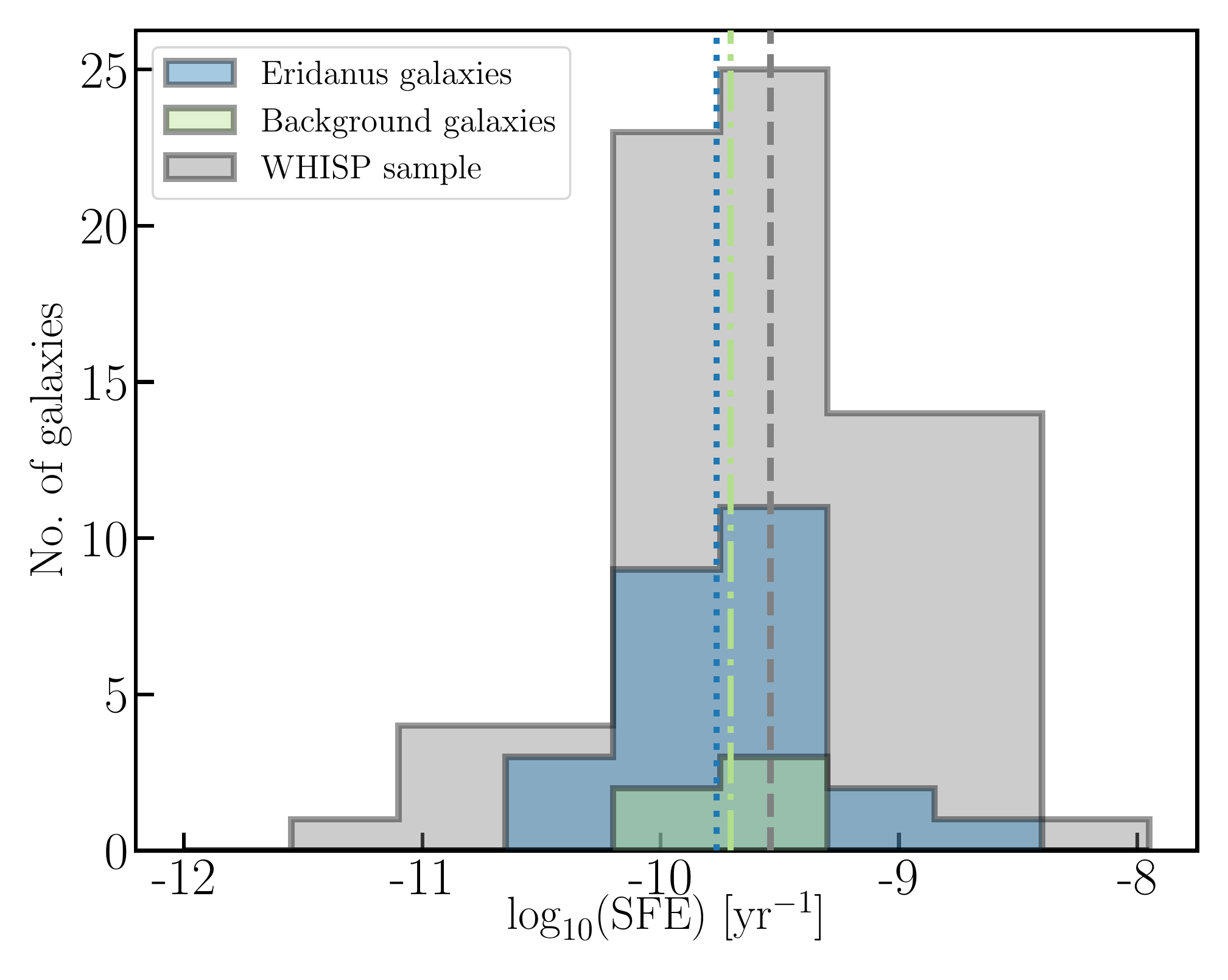}
  \caption{}
  \label{Fig.8.a}
\end{subfigure}%
\begin{subfigure}{.5\textwidth}
  \hspace{-0.3cm}
  \includegraphics[width=9.5cm,height=7.2cm]{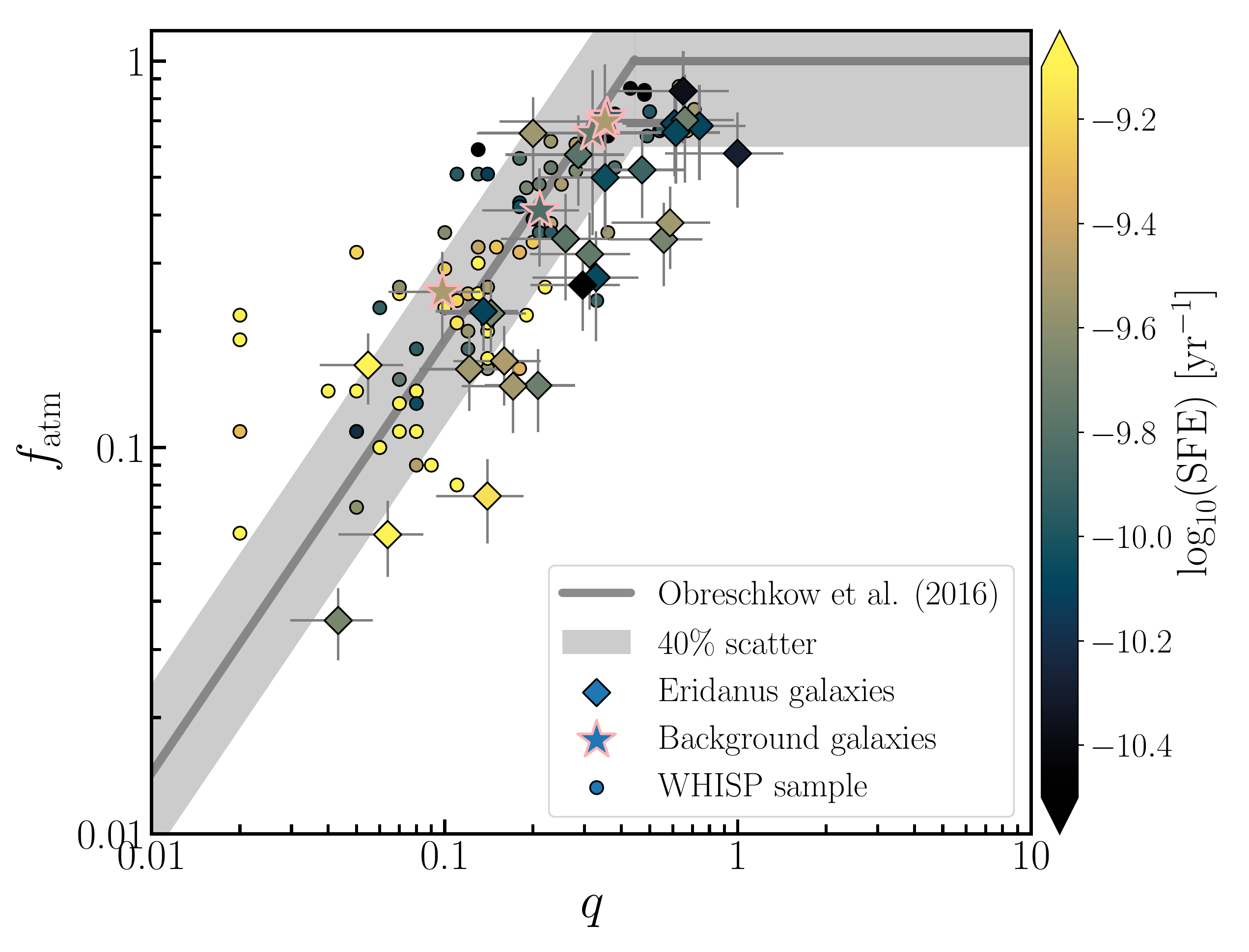}
  \caption{}
  \label{Fig.8.b}
\end{subfigure}
\caption{(a) Distribution of the SFE of the Eridanus, background and WHISP sample, with their mean values shown by the dotted blue, dash-dot green and dashed grey lines respectively. The distribution of the SFE of the three samples are consistent with each other and are observed to have a similar mean SFE (b) The $f_{\mathrm{atm}} - q -$ SFE parameter space for the Eridanus (diamonds) and background galaxies (stars), color coded by their $\log_{10}$(SFE) values. Normal star-forming late-type galaxies from the WHISP sample (\citetalias{Murugeshan20}) are also shown (circles). A general trend of decreasing SFE with increasing $q$ is observed, indicating that galaxies with a higher disc stability are inefficient in forming stars from their \h1 reservoir, consequently maintaining a higher \h1 gas fraction.}
\label{Fig.8}
\end{figure*}

In Fig.~\ref{Fig.8.a}, we show the distribution of the SFE of the Eridanus, background and WHISP galaxies. The mean (standard deviation within brackets) SFE of the Eridanus galaxies is found to be $\log_{10}$(SFE) [yr$^{-1}$] $\approx  -9.76 (0.41)$, while that of the background galaxies is $\log_{10}$(SFE) [yr$^{-1}$] $\approx -9.71 (0.18)$. In comparison the mean SFE of the WHISP sample is found to be $\log_{10}$(SFE) [yr$^{-1}$] $\approx -9.54 (0.60)$. It is interesting to note that the SFE of both the Eridanus and the background galaxies is comparable to the WHISP galaxies. This indicates that the SFE in galaxies may be regulated by the marginal disc stability model of late-type galaxies as postulated by \citet{Wong16}, where they find an average SFE of $\log_{10}$(SFE) [yr$^{-1}$] $\approx -9.65$ with a dispersion of 0.3 dex, for a large sample of galaxies (also see~\citealt{schiminovich10};~\citealt{Parkash18}). We find that the observed SFE of both the Eridanus and the WHISP galaxies are consistent with this model. However, \citet{Wong16} see a weak correlation between SFE and the surface brightness of gas and stars, which they attribute to the sAM of the galaxies. 

In this work, we suggest that the SFE in late-type star-forming galaxies is mainly driven by the integrated disc stability parameter $q \propto j_{\mathrm{b}}/M_{\mathrm{b}}$, where both baryonic mass and AM have the ability to regulate the disc stability (\citetalias{obreschkow16}). We see this effect in Fig.~\ref{Fig.8.b} for the different sample galaxies. We observe that galaxies with higher $q$ values are found to have low SFE and correspondingly high $f_{\mathrm{atm}}$ values, and galaxies with low $q$ values are found to have high SFE (and low $f_{\mathrm{atm}}$ values). This is due to the fact that galaxies with a higher disc stability prevent the collapse of \h1 gas to form molecular gas, and subsequently stars, and hence retain a high $f_{\mathrm{atm}}$ value as opposed to those galaxies with lower disc stability (\citetalias{obreschkow16};~\citealt{Stevens18}). The fact that the Eridanus galaxies (which are from high-density environments) also follow this trend in the \fatm parameter space, alludes to the point that mass and AM together play very important roles in determining the SFE of galaxies, via their ability to regulate the overall disc stability of late-type galaxies, irrespective of their environment. 

\begin{table}
    \caption{The star formation properties of the different samples. The sSFR and SFE values listed are the mean values for the different samples, with the standard deviation quoted within brackets.}
    \label{tab:SFR-properties}
    \begin{tabular}{lccc}       
    \hline \hline
        Sample & Size & $\log_{10}$(sSFR) & $\log_{10}$(SFE)  \\
             &  & [yr$^{-1}$] & [yr$^{-1}$] \\
    \hline         
        Eridanus & 26  & $-10.20 (0.45)$ & $-9.76 (0.41)$ \\
        Background & 5 & $-9.69 (0.34)$ & $-9.71 (0.18)$ \\
        WHISP & 86 & $-9.93 (0.49)$ & $-9.54 (0.60)$ \\
    \hline
    \end{tabular}
\end{table}

\section{Discussion}
\label{sec:Discussion}
\noindent We have shown that intrinsic properties such as mass and AM along with environmental processes play important roles in influencing both the \h1 gas and star formation properties of the Eridanus galaxies. In this section, we bring in some important discussion surrounding the observed results presented in Section~\ref{sec:results}, their interpretation, potential implications and associated caveats.

In Section~\ref{Subsec:jb-Mb relation} we presented the \jb relation for the Eridanus galaxies and compared the distribution of the Eridanus galaxies with galaxies from other high-resolution \h1 surveys targeting late-type galaxies, including high- and low-mass spirals and dwarfs. It was observed that the slope of the \jb relation for the Eridanus galaxies is statistically consistent with that of the larger sample of late-type galaxies. In addition, this slope is also consistent with the slope derived by previous works that have explored the $j - M$ space (\citealt{romanowsky12};~\citealt{sweet18};
~\citealt{Posti18};~\citetalias{Murugeshan20}). In a recent work, \citet{ManceraPina21} examine the \jb scaling relation for a sample of 157 galaxies ranging from dwarfs to high-mass spirals, and report a best fitting slope of $0.60 \pm 0.02$ with an intrinsic scatter 0.14 dex, consistent (within errors) with the fitted slopes in this work. The fact that independent studies using different samples and slightly different methods yield similar slopes is remarkable, and suggests that the \jb relation indeed follows an unbroken power-law over five orders of magnitude in $M_{\mathrm{b}}$. This has important implications for constraining our current cosmological models, and also serves as a great empirical benchmark for semi-analytic models to rely on. 

While studying the \jb relation for the Eridanus galaxies, we found that while the slope of the relation for the Eridanus galaxies and late-type galaxies from other surveys were statistically consistent, their intercepts were not. The $j_{\mathrm{b}}$ values for the Eridanus galaxies were on average lower (for the same $M_{\mathrm{b}}$) compared to late-type galaxies from other surveys (see Fig.~\ref{Fig.2}). It may be incorrect to associate the systematically lower $j_{\mathrm{b}}$ values of the Eridanus galaxies solely to environmentally driven processes. The observed discrepancy between the Eridanus sample and late-type and dwarf galaxies from other surveys may partly be arising due to selection biases. Most previous high-resolution \h1 surveys have selected their sample on the basis of the optical sizes and/or the stellar masses of the galaxies. They are often biased towards larger and more massive systems, implying that the selected galaxies are likely to be \h1-normal or -excess, and thus likely to be high AM systems to start with. WALLABY on the other hand is a untargeted survey, meaning all galaxies within the detection limit of the survey are detected and hence is not particularly biased towards high AM systems. Thus, the systematically lower $j_{\mathrm{b}}$ of the Eridanus galaxies may be partly explained by arguing that galaxies from previous surveys are biased towards high $j_{\mathrm{b}}$ systems for the same $M_{\mathrm{b}}$. 

That being said, we cannot completely ignore the effects of environment on the $j_{\mathrm{b}}$ values of the Eridanus galaxies. \citet{fall83} conducted the first empirical study that showed a tight correlation between the specific stellar specific AM ($j_{\star}$) and the stellar mass ($M_{\star}$) of galaxies. They found that both late- and early-type galaxies showed a linear correlation between $\log(j_{\star})$ and $\log(M_{\star})$, with a slope $\sim 2/3$. However, they observed that the $j_{\star}$ values for early-type galaxies were over five times lower for a given $M_{\star}$ compared to the late-type galaxies. In a follow-up work, \citet{romanowsky12} confirmed this for a large sample of late- and early-type galaxies. They link the observed lower $j_{\star}$ values among early-type galaxies to interactions and mergers prevalent in higher density environments. Tidal interactions, in extreme cases, may lead to stripping of significant fractions of gas and stars in the galactic discs leading to a reduction in $j_{\mathrm{b}}$ (see \citealt{Maller02}). In addition, tidal interactions between galaxies exert additional external torques on the \h1 gas discs of the interacting galaxies, making the \h1 gas lose some of its AM and migrate to the centre of the galaxy \citep{barnes96} to form stars, which eventually contributes to the growth of a (pseudo) bulge (e.g.~\citealt{Courteau96};~\citealt{Kormendy04}). Furthermore, hydrodynamic simulations have shown that dry mergers have the overall effect of reducing $j_{\mathrm{b}}$ \citep{Lagos18}. It is therefore possible that the systematically lower $j_{\mathrm{b}}$ values observed among the Eridanus galaxies is also driven by environmental effects such as tidal interactions and galaxy harassment.

It is also worth noting that the $j_{\mathrm{b}}$ values of the majority of the Virgo cluster galaxies is in the expected range for their baryonic mass and morphological type (pentagons in Fig.~\ref{Fig.2}). \citet{Li20} argue that ram-pressure is the most dominant environmental process that is stripping the \h1 gas among the Virgo cluster galaxies. Ram-pressure typically removes the diffuse \h1 gas in the outskirts of galaxies first, which does not significantly affect the cumulative (or total) $j_{\mathrm{b}}$ and $M_{\mathrm{b}}$ values of the galaxies. As a result of this, the $j_{\mathrm{b}}$, $M_{\mathrm{b}}$ and consequently the $q$ values of galaxies undergoing ram-pressure stripping are not significantly affected. 

It is well established that the cluster environment has an adverse effect on the \h1 gas fractions of galaxies, with a number of environmental processes affecting it (\citealt{davies73};~\citealt{giovanelli85};~\citealt{solanes01}). The effect of the group environment on the \h1 and AM properties of the constituent galaxies, however, is still not fully understood, and an area of active research. Over $\sim50$\% of galaxies reside in group-like environments (see for example~\citealt{Eke04}). Many studies have shown evidence of pre-processing in groups (\citealt{Fujita04};~\citealt{Just19}), which affect the observed global properties of the constituent galaxies, such as their \h1 gas content (\citealt{Cortese06};~\citealt{hess13};~\citealt{dzudzar19a},~\citeyear{dzudzar19b}), morphology, and star formation (\citealt{Roberts17};~\citealt{Barsanti18}). This implies that galaxies in groups likely experience a significant level of pre-processing before such groups merge to form larger clusters. Since the Eridanus supergroup is composed of three sub-groups in the process of merging to form a cluster, it presents us with an opportunity to study the level of pre-processing among the galaxies in the sub-groups. While tidal interactions and mergers may be the dominant environmental processes affecting galaxies in the group regime, the effects of ram-pressure, specifically from groups that show signs of hot diffuse ionized gas in their IGM cannot be neglected (see for example~\citealt{Marcolini03};~\citealt{Chamaraux04};~\citealt{sengupta06};~\citealt{Mayer06};~\citealt{Rasmussen08};~\citealt{denes16};~\citealt{Marasco16};~\citealt{brown17}). 

Furthermore, in the context of the \fatm framework, one cannot fully explain the behaviour of the Eridanus galaxies consistently falling below the \fatm relation by only invoking tidal interactions as the gas removing mechanism. Semi-analytic simulations by \citet{Stevens18} showed that ram-pressure and quasar driven winds from minor mergers are two important processes that can move galaxies below the \fatm relation. Given these arguments, the effects of ram-pressure stripping and minor mergers are possibly needed in addition to tidal stripping to explain the observed deviation of the Eridanus galaxies below the \fatm relation. As mentioned in Section~\ref{Subsec:Env. effect}, the \h1 maps of select galaxies in the Eridanus sample do indeed exhibit features observed in galaxies affected by ram-pressure, such as asymmetric and truncated \h1 discs, as well as compressed \h1 contours towards one side, which additionally come in support of the argument in favour of ram-pressure.
In addition, the majority of the Eridanus galaxies are dwarfs or low-mass spirals (with a median stellar mass $\log_{10}(M_{\star})$[M$_{\odot}] \sim 8.8$), making it easier for ram-pressure assisted by tidal interactions to strip off the \h1 gas. The behaviour of the Eridanus galaxies in the \fatm parameter space is also consistent with a previous study of the Virgo cluster galaxies by \citet{Li20}, where ram-pressure effects have been observed to significantly reduce the $f_{\mathrm{atm}}$ values of the galaxies, thereby making them lie consistently below the \fatm relation. 

It should however be noted that the degree of influence of ram-pressure on the \h1 gas of galaxies is strongly dependent on their orbit, as well as their orientation (\citealt{vollmer01};~\citealt{Jachym07};~\citealt{Stevens19}). The main caveat of the current analysis is that we do not have prior information regarding the orbits of the galaxies in the Eridanus sub-group, and have only used the projected distances of the galaxies from the central elliptical NGC 1395, to compute the level of ram-pressure experienced by the galaxies. The calculations also assume that the galaxies are moving face-on in the direction of motion. The current analysis is therefore only a first order approximation to probe the level of ram-pressure in the Eridanus sub-group. Contrary to our interpretations, \citet{Omar05b} argue that the \h1 deficiency among the Eridanus galaxies is mainly driven by tidal interactions. While this may be true, based on our observations, we conclude that other environmental processes are also likely playing a role in affecting the \h1 gas of the Eridanus galaxies.

\section{Conclusions}
\noindent We have used ASKAP \h1 observations of the Eridanus galaxies, to study their \h1, angular momentum (AM) and star formation properties, as part of the WALLABY pre-pilot commissioning survey. We summarise the important findings of this work, and also discuss some future prospects.
\begin{itemize}
    \item We have measured the total baryonic mass and total baryonic specific AM of the Eridanus and background galaxies, using robust rotation curves derived from 3D kinematic modelling of their \h1 gas distribution. We examined the \jb relation for the Eridanus galaxies and find that the slope of the \jb relation is $0.57 \pm 0.05$, consistent with other studies that have probed the relation for late-type galaxies. 
    \item We have explored the behaviour of the Eridanus and background galaxies in the \fatm parameter space (see \citetalias{obreschkow16}), and find that the Eridanus galaxies lie consistently off and below the \fatm relation. This is suggestive that environmental processes such as tidal interactions and ram-pressure stripping are prevalent in the Eridanus sub-group, which have removed a significant fraction of the \h1 gas from the galaxies, making them \h1 deficient for their $q$ values. In contrast, the background galaxies, which are from low-density environments follow the \fatm relation consistently.
    \item Further, we examined the effects of the local environment on the Eridanus galaxies. We estimated the local density using the projected 2nd nearest-neighbour metric ($\Sigma_2$ [Mpc$^{-2}$]), and find that galaxies within the Eridanus sub-group with higher $\Sigma_2$ values are seen to deviate more from the \fatm relation, again alluding to the fact that galaxies at higher local densities are influenced by their environment more than galaxies residing in lower densities. We also examined the \h1 and optical morphologies of the Eridanus galaxies and find that most galaxies in the Eridanus sample show a disturbed \h1 morphology. Some Eridanus galaxies exhibit asymmetric \h1 morphologies and truncated \h1 discs, but with no visible perturbations to their stellar component. These signatures are typical of ram-pressure effects. We computed the level of ram-pressure in the Eridanus sub-group following the traditional method introduced by \citet{gunn72} and find that over $\sim 48\%$ of the galaxies part of the Eridanus sub-group may have been influenced by ram-pressure. We thus conclude that both tidal interactions and ram-pressure are playing important roles in influencing the \h1 content of the Eridanus galaxies.
    \item We measure the \h1 deficiency among the Eridanus and the background galaxies in the context of their disc stability ($q$), by measuring the offset in their observed $f_{\mathrm{atm}}$ values with respect to the expected theoretical $f_{\mathrm{atm}}$. We find that a majority of the Eridanus galaxies are significantly \h1 deficient for their integrated disc stability. We also compare the \h1 deficiency of the Eridanus galaxies with normal star-forming galaxies in low-density environments (\citetalias{Murugeshan20}), and find that the Eridanus galaxies are indeed \h1 deficient. In contrast, the background galaxies are found to be \h1-normal.
    \item Finally, we examined the star formation properties of the sample galaxies by studying their sSFR and star formation efficiency (SFE), and their behaviour in the \fatm parameter space. We find that star formation in the Eridanus galaxies may be marginally suppressed due to environmental effects. We find that the SFE of galaxies, irrespective of their environments, tend to decrease with increasing $q$. This is consistent within the original theoretical framework of \citetalias{obreschkow16}, where galaxies with high $q$ values are expected to have low SFE and vice versa, due the ability of $q$ to regulate the collapse of \h1 gas to form molecular gas and eventually stars.
 \end{itemize}
 We have shown that pre-processing is prevalent in the Eridanus sub-group, which is in the processes of merging with at least two other sub-groups, and will eventually form a cluster \citep{Brough06}. While tidal interactions among galaxies is believed to be the most common environmental process driving the \h1 deficiency in group environments, we have shown here that in the Eridanus sub-group, which hosts a hot IGM, there is evidence of ram-pressure (also) causing a depletion of the galaxies' \h1 gas.
 
 The \fatm parameter space has shown great potential to probe the \h1 gas fractions of galaxies in the context of their mass and AM. It is proving to be a very useful parameter space, where trends in galaxies' SFR, sSFR, SFE and bulge-to-total ratio (a proxy for morphology) has been observed (\citetalias{Murugeshan20}), in addition to its ability to predict the expected \h1 saturation (or gas fraction) among rotationally dominated late-type galaxies. The \fatm relation is being increasingly used as a diagnostic tool to identify various environmental effects on the \h1 gas of galaxies. For instance, \citet{Li20} find that galaxies affected by ram-pressure, tend to consistently fall off the \fatm relation, while \citetalias{Murugeshan20} find that galaxies that are currently interacting and accreting \h1 gas tend to have higher $f_{\mathrm{atm}}$ for their $q$. This gives us a handle on the most dominant environmental process affecting the \h1 gas fractions in galaxies. Furthermore, the \fatm parameter space provides us with a new physically motivated estimate of the \h1 deficiency ($\Delta f_q$) among late-type galaxies as demonstrated by \citet{Li20}, where $\Delta f_q$ is simply the difference in the theoretically expected $f_{\mathrm{atm}}$ value (for a given $q$) and the observed $f_{\mathrm{atm}}$ of the galaxy. In this work, we have used this \h1 deficiency parameter to measure the extent of the \h1 deficiency among the Eridanus galaxies. We have additionally also compared the $\Delta f_q$ \h1 deficiency parameter with more traditionally used \h1 deficiency parameters, and find a good one-to-one agreement, further validating the potential and robustness of this parameter. 
 
 While several simulation studies have explored the \jb and \fatm relations for statistically large samples of galaxies (e.g.~\citealt{Stevens18};~\citealt{Lagos18};~\citealt{Stevens19}), empirical studies of the same are still limited to sample sizes of only a few hundred galaxies. WALLABY has the potential to resolve the gas structure of $\sim$ 5000 galaxies in high resolution \citep{Koribalski20}, thus positing the ability to accurately measure both the AM and \h1 properties for a large sample of galaxies. Using this wealth of data, we can then study the behaviour of galaxies across various environments on the \jb and \fatm parameter spaces. Such studies will be invaluable to our understanding of the degree of influence of internal parameters (such as mass and AM) and environmental factors, on the observed global properties of galaxies. We have demonstrated in this study that the reasonably high resolution (30 arcsec) \h1 observations, combined with a good velocity resolution (4~\kms) yield robust rotation curves, which is crucial for an accurate measurement of the AM in galaxies. In a future work, we aim to study the AM and \h1 properties of cluster galaxies, by observing their behaviour on the \fatm relation. Since these galaxies are from extreme environments, do we expect to see galaxies deviating from the \fatm relation as was observed by \citet{Li20} among the Virgo cluster galaxies? How do cluster environments affect the AM, \h1 and star formation properties of galaxies? Such follow-up studies are important to test the full potential of this new parameter space. With the WALLABY survey anticipated to commence full operations in 2021, we hope to answer some of these questions, which will allow us to further our understanding of galaxy evolution across environments.

\section*{Acknowledgements}
\label{sec:acknowledgements}
We thank the anonymous referee for their useful comments which considerably improved the quality of the paper. CM would like to thank Christopher Fluke, Robert D{\v{z}}ud{\v{z}}ar and Elaine Sadler for useful discussions and comments. CM is supported by the Swinburne University Postgraduate Award (SUPRA). ARHS acknowledges receipt of the Jim Buckee Fellowship at ICRAR-UWA. PK is partially supported by the BMBF project 05A17PC2 for D-MeerKAT. MEC is a recipient of an Australian Research Council Future Fellowship (project No. FT170100273) funded by the Australian Government. SHO acknowledges a support from the National Research Foundation of Korea (NRF) grant funded by the Korea government (Ministry of Science and ICT: MSIT) (No. NRF-2020R1A2C1008706).

This research was supported by the Australian Research Council Centre of Excellence for All Sky Astrophysics in 3 Dimensions (ASTRO 3D), through project number CE170100013.

The Australian SKA Pathfinder is part of the Australia Telescope National Facility which is managed by CSIRO. Operation of ASKAP is funded by the Australian Government with support from the National Collaborative Research Infrastructure Strategy. ASKAP uses the resources of the Pawsey Supercomputing Centre. Establishment of ASKAP, the Murchison Radio-astronomy Observatory and the Pawsey Supercomputing Centre are initiatives of the Australian Government, with support from the Government of Western Australia and the Science and Industry Endowment Fund. We acknowledge the Wajarri Yamatji people as the traditional owners of the Observatory site.

This publication makes use of data products from the Two Micron
All Sky Survey, which is a joint project of the University
of Massachusetts and the Infrared Processing and Analysis Center/
California Institute of Technology, funded by the National Aeronautics
and Space Administration and the National Science Foundation. 

This research has made use of the NASA/IPAC Extragalactic
Database (NED), which is operated by the Jet Propulsion Laboratory,
California Institute of Technology, under contract with the
National Aeronautics and Space Administration. 

This publication makes use of data products from the Wide-field Infrared Survey Explorer, which is a joint project of the University of California, Los Angeles, and the Jet Propulsion 
Laboratory/California Institute of Technology, funded by the National Aeronautics and Space Administration.

Parts of the results in this work make use of the colourmaps in the CMasher package (van der Velden~\citeyear{ellert20}).
\section*{Data Availability}
\noindent All data underlying this article is available within the article and enlisted in Table \ref{tab:sample} in Appendix \ref{appendix:properties_table}. In addition, the calibrated visibility data and the processed image cubes of the Eridanus footprints are available to the public through the CSIRO ASKAP Science Data Archive (CASDA), via this DOI:~\url{https://dx.doi.org/10.25919/0yc5-f769}




\bibliographystyle{mnras}
\bibliography{ref} 

\begin{thebibliography}{}
\makeatletter
\relax
\def\mn@urlcharsother{\let\do\@makeother \do\$\do\&\do\#\do\^\do\_\do\%\do\~}
\def\mn@doi{\begingroup\mn@urlcharsother \@ifnextchar [ {\mn@doi@}
  {\mn@doi@[]}}
\def\mn@doi@[#1]#2{\def\@tempa{#1}\ifx\@tempa\@empty \href
  {http://dx.doi.org/#2} {doi:#2}\else \href {http://dx.doi.org/#2} {#1}\fi
  \endgroup}
\def\mn@eprint#1#2{\mn@eprint@#1:#2::\@nil}
\def\mn@eprint@arXiv#1{\href {http://arxiv.org/abs/#1} {{\tt arXiv:#1}}}
\def\mn@eprint@dblp#1{\href {http://dblp.uni-trier.de/rec/bibtex/#1.xml}
  {dblp:#1}}
\def\mn@eprint@#1:#2:#3:#4\@nil{\def\@tempa {#1}\def\@tempb {#2}\def\@tempc
  {#3}\ifx \@tempc \@empty \let \@tempc \@tempb \let \@tempb \@tempa \fi \ifx
  \@tempb \@empty \def\@tempb {arXiv}\fi \@ifundefined
  {mn@eprint@\@tempb}{\@tempb:\@tempc}{\expandafter \expandafter \csname
  mn@eprint@\@tempb\endcsname \expandafter{\@tempc}}}

\bibitem[\protect\citeauthoryear{{Angiras}, {Jog}, {Omar}  \&
  {Dwarakanath}}{{Angiras} et~al.}{2006}]{Angiras06}
{Angiras} R.~A.,  {Jog} C.~J.,  {Omar} A.,   {Dwarakanath} K.~S.,  2006,
  \mn@doi [\mnras] {10.1111/j.1365-2966.2006.10418.x}, \href
  {https://ui-adsabs-harvard-edu.ezproxy.lib.swin.edu.au/abs/2006MNRAS.369.1849A}
  {369, 1849}

\bibitem[\protect\citeauthoryear{{Barnes} \& {Hernquist}}{{Barnes} \&
  {Hernquist}}{1996}]{barnes96}
{Barnes} J.~E.,  {Hernquist} L.,  1996, \mn@doi [\apj] {10.1086/177957}, \href
  {https://ui.adsabs.harvard.edu/abs/1996ApJ...471..115B} {471, 115}

\bibitem[\protect\citeauthoryear{{Barsanti} et~al.,}{{Barsanti}
  et~al.}{2018}]{Barsanti18}
{Barsanti} S.,  et~al., 2018, \mn@doi [\apj] {10.3847/1538-4357/aab61a}, \href
  {https://ui.adsabs.harvard.edu/abs/2018ApJ...857...71B} {857, 71}

\bibitem[\protect\citeauthoryear{{Barton}, {Geller}, {Ramella}, {Marzke}  \&
  {da Costa}}{{Barton} et~al.}{1996}]{Barton96}
{Barton} E.,  {Geller} M.,  {Ramella} M.,  {Marzke} R.~O.,   {da Costa} L.~N.,
  1996, \mn@doi [\aj] {10.1086/118060}, \href
  {https://ui-adsabs-harvard-edu.ezproxy.lib.swin.edu.au/abs/1996AJ....112..871B}
  {112, 871}

\bibitem[\protect\citeauthoryear{{Brough}, {Forbes}, {Kilborn}, {Couch}  \&
  {Colless}}{{Brough} et~al.}{2006}]{Brough06}
{Brough} S.,  {Forbes} D.~A.,  {Kilborn} V.~A.,  {Couch} W.,   {Colless} M.,
  2006, \mn@doi [\mnras] {10.1111/j.1365-2966.2006.10387.x}, \href
  {https://ui-adsabs-harvard-edu.ezproxy.lib.swin.edu.au/abs/2006MNRAS.369.1351B}
  {369, 1351}

\bibitem[\protect\citeauthoryear{{Brown}, {Catinella}, {Cortese}, {Kilborn},
  {Haynes}  \& {Giovanelli}}{{Brown} et~al.}{2015}]{brown15}
{Brown} T.,  {Catinella} B.,  {Cortese} L.,  {Kilborn} V.,  {Haynes} M.~P.,
  {Giovanelli} R.,  2015, \mn@doi [\mnras] {10.1093/mnras/stv1311}, \href
  {http://adsabs.harvard.edu/abs/2015MNRAS.452.2479B} {452, 2479}

\bibitem[\protect\citeauthoryear{{Brown} et~al.,}{{Brown}
  et~al.}{2017}]{brown17}
{Brown} T.,  et~al., 2017, \mn@doi [\mnras] {10.1093/mnras/stw2991}, \href
  {http://adsabs.harvard.edu/abs/2017MNRAS.466.1275B} {466, 1275}

\bibitem[\protect\citeauthoryear{{Butler}, {Obreschkow}  \& {Oh}}{{Butler}
  et~al.}{2017}]{butler17}
{Butler} K.~M.,  {Obreschkow} D.,   {Oh} S.-H.,  2017, \mn@doi [\apjl]
  {10.3847/2041-8213/834/1/L4}, \href
  {http://adsabs.harvard.edu/abs/2017ApJ...834L...4B} {834, L4}

\bibitem[\protect\citeauthoryear{{Byrd} \& {Valtonen}}{{Byrd} \&
  {Valtonen}}{1990}]{Byrd90}
{Byrd} G.,  {Valtonen} M.,  1990, \mn@doi [\apj] {10.1086/168362}, \href
  {https://ui-adsabs-harvard-edu.ezproxy.lib.swin.edu.au/abs/1990ApJ...350...89B}
  {350, 89}

\bibitem[\protect\citeauthoryear{{Catinella} et~al.,}{{Catinella}
  et~al.}{2010}]{catinella10}
{Catinella} B.,  et~al., 2010, \mn@doi [\mnras]
  {10.1111/j.1365-2966.2009.16180.x}, \href
  {http://adsabs.harvard.edu/abs/2010MNRAS.403..683C} {403, 683}

\bibitem[\protect\citeauthoryear{{Catinella} et~al.,}{{Catinella}
  et~al.}{2012}]{catinella12}
{Catinella} B.,  et~al., 2012, \mn@doi [\aap] {10.1051/0004-6361/201219261},
  \href {http://adsabs.harvard.edu/abs/2012A%26A...544A..65C} {544, A65}

\bibitem[\protect\citeauthoryear{{Catinella} et~al.,}{{Catinella}
  et~al.}{2018}]{catinella18}
{Catinella} B.,  et~al., 2018, \mn@doi [\mnras] {10.1093/mnras/sty089}, \href
  {http://adsabs.harvard.edu/abs/2018MNRAS.476..875C} {476, 875}

\bibitem[\protect\citeauthoryear{{Cayatte}, {Kotanyi}, {Balkowski}  \& {van
  Gorkom}}{{Cayatte} et~al.}{1994}]{cayatte94}
{Cayatte} V.,  {Kotanyi} C.,  {Balkowski} C.,   {van Gorkom} J.~H.,  1994,
  \mn@doi [\aj] {10.1086/116913}, \href
  {http://adsabs.harvard.edu/abs/1994AJ....107.1003C} {107, 1003}

\bibitem[\protect\citeauthoryear{{Chamaraux} \& {Masnou}}{{Chamaraux} \&
  {Masnou}}{2004}]{Chamaraux04}
{Chamaraux} P.,  {Masnou} J.-L.,  2004, \mn@doi [\mnras]
  {10.1111/j.1365-2966.2004.07226.x}, \href
  {https://ui-adsabs-harvard-edu.ezproxy.lib.swin.edu.au/abs/2004MNRAS.347..541C}
  {347, 541}

\bibitem[\protect\citeauthoryear{{Chamaraux}, {Balkowski}  \&
  {Fontanelli}}{{Chamaraux} et~al.}{1986}]{chamaraux86}
{Chamaraux} P.,  {Balkowski} C.,   {Fontanelli} P.,  1986, \aap, \href
  {http://adsabs.harvard.edu/abs/1986A%26A...165...15C} {165, 15}

\bibitem[\protect\citeauthoryear{{Chippendale}, {O'Sullivan}, {Reynolds},
  {Gough}, {Hayman}  \& {Hay}}{{Chippendale} et~al.}{2010}]{Chippendale10}
{Chippendale} A.~P.,  {O'Sullivan} J.,  {Reynolds} J.,  {Gough} R.,  {Hayman}
  D.,   {Hay} S.,  2010, in Phased Array Systems and Technology (ARRAY. pp
  648--652, \mn@doi{10.1109/ARRAY.2010.5613298}

\bibitem[\protect\citeauthoryear{{Chung}, {van Gorkom}, {Kenney}, {Crowl}  \&
  {Vollmer}}{{Chung} et~al.}{2009}]{Chung09}
{Chung} A.,  {van Gorkom} J.~H.,  {Kenney} J. D.~P.,  {Crowl} H.,   {Vollmer}
  B.,  2009, \mn@doi [\aj] {10.1088/0004-6256/138/6/1741}, \href
  {https://ui.adsabs.harvard.edu/abs/2009AJ....138.1741C} {138, 1741}

\bibitem[\protect\citeauthoryear{{Cortese}, {Gavazzi}, {Boselli}, {Franzetti},
  {Kennicutt}, {O'Neil}  \& {Sakai}}{{Cortese} et~al.}{2006}]{Cortese06}
{Cortese} L.,  {Gavazzi} G.,  {Boselli} A.,  {Franzetti} P.,  {Kennicutt}
  R.~C.,  {O'Neil} K.,   {Sakai} S.,  2006, \mn@doi [\aap]
  {10.1051/0004-6361:20064873}, \href
  {https://ui.adsabs.harvard.edu/abs/2006A&A...453..847C} {453, 847}

\bibitem[\protect\citeauthoryear{{Cortese}, {Catinella}  \& {Smith}}{{Cortese}
  et~al.}{2021}]{Cortese21}
{Cortese} L.,  {Catinella} B.,   {Smith} R.,  2021, arXiv e-prints, \href
  {https://ui.adsabs.harvard.edu/abs/2021arXiv210402193C} {p. arXiv:2104.02193}

\bibitem[\protect\citeauthoryear{{Courteau}, {de Jong}  \&
  {Broeils}}{{Courteau} et~al.}{1996}]{Courteau96}
{Courteau} S.,  {de Jong} R.~S.,   {Broeils} A.~H.,  1996, \mn@doi [\apjl]
  {10.1086/309906}, \href
  {https://ui.adsabs.harvard.edu/abs/1996ApJ...457L..73C} {457, L73}

\bibitem[\protect\citeauthoryear{{Cowie} \& {Songaila}}{{Cowie} \&
  {Songaila}}{1977}]{cowie77}
{Cowie} L.~L.,  {Songaila} A.,  1977, \mn@doi [\nat] {10.1038/266501a0}, \href
  {http://adsabs.harvard.edu/abs/1977Natur.266..501C} {266, 501}

\bibitem[\protect\citeauthoryear{{Davies} \& {Lewis}}{{Davies} \&
  {Lewis}}{1973}]{davies73}
{Davies} R.~D.,  {Lewis} B.~M.,  1973, \mn@doi [\mnras]
  {10.1093/mnras/165.2.231}, \href
  {http://adsabs.harvard.edu/abs/1973MNRAS.165..231D} {165, 231}

\bibitem[\protect\citeauthoryear{{DeBoer} et~al.,}{{DeBoer}
  et~al.}{2009}]{DeBoer09}
{DeBoer} D.~R.,  et~al., 2009, Proceedings of the IEEE, 97, 1507

\bibitem[\protect\citeauthoryear{{D{\'e}nes}, {Kilborn}  \&
  {Koribalski}}{{D{\'e}nes} et~al.}{2014}]{denes14}
{D{\'e}nes} H.,  {Kilborn} V.~A.,   {Koribalski} B.~S.,  2014, \mn@doi [\mnras]
  {10.1093/mnras/stu1337}, \href
  {http://adsabs.harvard.edu/abs/2014MNRAS.444..667D} {444, 667}

\bibitem[\protect\citeauthoryear{{D{\'e}nes}, {Kilborn}, {Koribalski}  \&
  {Wong}}{{D{\'e}nes} et~al.}{2016}]{denes16}
{D{\'e}nes} H.,  {Kilborn} V.~A.,  {Koribalski} B.~S.,   {Wong} O.~I.,  2016,
  \mn@doi [\mnras] {10.1093/mnras/stv2391}, \href
  {http://adsabs.harvard.edu/abs/2016MNRAS.455.1294D} {455, 1294}

\bibitem[\protect\citeauthoryear{{Di Teodoro} \& {Fraternali}}{{Di Teodoro} \&
  {Fraternali}}{2015}]{teodoro15}
{Di Teodoro} E.~M.,  {Fraternali} F.,  2015, \mn@doi [\mnras]
  {10.1093/mnras/stv1213}, \href
  {http://adsabs.harvard.edu/abs/2015MNRAS.451.3021D} {451, 3021}

\bibitem[\protect\citeauthoryear{{D{\v{z}}ud{\v{z}}ar}
  et~al.,}{{D{\v{z}}ud{\v{z}}ar} et~al.}{2019a}]{dzudzar19a}
{D{\v{z}}ud{\v{z}}ar} R.,  et~al., 2019a, \mn@doi [\mnras]
  {10.1093/mnras/sty3500}, \href
  {https://ui.adsabs.harvard.edu/abs/2019MNRAS.483.5409D} {483, 5409}

\bibitem[\protect\citeauthoryear{{D{\v{z}}ud{\v{z}}ar}, {Kilborn},
  {Murugeshan}, {Meurer}, {Sweet}  \& {Putman}}{{D{\v{z}}ud{\v{z}}ar}
  et~al.}{2019b}]{dzudzar19b}
{D{\v{z}}ud{\v{z}}ar} R.,  {Kilborn} V.,  {Murugeshan} C.,  {Meurer} G.,
  {Sweet} S.~M.,   {Putman} M.,  2019b, \mn@doi [\mnras]
  {10.1093/mnrasl/slz139}, \href
  {https://ui.adsabs.harvard.edu/abs/2019MNRAS.490L...6D} {490, L6}

\bibitem[\protect\citeauthoryear{{Eke} et~al.,}{{Eke} et~al.}{2004}]{Eke04}
{Eke} V.~R.,  et~al., 2004, \mn@doi [\mnras]
  {10.1111/j.1365-2966.2004.08354.x}, \href
  {https://ui-adsabs-harvard-edu.ezproxy.lib.swin.edu.au/abs/2004MNRAS.355..769E}
  {355, 769}

\bibitem[\protect\citeauthoryear{{Elagali} et~al.,}{{Elagali}
  et~al.}{2019}]{Elagali19}
{Elagali} A.,  et~al., 2019, \mn@doi [\mnras] {10.1093/mnras/stz1448}, \href
  {https://ui-adsabs-harvard-edu.ezproxy.lib.swin.edu.au/abs/2019MNRAS.487.2797E}
  {487, 2797}

\bibitem[\protect\citeauthoryear{{Ellison}, {Patton}, {Simard}, {McConnachie},
  {Baldry}  \& {Mendel}}{{Ellison} et~al.}{2010}]{ellison10}
{Ellison} S.~L.,  {Patton} D.~R.,  {Simard} L.,  {McConnachie} A.~W.,  {Baldry}
  I.~K.,   {Mendel} J.~T.,  2010, \mn@doi [\mnras]
  {10.1111/j.1365-2966.2010.17076.x}, \href
  {https://ui.adsabs.harvard.edu/abs/2010MNRAS.407.1514E} {407, 1514}

\bibitem[\protect\citeauthoryear{{Ellison}, {Nair}, {Patton}, {Scudder},
  {Mendel}  \& {Simard}}{{Ellison} et~al.}{2011}]{ellison11}
{Ellison} S.~L.,  {Nair} P.,  {Patton} D.~R.,  {Scudder} J.~M.,  {Mendel}
  J.~T.,   {Simard} L.,  2011, \mn@doi [\mnras]
  {10.1111/j.1365-2966.2011.19195.x}, \href
  {https://ui.adsabs.harvard.edu/abs/2011MNRAS.416.2182E} {416, 2182}

\bibitem[\protect\citeauthoryear{{Ellison}, {Catinella}  \&
  {Cortese}}{{Ellison} et~al.}{2018}]{ellison18}
{Ellison} S.~L.,  {Catinella} B.,   {Cortese} L.,  2018, \mn@doi [\mnras]
  {10.1093/mnras/sty1247}, \href
  {https://ui.adsabs.harvard.edu/abs/2018MNRAS.478.3447E} {478, 3447}

\bibitem[\protect\citeauthoryear{{Fall}}{{Fall}}{1983}]{fall83}
{Fall} S.~M.,  1983, in {Athanassoula} E.,  ed.,  IAU Symposium Vol. 100,
  Internal Kinematics and Dynamics of Galaxies. pp 391--398

\bibitem[\protect\citeauthoryear{{For} et~al.,}{{For} et~al.}{2019}]{For19}
{For} B.~Q.,  et~al., 2019, \mn@doi [\mnras] {10.1093/mnras/stz2501}, \href
  {https://ui-adsabs-harvard-edu.ezproxy.lib.swin.edu.au/abs/2019MNRAS.489.5723F}
  {489, 5723}

\bibitem[\protect\citeauthoryear{{For} et~al.}{{For} et~al.}{2021}]{For21}
{For} B.-Q.,  et~al., 2021, \mnras

\bibitem[\protect\citeauthoryear{{Fujita}}{{Fujita}}{2004}]{Fujita04}
{Fujita} Y.,  2004, \mn@doi [\pasj] {10.1093/pasj/56.1.29}, \href
  {https://ui.adsabs.harvard.edu/abs/2004PASJ...56...29F} {56, 29}

\bibitem[\protect\citeauthoryear{{Fukazawa}, {Makishima}  \&
  {Ohashi}}{{Fukazawa} et~al.}{2004}]{Fukazawa04}
{Fukazawa} Y.,  {Makishima} K.,   {Ohashi} T.,  2004, \mn@doi [\pasj]
  {10.1093/pasj/56.6.965}, \href
  {https://ui.adsabs.harvard.edu/abs/2004PASJ...56..965F} {56, 965}

\bibitem[\protect\citeauthoryear{{Fumagalli} \& {Gavazzi}}{{Fumagalli} \&
  {Gavazzi}}{2008}]{Fumagalli08}
{Fumagalli} M.,  {Gavazzi} G.,  2008, \mn@doi [\aap]
  {10.1051/0004-6361:200810604}, \href
  {https://ui-adsabs-harvard-edu.ezproxy.lib.swin.edu.au/abs/2008A&A...490..571F}
  {490, 571}

\bibitem[\protect\citeauthoryear{{Garcia}, {Paturel}, {Bottinelli}  \&
  {Gouguenheim}}{{Garcia} et~al.}{1993}]{Garcia93}
{Garcia} A.~M.,  {Paturel} G.,  {Bottinelli} L.,   {Gouguenheim} L.,  1993,
  \aaps, \href
  {https://ui-adsabs-harvard-edu.ezproxy.lib.swin.edu.au/abs/1993A&AS...98....7G}
  {98, 7}

\bibitem[\protect\citeauthoryear{{Giovanelli} \& {Haynes}}{{Giovanelli} \&
  {Haynes}}{1985}]{giovanelli85}
{Giovanelli} R.,  {Haynes} M.~P.,  1985, \mn@doi [\apj] {10.1086/163170}, \href
  {http://adsabs.harvard.edu/abs/1985ApJ...292..404G} {292, 404}

\bibitem[\protect\citeauthoryear{{Gunn} \& {Gott}}{{Gunn} \&
  {Gott}}{1972}]{gunn72}
{Gunn} J.~E.,  {Gott} III J.~R.,  1972, \mn@doi [\apj] {10.1086/151605}, \href
  {http://adsabs.harvard.edu/abs/1972ApJ...176....1G} {176, 1}

\bibitem[\protect\citeauthoryear{{Haynes} \& {Giovanelli}}{{Haynes} \&
  {Giovanelli}}{1984}]{haynes84}
{Haynes} M.~P.,  {Giovanelli} R.,  1984, \mn@doi [\aj] {10.1086/113573}, \href
  {http://adsabs.harvard.edu/abs/1984AJ.....89..758H} {89, 758}

\bibitem[\protect\citeauthoryear{{Hess} \& {Wilcots}}{{Hess} \&
  {Wilcots}}{2013}]{hess13}
{Hess} K.~M.,  {Wilcots} E.~M.,  2013, \mn@doi [\aj]
  {10.1088/0004-6256/146/5/124}, \href
  {http://adsabs.harvard.edu/abs/2013AJ....146..124H} {146, 124}

\bibitem[\protect\citeauthoryear{{Hibbard} \& {van Gorkom}}{{Hibbard} \& {van
  Gorkom}}{1996}]{hibbard96}
{Hibbard} J.~E.,  {van Gorkom} J.~H.,  1996, \mn@doi [\aj] {10.1086/117815},
  \href {http://adsabs.harvard.edu/abs/1996AJ....111..655H} {111, 655}

\bibitem[\protect\citeauthoryear{Hotan et~al.,}{Hotan et~al.}{2014}]{hotan14}
Hotan A.~W.,  et~al., 2014, \mn@doi [Publications of the Astronomical Society
  of Australia] {10.1017/pasa.2014.36}, 31, e041

\bibitem[\protect\citeauthoryear{{Hotan} et~al.,}{{Hotan}
  et~al.}{2021}]{Hotan21}
{Hotan} A.~W.,  et~al., 2021, \mn@doi [\pasa] {10.1017/pasa.2021.1}, \href
  {https://ui.adsabs.harvard.edu/abs/2021PASA...38....9H} {38, e009}

\bibitem[\protect\citeauthoryear{{Huang}, {Haynes}, {Giovanelli}  \&
  {Brinchmann}}{{Huang} et~al.}{2012}]{huang12}
{Huang} S.,  {Haynes} M.~P.,  {Giovanelli} R.,   {Brinchmann} J.,  2012,
  \mn@doi [\apj] {10.1088/0004-637X/756/2/113}, \href
  {https://ui.adsabs.harvard.edu/abs/2012ApJ...756..113H} {756, 113}

\bibitem[\protect\citeauthoryear{{Huchra} et~al.,}{{Huchra}
  et~al.}{2012}]{huchra12}
{Huchra} J.~P.,  et~al., 2012, \mn@doi [\apjs] {10.1088/0067-0049/199/2/26},
  \href {https://ui.adsabs.harvard.edu/abs/2012ApJS..199...26H} {199, 26}

\bibitem[\protect\citeauthoryear{{Hunter} et~al.,}{{Hunter}
  et~al.}{2012}]{littlethings12}
{Hunter} D.~A.,  et~al., 2012, \mn@doi [\aj] {10.1088/0004-6256/144/5/134},
  \href {http://adsabs.harvard.edu/abs/2012AJ....144..134H} {144, 134}

\bibitem[\protect\citeauthoryear{{Isbell}, {Xue}  \& {Fu}}{{Isbell}
  et~al.}{2018}]{Isbell18}
{Isbell} J.~W.,  {Xue} R.,   {Fu} H.,  2018, \mn@doi [\apjl]
  {10.3847/2041-8213/aaf872}, \href
  {https://ui.adsabs.harvard.edu/abs/2018ApJ...869L..37I} {869, L37}

\bibitem[\protect\citeauthoryear{{J{\'a}chym}, {Palou{\v{s}}}, {K{\"o}ppen}  \&
  {Combes}}{{J{\'a}chym} et~al.}{2007}]{Jachym07}
{J{\'a}chym} P.,  {Palou{\v{s}}} J.,  {K{\"o}ppen} J.,   {Combes} F.,  2007,
  \mn@doi [\aap] {10.1051/0004-6361:20066442}, \href
  {https://ui.adsabs.harvard.edu/abs/2007A&A...472....5J} {472, 5}

\bibitem[\protect\citeauthoryear{{Just} et~al.,}{{Just} et~al.}{2019}]{Just19}
{Just} D.~W.,  et~al., 2019, \mn@doi [\apj] {10.3847/1538-4357/ab44a0}, \href
  {https://ui.adsabs.harvard.edu/abs/2019ApJ...885....6J} {885, 6}

\bibitem[\protect\citeauthoryear{{Kennicutt}, {Keel}, {van der Hulst}, {Hummel}
   \& {Roettiger}}{{Kennicutt} et~al.}{1987}]{Kennicutt87}
{Kennicutt} Robert~C. J.,  {Keel} W.~C.,  {van der Hulst} J.~M.,  {Hummel} E.,
   {Roettiger} K.~A.,  1987, \mn@doi [\aj] {10.1086/114384}, \href
  {https://ui.adsabs.harvard.edu/abs/1987AJ.....93.1011K} {93, 1011}

\bibitem[\protect\citeauthoryear{{Kilborn}, {Koribalski}, {Forbes}, {Barnes}
  \& {Musgrave}}{{Kilborn} et~al.}{2005}]{kilborn05}
{Kilborn} V.~A.,  {Koribalski} B.~S.,  {Forbes} D.~A.,  {Barnes} D.~G.,
  {Musgrave} R.~C.,  2005, \mn@doi [\mnras] {10.1111/j.1365-2966.2004.08450.x},
  \href {http://adsabs.harvard.edu/abs/2005MNRAS.356...77K} {356, 77}

\bibitem[\protect\citeauthoryear{{Kilborn}, {Forbes}, {Barnes}, {Koribalski},
  {Brough}  \& {Kern}}{{Kilborn} et~al.}{2009}]{Kilborn09}
{Kilborn} V.~A.,  {Forbes} D.~A.,  {Barnes} D.~G.,  {Koribalski} B.~S.,
  {Brough} S.,   {Kern} K.,  2009, \mn@doi [\mnras]
  {10.1111/j.1365-2966.2009.15587.x}, \href
  {https://ui-adsabs-harvard-edu.ezproxy.lib.swin.edu.au/abs/2009MNRAS.400.1962K}
  {400, 1962}

\bibitem[\protect\citeauthoryear{{Kleiner} et~al.,}{{Kleiner}
  et~al.}{2019}]{Kleiner19}
{Kleiner} D.,  et~al., 2019, \mn@doi [\mnras] {10.1093/mnras/stz2063}, \href
  {https://ui-adsabs-harvard-edu.ezproxy.lib.swin.edu.au/abs/2019MNRAS.488.5352K}
  {488, 5352}

\bibitem[\protect\citeauthoryear{{Koribalski} et~al.,}{{Koribalski}
  et~al.}{2004}]{Koribalski04}
{Koribalski} B.~S.,  et~al., 2004, \mn@doi [\aj] {10.1086/421744}, \href
  {https://ui.adsabs.harvard.edu/abs/2004AJ....128...16K} {128, 16}

\bibitem[\protect\citeauthoryear{{Koribalski} et~al.,}{{Koribalski}
  et~al.}{2020}]{Koribalski20}
{Koribalski} B.~S.,  et~al., 2020, \mn@doi [\apss]
  {10.1007/s10509-020-03831-4}, \href
  {https://ui-adsabs-harvard-edu.ezproxy.lib.swin.edu.au/abs/2020Ap&SS.365..118K}
  {365, 118}

\bibitem[\protect\citeauthoryear{{Kormendy} \& {Kennicutt}}{{Kormendy} \&
  {Kennicutt}}{2004}]{Kormendy04}
{Kormendy} J.,  {Kennicutt} Robert~C. J.,  2004, \mn@doi [\araa]
  {10.1146/annurev.astro.42.053102.134024}, \href
  {https://ui.adsabs.harvard.edu/abs/2004ARA&A..42..603K} {42, 603}

\bibitem[\protect\citeauthoryear{{Krumholz}, {Burkhart}, {Forbes}  \&
  {Crocker}}{{Krumholz} et~al.}{2018}]{Krumholz18}
{Krumholz} M.~R.,  {Burkhart} B.,  {Forbes} J.~C.,   {Crocker} R.~M.,  2018,
  \mn@doi [\mnras] {10.1093/mnras/sty852}, \href
  {https://ui-adsabs-harvard-edu.ezproxy.lib.swin.edu.au/abs/2018MNRAS.477.2716K}
  {477, 2716}

\bibitem[\protect\citeauthoryear{{Kurapati}, {Chengalur}, {Pustilnik}  \&
  {Kamphuis}}{{Kurapati} et~al.}{2018}]{Kurapati18}
{Kurapati} S.,  {Chengalur} J.~N.,  {Pustilnik} S.,   {Kamphuis} P.,  2018,
  \mn@doi [\mnras] {10.1093/mnras/sty1397}, \href
  {https://ui.adsabs.harvard.edu/abs/2018MNRAS.479..228K} {479, 228}

\bibitem[\protect\citeauthoryear{{Lagos}, {Theuns}, {Stevens}, {Cortese},
  {Padilla}, {Davis}, {Contreras}  \& {Croton}}{{Lagos} et~al.}{2017}]{lagos17}
{Lagos} C. d.~P.,  {Theuns} T.,  {Stevens} A. R.~H.,  {Cortese} L.,  {Padilla}
  N.~D.,  {Davis} T.~A.,  {Contreras} S.,   {Croton} D.,  2017, \mn@doi
  [\mnras] {10.1093/mnras/stw2610}, \href
  {https://ui.adsabs.harvard.edu/abs/2017MNRAS.464.3850L} {464, 3850}

\bibitem[\protect\citeauthoryear{{Lagos} et~al.,}{{Lagos}
  et~al.}{2018}]{Lagos18}
{Lagos} C. d.~P.,  et~al., 2018, \mn@doi [\mnras] {10.1093/mnras/stx2667},
  \href
  {https://ui-adsabs-harvard-edu.ezproxy.lib.swin.edu.au/abs/2018MNRAS.473.4956L}
  {473, 4956}

\bibitem[\protect\citeauthoryear{{Larson}, {Tinsley}  \& {Caldwell}}{{Larson}
  et~al.}{1980}]{larson80}
{Larson} R.~B.,  {Tinsley} B.~M.,   {Caldwell} C.~N.,  1980, \mn@doi [\apj]
  {10.1086/157917}, \href {http://adsabs.harvard.edu/abs/1980ApJ...237..692L}
  {237, 692}

\bibitem[\protect\citeauthoryear{{Lauberts} \& {Valentijn}}{{Lauberts} \&
  {Valentijn}}{1989}]{Lauberts89}
{Lauberts} A.,  {Valentijn} E.~A.,  1989, {The surface photometry catalogue of
  the ESO-Uppsala galaxies}

\bibitem[\protect\citeauthoryear{{Lee-Waddell} et~al.,}{{Lee-Waddell}
  et~al.}{2019}]{Lee-Waddell19}
{Lee-Waddell} K.,  et~al., 2019, \mn@doi [\mnras] {10.1093/mnras/stz017}, \href
  {https://ui-adsabs-harvard-edu.ezproxy.lib.swin.edu.au/abs/2019MNRAS.487.5248L}
  {487, 5248}

\bibitem[\protect\citeauthoryear{{Leroy}, {Walter}, {Brinks}, {Bigiel}, {de
  Blok}, {Madore}  \& {Thornley}}{{Leroy} et~al.}{2008}]{leroy08}
{Leroy} A.~K.,  {Walter} F.,  {Brinks} E.,  {Bigiel} F.,  {de Blok} W.~J.~G.,
  {Madore} B.,   {Thornley} M.~D.,  2008, \mn@doi [\aj]
  {10.1088/0004-6256/136/6/2782}, \href
  {http://adsabs.harvard.edu/abs/2008AJ....136.2782L} {136, 2782}

\bibitem[\protect\citeauthoryear{{Li}, {Obreschkow}, {Lagos}, {Cortese},
  {Welker}  \& {D{\v{z}}ud{\v{z}}ar}}{{Li} et~al.}{2020}]{Li20}
{Li} J.,  {Obreschkow} D.,  {Lagos} C.,  {Cortese} L.,  {Welker} C.,
  {D{\v{z}}ud{\v{z}}ar} R.,  2020, arXiv e-prints, \href
  {https://ui.adsabs.harvard.edu/abs/2020arXiv200209083L} {p. arXiv:2002.09083}

\bibitem[\protect\citeauthoryear{{Lutz} et~al.,}{{Lutz} et~al.}{2017}]{lutz17}
{Lutz} K.~A.,  et~al., 2017, \mn@doi [\mnras] {10.1093/mnras/stx053}, \href
  {http://adsabs.harvard.edu/abs/2017MNRAS.467.1083L} {467, 1083}

\bibitem[\protect\citeauthoryear{{Maddox}, {Hess}, {Obreschkow}, {Jarvis}  \&
  {Blyth}}{{Maddox} et~al.}{2015}]{maddox15}
{Maddox} N.,  {Hess} K.~M.,  {Obreschkow} D.,  {Jarvis} M.~J.,   {Blyth} S.-L.,
   2015, \mn@doi [\mnras] {10.1093/mnras/stu2532}, \href
  {http://adsabs.harvard.edu/abs/2015MNRAS.447.1610M} {447, 1610}

\bibitem[\protect\citeauthoryear{{Maller} \& {Dekel}}{{Maller} \&
  {Dekel}}{2002}]{Maller02}
{Maller} A.~H.,  {Dekel} A.,  2002, \mn@doi [\mnras]
  {10.1046/j.1365-8711.2002.05646.x}, \href
  {https://ui.adsabs.harvard.edu/abs/2002MNRAS.335..487M} {335, 487}

\bibitem[\protect\citeauthoryear{{Mancera Pi{\~n}a}, {Posti}, {Pezzulli},
  {Fraternali}, {Fall}, {Oosterloo}  \& {Adams}}{{Mancera Pi{\~n}a}
  et~al.}{2021a}]{ManceraPina21b}
{Mancera Pi{\~n}a} P.~E.,  {Posti} L.,  {Pezzulli} G.,  {Fraternali} F.,
  {Fall} S.~M.,  {Oosterloo} T.,   {Adams} E. A.~K.,  2021a, arXiv e-prints,
  \href {https://ui.adsabs.harvard.edu/abs/2021arXiv210702809M} {p.
  arXiv:2107.02809}

\bibitem[\protect\citeauthoryear{{Mancera Pi{\~n}a}, {Posti}, {Fraternali},
  {Adams}  \& {Oosterloo}}{{Mancera Pi{\~n}a} et~al.}{2021b}]{ManceraPina21}
{Mancera Pi{\~n}a} P.~E.,  {Posti} L.,  {Fraternali} F.,  {Adams} E. A.~K.,
  {Oosterloo} T.,  2021b, \mn@doi [\aap] {10.1051/0004-6361/202039340}, \href
  {https://ui.adsabs.harvard.edu/abs/2021A&A...647A..76M} {647, A76}

\bibitem[\protect\citeauthoryear{{Marasco}, {Crain}, {Schaye}, {Bah{\'e}}, {van
  der Hulst}, {Theuns}  \& {Bower}}{{Marasco} et~al.}{2016}]{Marasco16}
{Marasco} A.,  {Crain} R.~A.,  {Schaye} J.,  {Bah{\'e}} Y.~M.,  {van der Hulst}
  T.,  {Theuns} T.,   {Bower} R.~G.,  2016, \mn@doi [\mnras]
  {10.1093/mnras/stw1498}, \href
  {https://ui.adsabs.harvard.edu/abs/2016MNRAS.461.2630M} {461, 2630}

\bibitem[\protect\citeauthoryear{{Marcolini}, {Brighenti}  \&
  {D'Ercole}}{{Marcolini} et~al.}{2003}]{Marcolini03}
{Marcolini} A.,  {Brighenti} F.,   {D'Ercole} A.,  2003, \mn@doi [\mnras]
  {10.1046/j.1365-2966.2003.07054.x}, \href
  {https://ui.adsabs.harvard.edu/abs/2003MNRAS.345.1329M} {345, 1329}

\bibitem[\protect\citeauthoryear{{Martin} et~al.,}{{Martin}
  et~al.}{2005}]{Martin05}
{Martin} D.~C.,  et~al., 2005, \mn@doi [\apjl] {10.1086/426387}, \href
  {https://ui-adsabs-harvard-edu.ezproxy.lib.swin.edu.au/abs/2005ApJ...619L...1M}
  {619, L1}

\bibitem[\protect\citeauthoryear{{Martinez-Delgado} et~al.,}{{Martinez-Delgado}
  et~al.}{2010}]{MartinezDelgado10}
{Martinez-Delgado} D.,  et~al., 2010, \mn@doi [\aj]
  {10.1088/0004-6256/140/4/962}, \href
  {https://ui-adsabs-harvard-edu.ezproxy.lib.swin.edu.au/abs/2010AJ....140..962M}
  {140, 962}

\bibitem[\protect\citeauthoryear{{Mayer}, {Mastropietro}, {Wadsley}, {Stadel}
  \& {Moore}}{{Mayer} et~al.}{2006}]{Mayer06}
{Mayer} L.,  {Mastropietro} C.,  {Wadsley} J.,  {Stadel} J.,   {Moore} B.,
  2006, \mn@doi [\mnras] {10.1111/j.1365-2966.2006.10403.x}, \href
  {https://ui-adsabs-harvard-edu.ezproxy.lib.swin.edu.au/abs/2006MNRAS.369.1021M}
  {369, 1021}

\bibitem[\protect\citeauthoryear{{Merritt}}{{Merritt}}{1983}]{Merritt83}
{Merritt} D.,  1983, \mn@doi [\apj] {10.1086/160571}, \href
  {https://ui-adsabs-harvard-edu.ezproxy.lib.swin.edu.au/abs/1983ApJ...264...24M}
  {264, 24}

\bibitem[\protect\citeauthoryear{{Mogotsi}, {de Blok}, {Cald{\'u}-Primo},
  {Walter}, {Ianjamasimanana}  \& {Leroy}}{{Mogotsi} et~al.}{2016}]{Mogotsi16}
{Mogotsi} K.~M.,  {de Blok} W.~J.~G.,  {Cald{\'u}-Primo} A.,  {Walter} F.,
  {Ianjamasimanana} R.,   {Leroy} A.~K.,  2016, \mn@doi [\aj]
  {10.3847/0004-6256/151/1/15}, \href
  {https://ui-adsabs-harvard-edu.ezproxy.lib.swin.edu.au/abs/2016AJ....151...15M}
  {151, 15}

\bibitem[\protect\citeauthoryear{{Moore}, {Katz}, {Lake}, {Dressler}  \&
  {Oemler}}{{Moore} et~al.}{1996}]{moore96}
{Moore} B.,  {Katz} N.,  {Lake} G.,  {Dressler} A.,   {Oemler} A.,  1996,
  \mn@doi [\nat] {10.1038/379613a0}, \href
  {http://adsabs.harvard.edu/abs/1996Natur.379..613M} {379, 613}

\bibitem[\protect\citeauthoryear{{Moorman}, {Moreno}, {White}, {Vogeley},
  {Hoyle}, {Giovanelli}  \& {Haynes}}{{Moorman} et~al.}{2016}]{Moorman16}
{Moorman} C.~M.,  {Moreno} J.,  {White} A.,  {Vogeley} M.~S.,  {Hoyle} F.,
  {Giovanelli} R.,   {Haynes} M.~P.,  2016, \mn@doi [\apj]
  {10.3847/0004-637X/831/2/118}, \href
  {https://ui.adsabs.harvard.edu/abs/2016ApJ...831..118M} {831, 118}

\bibitem[\protect\citeauthoryear{{Murugeshan}, {Kilborn}, {Obreschkow},
  {Glazebrook}, {Lutz}, {D{\v{z}}ud{\v{z}}ar}  \& {D{\'e}nes}}{{Murugeshan}
  et~al.}{2019}]{murugeshan2019}
{Murugeshan} C.,  {Kilborn} V.,  {Obreschkow} D.,  {Glazebrook} K.,  {Lutz} K.,
   {D{\v{z}}ud{\v{z}}ar} R.,   {D{\'e}nes} H.,  2019, \mn@doi [\mnras]
  {10.1093/mnras/sty3265}, \href
  {https://ui.adsabs.harvard.edu/abs/2019MNRAS.483.2398M} {483, 2398}

\bibitem[\protect\citeauthoryear{{Murugeshan}, {Kilborn}, {Jarrett}, {Wong},
  {Obreschkow}, {Glazebrook}, {Cluver}  \& {Fluke}}{{Murugeshan}
  et~al.}{2020}]{Murugeshan20}
{Murugeshan} C.,  {Kilborn} V.,  {Jarrett} T.,  {Wong} O.~I.,  {Obreschkow} D.,
   {Glazebrook} K.,  {Cluver} M.~E.,   {Fluke} C.~J.,  2020, \mn@doi [\mnras]
  {10.1093/mnras/staa1731}, \href
  {https://ui-adsabs-harvard-edu.ezproxy.lib.swin.edu.au/abs/2020MNRAS.496.2516M}
  {496, 2516}

\bibitem[\protect\citeauthoryear{{Nikolic}, {Cullen}  \& {Alexander}}{{Nikolic}
  et~al.}{2004}]{Nikolic04}
{Nikolic} B.,  {Cullen} H.,   {Alexander} P.,  2004, \mn@doi [\mnras]
  {10.1111/j.1365-2966.2004.08366.x}, \href
  {https://ui.adsabs.harvard.edu/abs/2004MNRAS.355..874N} {355, 874}

\bibitem[\protect\citeauthoryear{{Obreschkow} \& {Glazebrook}}{{Obreschkow} \&
  {Glazebrook}}{2014}]{obreschkow14}
{Obreschkow} D.,  {Glazebrook} K.,  2014, \mn@doi [\apj]
  {10.1088/0004-637X/784/1/26}, \href
  {http://adsabs.harvard.edu/abs/2014ApJ...784...26O} {784, 26}

\bibitem[\protect\citeauthoryear{{Obreschkow}, {Glazebrook}, {Kilborn}  \&
  {Lutz}}{{Obreschkow} et~al.}{2016}]{obreschkow16}
{Obreschkow} D.,  {Glazebrook} K.,  {Kilborn} V.,   {Lutz} K.,  2016, \mn@doi
  [\apjl] {10.3847/2041-8205/824/2/L26}, \href
  {http://adsabs.harvard.edu/abs/2016ApJ...824L..26O} {824, L26}

\bibitem[\protect\citeauthoryear{{Omar} \& {Dwarakanath}}{{Omar} \&
  {Dwarakanath}}{2005a}]{Omar05a}
{Omar} A.,  {Dwarakanath} K.~S.,  2005a, \mn@doi [Journal of Astrophysics and
  Astronomy] {10.1007/BF02702451}, \href
  {https://ui-adsabs-harvard-edu.ezproxy.lib.swin.edu.au/abs/2005JApA...26....1O}
  {26, 1}

\bibitem[\protect\citeauthoryear{{Omar} \& {Dwarakanath}}{{Omar} \&
  {Dwarakanath}}{2005b}]{Omar05b}
{Omar} A.,  {Dwarakanath} K.~S.,  2005b, \mn@doi [Journal of Astrophysics and
  Astronomy] {10.1007/BF02702452}, \href
  {https://ui-adsabs-harvard-edu.ezproxy.lib.swin.edu.au/abs/2005JApA...26...71O}
  {26, 71}

\bibitem[\protect\citeauthoryear{{Osmond} \& {Ponman}}{{Osmond} \&
  {Ponman}}{2004}]{osmond04}
{Osmond} J.~P.~F.,  {Ponman} T.~J.,  2004, \mn@doi [\mnras]
  {10.1111/j.1365-2966.2004.07742.x}, \href
  {http://adsabs.harvard.edu/abs/2004MNRAS.350.1511O} {350, 1511}

\bibitem[\protect\citeauthoryear{{Parkash}, {Brown}, {Jarrett}  \&
  {Bonne}}{{Parkash} et~al.}{2018}]{Parkash18}
{Parkash} V.,  {Brown} M. J.~I.,  {Jarrett} T.~H.,   {Bonne} N.~J.,  2018,
  \mn@doi [\apj] {10.3847/1538-4357/aad3b9}, \href
  {https://ui-adsabs-harvard-edu.ezproxy.lib.swin.edu.au/abs/2018ApJ...864...40P}
  {864, 40}

\bibitem[\protect\citeauthoryear{{Posti}, {Fraternali}, {Di Teodoro}  \&
  {Pezzulli}}{{Posti} et~al.}{2018}]{Posti18}
{Posti} L.,  {Fraternali} F.,  {Di Teodoro} E.~M.,   {Pezzulli} G.,  2018,
  \mn@doi [\aap] {10.1051/0004-6361/201833091}, \href
  {https://ui.adsabs.harvard.edu/abs/2018A&A...612L...6P} {612, L6}

\bibitem[\protect\citeauthoryear{{Putman} et~al.,}{{Putman}
  et~al.}{1998}]{Putman98}
{Putman} M.~E.,  et~al., 1998, \mn@doi [\nat] {10.1038/29466}, \href
  {https://ui-adsabs-harvard-edu.ezproxy.lib.swin.edu.au/abs/1998Natur.394..752P}
  {394, 752}

\bibitem[\protect\citeauthoryear{{Rasmussen}, {Ponman}, {Verdes-Montenegro},
  {Yun}  \& {Borthakur}}{{Rasmussen} et~al.}{2008}]{Rasmussen08}
{Rasmussen} J.,  {Ponman} T.~J.,  {Verdes-Montenegro} L.,  {Yun} M.~S.,
  {Borthakur} S.,  2008, \mn@doi [\mnras] {10.1111/j.1365-2966.2008.13451.x},
  \href
  {https://ui-adsabs-harvard-edu.ezproxy.lib.swin.edu.au/abs/2008MNRAS.388.1245R}
  {388, 1245}

\bibitem[\protect\citeauthoryear{{Raymond} \& {Smith}}{{Raymond} \&
  {Smith}}{1977}]{Raymond77}
{Raymond} J.~C.,  {Smith} B.~W.,  1977, \mn@doi [\apjs] {10.1086/190486}, \href
  {https://ui.adsabs.harvard.edu/abs/1977ApJS...35..419R} {35, 419}

\bibitem[\protect\citeauthoryear{{Reynolds} et~al.,}{{Reynolds}
  et~al.}{2019}]{Reynolds19}
{Reynolds} T.~N.,  et~al., 2019, \mn@doi [\mnras] {10.1093/mnras/sty2930},
  \href
  {https://ui-adsabs-harvard-edu.ezproxy.lib.swin.edu.au/abs/2019MNRAS.482.3591R}
  {482, 3591}

\bibitem[\protect\citeauthoryear{{Roberts} \& {Parker}}{{Roberts} \&
  {Parker}}{2017}]{Roberts17}
{Roberts} I.~D.,  {Parker} L.~C.,  2017, \mn@doi [\mnras]
  {10.1093/mnras/stx317}, \href
  {https://ui.adsabs.harvard.edu/abs/2017MNRAS.467.3268R} {467, 3268}

\bibitem[\protect\citeauthoryear{{Robotham} \& {Obreschkow}}{{Robotham} \&
  {Obreschkow}}{2015}]{Robotham15}
{Robotham} A.~S.~G.,  {Obreschkow} D.,  2015, \mn@doi [\pasa]
  {10.1017/pasa.2015.33}, \href
  {https://ui.adsabs.harvard.edu/abs/2015PASA...32...33R} {32, e033}

\bibitem[\protect\citeauthoryear{{Rogstad}, {Lockhart}  \& {Wright}}{{Rogstad}
  et~al.}{1974}]{rogstad74}
{Rogstad} D.~H.,  {Lockhart} I.~A.,   {Wright} M.~C.~H.,  1974, \mn@doi [\apj]
  {10.1086/153164}, \href {http://adsabs.harvard.edu/abs/1974ApJ...193..309R}
  {193, 309}

\bibitem[\protect\citeauthoryear{{Romanowsky} \& {Fall}}{{Romanowsky} \&
  {Fall}}{2012}]{romanowsky12}
{Romanowsky} A.~J.,  {Fall} S.~M.,  2012, \mn@doi [\apjs]
  {10.1088/0067-0049/203/2/17}, \href
  {http://adsabs.harvard.edu/abs/2012ApJS..203...17R} {203, 17}

\bibitem[\protect\citeauthoryear{{Rots}, {Bosma}, {van der Hulst},
  {Athanassoula}  \& {Crane}}{{Rots} et~al.}{1990}]{Rots90}
{Rots} A.~H.,  {Bosma} A.,  {van der Hulst} J.~M.,  {Athanassoula} E.,
  {Crane} P.~C.,  1990, \mn@doi [\aj] {10.1086/115522}, \href
  {https://ui.adsabs.harvard.edu/abs/1990AJ....100..387R} {100, 387}

\bibitem[\protect\citeauthoryear{{Saintonge} et~al.,}{{Saintonge}
  et~al.}{2017}]{Saintonge17}
{Saintonge} A.,  et~al., 2017, \mn@doi [\apjs] {10.3847/1538-4365/aa97e0},
  \href {https://ui.adsabs.harvard.edu/abs/2017ApJS..233...22S} {233, 22}

\bibitem[\protect\citeauthoryear{{Schaye} et~al.,}{{Schaye}
  et~al.}{2015}]{Schaye15}
{Schaye} J.,  et~al., 2015, \mn@doi [\mnras] {10.1093/mnras/stu2058}, \href
  {https://ui.adsabs.harvard.edu/abs/2015MNRAS.446..521S} {446, 521}

\bibitem[\protect\citeauthoryear{{Schiminovich} et~al.,}{{Schiminovich}
  et~al.}{2010}]{schiminovich10}
{Schiminovich} D.,  et~al., 2010, \mn@doi [\mnras]
  {10.1111/j.1365-2966.2010.17210.x}, \href
  {https://ui-adsabs-harvard-edu.ezproxy.lib.swin.edu.au/abs/2010MNRAS.408..919S}
  {408, 919}

\bibitem[\protect\citeauthoryear{{Sengupta} \& {Balasubramanyam}}{{Sengupta} \&
  {Balasubramanyam}}{2006}]{sengupta06}
{Sengupta} C.,  {Balasubramanyam} R.,  2006, \mn@doi [\mnras]
  {10.1111/j.1365-2966.2006.10307.x}, \href
  {http://adsabs.harvard.edu/abs/2006MNRAS.369..360S} {369, 360}

\bibitem[\protect\citeauthoryear{{Serra} et~al.,}{{Serra}
  et~al.}{2015}]{Serra15}
{Serra} P.,  et~al., 2015, \mn@doi [\mnras] {10.1093/mnras/stv079}, \href
  {https://ui-adsabs-harvard-edu.ezproxy.lib.swin.edu.au/abs/2015MNRAS.448.1922S}
  {448, 1922}

\bibitem[\protect\citeauthoryear{{Skrutskie} et~al.,}{{Skrutskie}
  et~al.}{2006}]{2mass06}
{Skrutskie} M.~F.,  et~al., 2006, \mn@doi [\aj] {10.1086/498708}, \href
  {http://adsabs.harvard.edu/abs/2006AJ....131.1163S} {131, 1163}

\bibitem[\protect\citeauthoryear{{Solanes}, {Giovanelli}  \&
  {Haynes}}{{Solanes} et~al.}{1996}]{solanes96}
{Solanes} J.~M.,  {Giovanelli} R.,   {Haynes} M.~P.,  1996, \mn@doi [\apj]
  {10.1086/177089}, \href {http://adsabs.harvard.edu/abs/1996ApJ...461..609S}
  {461, 609}

\bibitem[\protect\citeauthoryear{{Solanes}, {Manrique}, {Garcia-Gomez},
  {Gonzalez-Casado}, {Giovanelli}  \& {Haynes}}{{Solanes}
  et~al.}{2001}]{solanes01}
{Solanes} J.~M.,  {Manrique} A.,  {Garcia-Gomez} C.,  {Gonzalez-Casado} G.,
  {Giovanelli} R.,   {Haynes} M.~P.,  2001, \mn@doi [\apj] {10.1086/318672},
  \href {http://adsabs.harvard.edu/abs/2001ApJ...548...97S} {548, 97}

\bibitem[\protect\citeauthoryear{{Stevens} \& {Brown}}{{Stevens} \&
  {Brown}}{2017}]{stevens17}
{Stevens} A.~R.~H.,  {Brown} T.,  2017, \mn@doi [\mnras]
  {10.1093/mnras/stx1596}, \href
  {http://adsabs.harvard.edu/abs/2017MNRAS.471..447S} {471, 447}

\bibitem[\protect\citeauthoryear{{Stevens}, {Lagos}, {Obreschkow}  \&
  {Sinha}}{{Stevens} et~al.}{2018}]{Stevens18}
{Stevens} A. R.~H.,  {Lagos} C. d.~P.,  {Obreschkow} D.,   {Sinha} M.,  2018,
  \mn@doi [\mnras] {10.1093/mnras/sty2650}, \href
  {https://ui.adsabs.harvard.edu/abs/2018MNRAS.481.5543S} {481, 5543}

\bibitem[\protect\citeauthoryear{{Stevens} et~al.,}{{Stevens}
  et~al.}{2019}]{Stevens19}
{Stevens} A. R.~H.,  et~al., 2019, \mn@doi [\mnras] {10.1093/mnras/sty3451},
  \href {https://ui.adsabs.harvard.edu/abs/2019MNRAS.483.5334S} {483, 5334}

\bibitem[\protect\citeauthoryear{{Swaters}, {van Albada}, {van der Hulst}  \&
  {Sancisi}}{{Swaters} et~al.}{2002}]{whisp02}
{Swaters} R.~A.,  {van Albada} T.~S.,  {van der Hulst} J.~M.,   {Sancisi} R.,
  2002, \mn@doi [\aap] {10.1051/0004-6361:20011755}, \href
  {http://adsabs.harvard.edu/abs/2002A%26A...390..829S} {390, 829}

\bibitem[\protect\citeauthoryear{{Sweet}, {Fisher}, {Glazebrook}, {Obreschkow},
  {Lagos}  \& {Wang}}{{Sweet} et~al.}{2018}]{sweet18}
{Sweet} S.~M.,  {Fisher} D.,  {Glazebrook} K.,  {Obreschkow} D.,  {Lagos} C.,
  {Wang} L.,  2018, \mn@doi [\apj] {10.3847/1538-4357/aabfc4}, \href
  {http://adsabs.harvard.edu/abs/2018ApJ...860...37S} {860, 37}

\bibitem[\protect\citeauthoryear{{Tamburro}, {Rix}, {Leroy}, {Mac Low},
  {Walter}, {Kennicutt}, {Brinks}  \& {de Blok}}{{Tamburro}
  et~al.}{2009}]{Tamburro09}
{Tamburro} D.,  {Rix} H.~W.,  {Leroy} A.~K.,  {Mac Low} M.~M.,  {Walter} F.,
  {Kennicutt} R.~C.,  {Brinks} E.,   {de Blok} W.~J.~G.,  2009, \mn@doi [\aj]
  {10.1088/0004-6256/137/5/4424}, \href
  {https://ui.adsabs.harvard.edu/abs/2009AJ....137.4424T} {137, 4424}

\bibitem[\protect\citeauthoryear{{Toomre}}{{Toomre}}{1964}]{toomre64}
{Toomre} A.,  1964, \mn@doi [\apj] {10.1086/147861}, \href
  {http://adsabs.harvard.edu/abs/1964ApJ...139.1217T} {139, 1217}

\bibitem[\protect\citeauthoryear{{Trinchieri}, {Fabbiano}  \&
  {Canizares}}{{Trinchieri} et~al.}{1986}]{Trinchieri86}
{Trinchieri} G.,  {Fabbiano} G.,   {Canizares} C.~R.,  1986, \mn@doi [\apj]
  {10.1086/164716}, \href
  {https://ui.adsabs.harvard.edu/abs/1986ApJ...310..637T} {310, 637}

\bibitem[\protect\citeauthoryear{{Trinchieri}, {Pellegrini}, {Wolter},
  {Fabbiano}  \& {Fiore}}{{Trinchieri} et~al.}{2000}]{Trinchieri2000}
{Trinchieri} G.,  {Pellegrini} S.,  {Wolter} A.,  {Fabbiano} G.,   {Fiore} F.,
  2000, \aap, \href
  {https://ui-adsabs-harvard-edu.ezproxy.lib.swin.edu.au/abs/2000A&A...364...53T}
  {364, 53}

\bibitem[\protect\citeauthoryear{{Tully}}{{Tully}}{2015}]{Tully15}
{Tully} R.~B.,  2015, \mn@doi [\aj] {10.1088/0004-6256/149/5/171}, \href
  {https://ui.adsabs.harvard.edu/abs/2015AJ....149..171T} {149, 171}

\bibitem[\protect\citeauthoryear{{Verheijen} \& {Sancisi}}{{Verheijen} \&
  {Sancisi}}{2001}]{verheijen01}
{Verheijen} M.~A.~W.,  {Sancisi} R.,  2001, \mn@doi [\aap]
  {10.1051/0004-6361:20010090}, \href
  {http://adsabs.harvard.edu/abs/2001A%26A...370..765V} {370, 765}

\bibitem[\protect\citeauthoryear{{Vollmer}, {Cayatte}, {Balkowski}  \&
  {Duschl}}{{Vollmer} et~al.}{2001}]{vollmer01}
{Vollmer} B.,  {Cayatte} V.,  {Balkowski} C.,   {Duschl} W.~J.,  2001, \mn@doi
  [\apj] {10.1086/323368}, \href
  {http://adsabs.harvard.edu/abs/2001ApJ...561..708V} {561, 708}

\bibitem[\protect\citeauthoryear{{Vulcani}, {Poggianti}, {Finn}, {Rudnick},
  {Desai}  \& {Bamford}}{{Vulcani} et~al.}{2010}]{Vulcani10}
{Vulcani} B.,  {Poggianti} B.~M.,  {Finn} R.~A.,  {Rudnick} G.,  {Desai} V.,
  {Bamford} S.,  2010, \mn@doi [\apjl] {10.1088/2041-8205/710/1/L1}, \href
  {https://ui-adsabs-harvard-edu.ezproxy.lib.swin.edu.au/abs/2010ApJ...710L...1V}
  {710, L1}

\bibitem[\protect\citeauthoryear{{Walter}, {Brinks}, {de Blok}, {Bigiel},
  {Kennicutt}, {Thornley}  \& {Leroy}}{{Walter} et~al.}{2008}]{things08}
{Walter} F.,  {Brinks} E.,  {de Blok} W.~J.~G.,  {Bigiel} F.,  {Kennicutt} Jr.
  R.~C.,  {Thornley} M.~D.,   {Leroy} A.,  2008, \mn@doi [\aj]
  {10.1088/0004-6256/136/6/2563}, \href
  {http://adsabs.harvard.edu/abs/2008AJ....136.2563W} {136, 2563}

\bibitem[\protect\citeauthoryear{{Wang} et~al.,}{{Wang} et~al.}{2017}]{Wang17}
{Wang} J.,  et~al., 2017, \mn@doi [\mnras] {10.1093/mnras/stx2073}, \href
  {https://ui-adsabs-harvard-edu.ezproxy.lib.swin.edu.au/abs/2017MNRAS.472.3029W}
  {472, 3029}

\bibitem[\protect\citeauthoryear{{Wang} et~al.,}{{Wang} et~al.}{2018}]{Wang18}
{Wang} L.,  et~al., 2018, \mn@doi [\apj] {10.3847/1538-4357/aae8de}, \href
  {https://ui.adsabs.harvard.edu/abs/2018ApJ...868...93W} {868, 93}

\bibitem[\protect\citeauthoryear{{Wen}, {Wu}, {Zhu}, {Lam}, {Wu}, {Wicker}  \&
  {Zhao}}{{Wen} et~al.}{2013}]{wen13}
{Wen} X.-Q.,  {Wu} H.,  {Zhu} Y.-N.,  {Lam} M.~I.,  {Wu} C.-J.,  {Wicker} J.,
  {Zhao} Y.-H.,  2013, \mn@doi [\mnras] {10.1093/mnras/stt939}, \href
  {http://adsabs.harvard.edu/abs/2013MNRAS.433.2946W} {433, 2946}

\bibitem[\protect\citeauthoryear{{Westmeier}, {Braun}  \&
  {Koribalski}}{{Westmeier} et~al.}{2011}]{westmeier11}
{Westmeier} T.,  {Braun} R.,   {Koribalski} B.~S.,  2011, \mn@doi [\mnras]
  {10.1111/j.1365-2966.2010.17596.x}, \href
  {http://adsabs.harvard.edu/abs/2011MNRAS.410.2217W} {410, 2217}

\bibitem[\protect\citeauthoryear{{Whiting}}{{Whiting}}{2020}]{Whiting20}
{Whiting} M.~T.,  2020, in {Ballester} P.,  {Ibsen} J.,  {Solar} M.,
  {Shortridge} K.,  eds,  Astronomical Society of the Pacific Conference Series
  Vol. 522, Astronomical Data Analysis Software and Systems XXVII. p.~469

\bibitem[\protect\citeauthoryear{{Willmer}, {Focardi}, {da Costa}  \&
  {Pellegrini}}{{Willmer} et~al.}{1989}]{Willmer89}
{Willmer} C.~N.~A.,  {Focardi} P.,  {da Costa} L.~N.,   {Pellegrini} P.~S.,
  1989, \mn@doi [\aj] {10.1086/115236}, \href
  {https://ui-adsabs-harvard-edu.ezproxy.lib.swin.edu.au/abs/1989AJ.....98.1531W}
  {98, 1531}

\bibitem[\protect\citeauthoryear{{Wong}, {Meurer}, {Zheng}, {Heckman},
  {Thilker}  \& {Zwaan}}{{Wong} et~al.}{2016}]{Wong16}
{Wong} O.~I.,  {Meurer} G.~R.,  {Zheng} Z.,  {Heckman} T.~M.,  {Thilker} D.~A.,
    {Zwaan} M.~A.,  2016, \mn@doi [\mnras] {10.1093/mnras/stw993}, \href
  {https://ui.adsabs.harvard.edu/abs/2016MNRAS.460.1106W} {460, 1106}

\bibitem[\protect\citeauthoryear{{Wong} et~al.}{{Wong} et~al.}{2021}]{Wong21}
{Wong} O.~I.,  et~al., 2021, \mnras

\bibitem[\protect\citeauthoryear{{Wright} et~al.,}{{Wright}
  et~al.}{2010}]{Wright10}
{Wright} E.~L.,  et~al., 2010, \mn@doi [\aj] {10.1088/0004-6256/140/6/1868},
  \href {https://ui.adsabs.harvard.edu/abs/2010AJ....140.1868W} {140, 1868}

\bibitem[\protect\citeauthoryear{{Zasov} \& {Rubtsova}}{{Zasov} \&
  {Rubtsova}}{1989}]{zasov89}
{Zasov} A.~V.,  {Rubtsova} T.~V.,  1989, Soviet Astronomy Letters, \href
  {https://ui.adsabs.harvard.edu/abs/1989SvAL...15...51Z} {15, 51}

\bibitem[\protect\citeauthoryear{{Zhou}, {Wu}, {Zhou}  \& {Ma}}{{Zhou}
  et~al.}{2018}]{Zhou18}
{Zhou} Z.,  {Wu} H.,  {Zhou} X.,   {Ma} J.,  2018, \mn@doi [\pasp]
  {10.1088/1538-3873/aad407}, \href
  {https://ui.adsabs.harvard.edu/abs/2018PASP..130i4101Z} {130, 094101}

\bibitem[\protect\citeauthoryear{{Zoldan}, {De Lucia}, {Xie}, {Fontanot}  \&
  {Hirschmann}}{{Zoldan} et~al.}{2018}]{Zoldan18}
{Zoldan} A.,  {De Lucia} G.,  {Xie} L.,  {Fontanot} F.,   {Hirschmann} M.,
  2018, \mn@doi [\mnras] {10.1093/mnras/sty2343}, \href
  {https://ui.adsabs.harvard.edu/abs/2018MNRAS.481.1376Z} {481, 1376}

\bibitem[\protect\citeauthoryear{{da Costa} et~al.,}{{da Costa}
  et~al.}{1988}]{dacosta88}
{da Costa} L.~N.,  et~al., 1988, \mn@doi [\apj] {10.1086/166215}, \href
  {https://ui-adsabs-harvard-edu.ezproxy.lib.swin.edu.au/abs/1988ApJ...327..544D}
  {327, 544}

\bibitem[\protect\citeauthoryear{{van der Velden}}{{van der
  Velden}}{2020}]{ellert20}
{van der Velden} E.,  2020, \mn@doi [The Journal of Open Source Software]
  {10.21105/joss.02004}, \href
  {https://ui.adsabs.harvard.edu/abs/2020JOSS....5.2004V} {5, 2004}

\makeatother
\end{thebibliography}




\appendix

\section{Tilted-ring model parameters and PV diagram}
\label{appendix:fitting_models}

\begin{figure*}
\centering
\begin{subfigure}{.5\textwidth}
  \centering
  \includegraphics[width=0.9\textwidth]{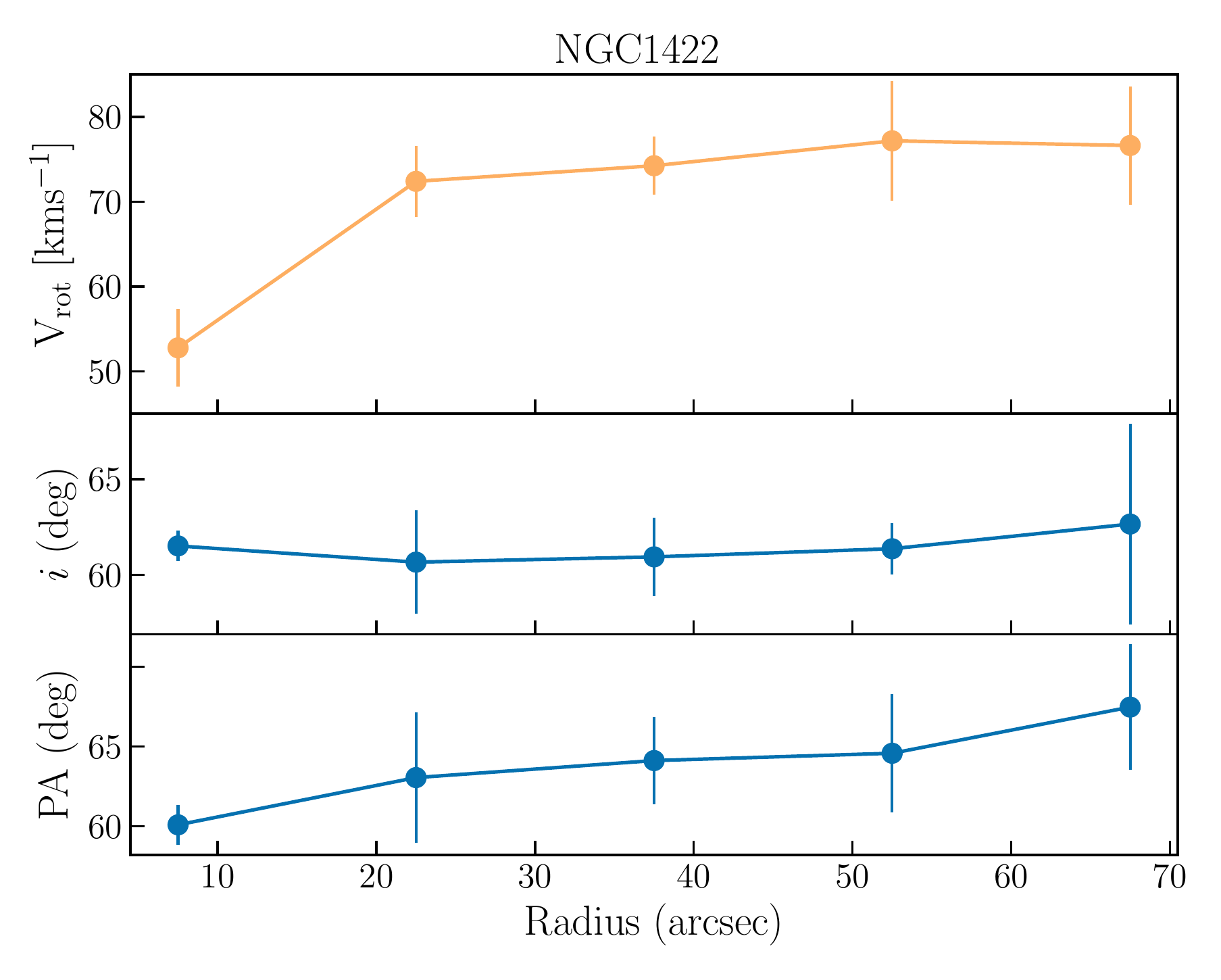}
  \caption{}
\end{subfigure}%
\begin{subfigure}{.5\textwidth}
  \hspace{-0.3cm}
  \includegraphics[width=0.9\textwidth]{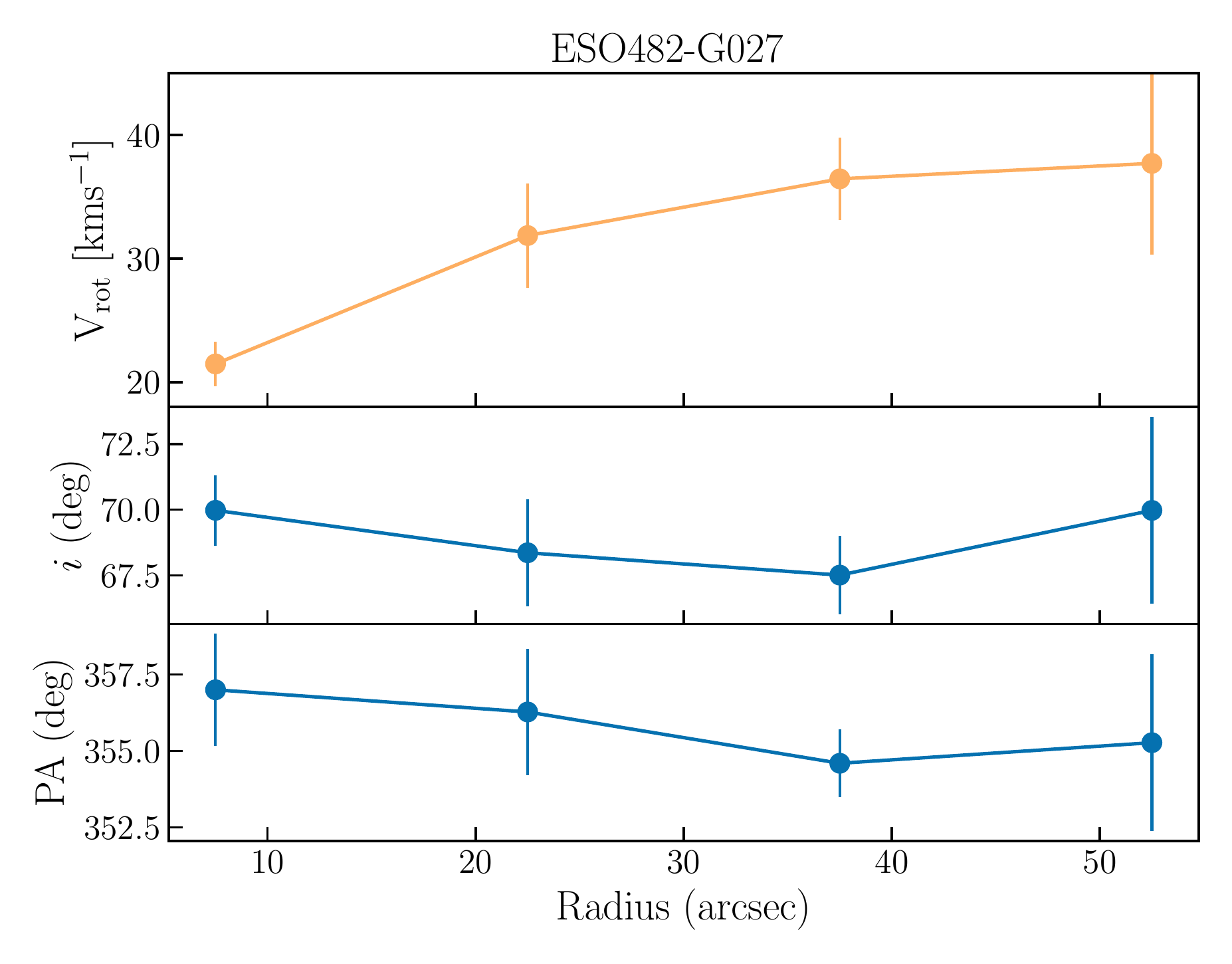}
  \caption{}
\end{subfigure}
\caption{(a) The rotation curve, radially varying inclination ($i$) and position angle (PA) from the 3D model fitting using \texttt{3DBarolo} for the NGC 1422 galaxy (b) Similar plot for the ESO482-G027 galaxy.}
\label{Fig.A1}
\end{figure*}

In order to show the robustness of the model fitting to the galaxies, we take specific examples of two galaxies -- NGC 1422 and ESO482-G027 (see Fig.~\ref{Fig.5}) -- for which the fitted model parameters such as their rotation curves (v$_{\mathrm{rot}}$), inclination ($i$) and position angles (PA) are shown in Fig.~\ref{Fig.A1}. In addition, we also show the position-velocity (PV) diagram along the major and minor axis for the two galaxies. Overall, we find a good agreement between the fitted models and the data for the two galaxies.
 
\begin{figure*}
    \centering
    \includegraphics[width=18cm,height=12cm]{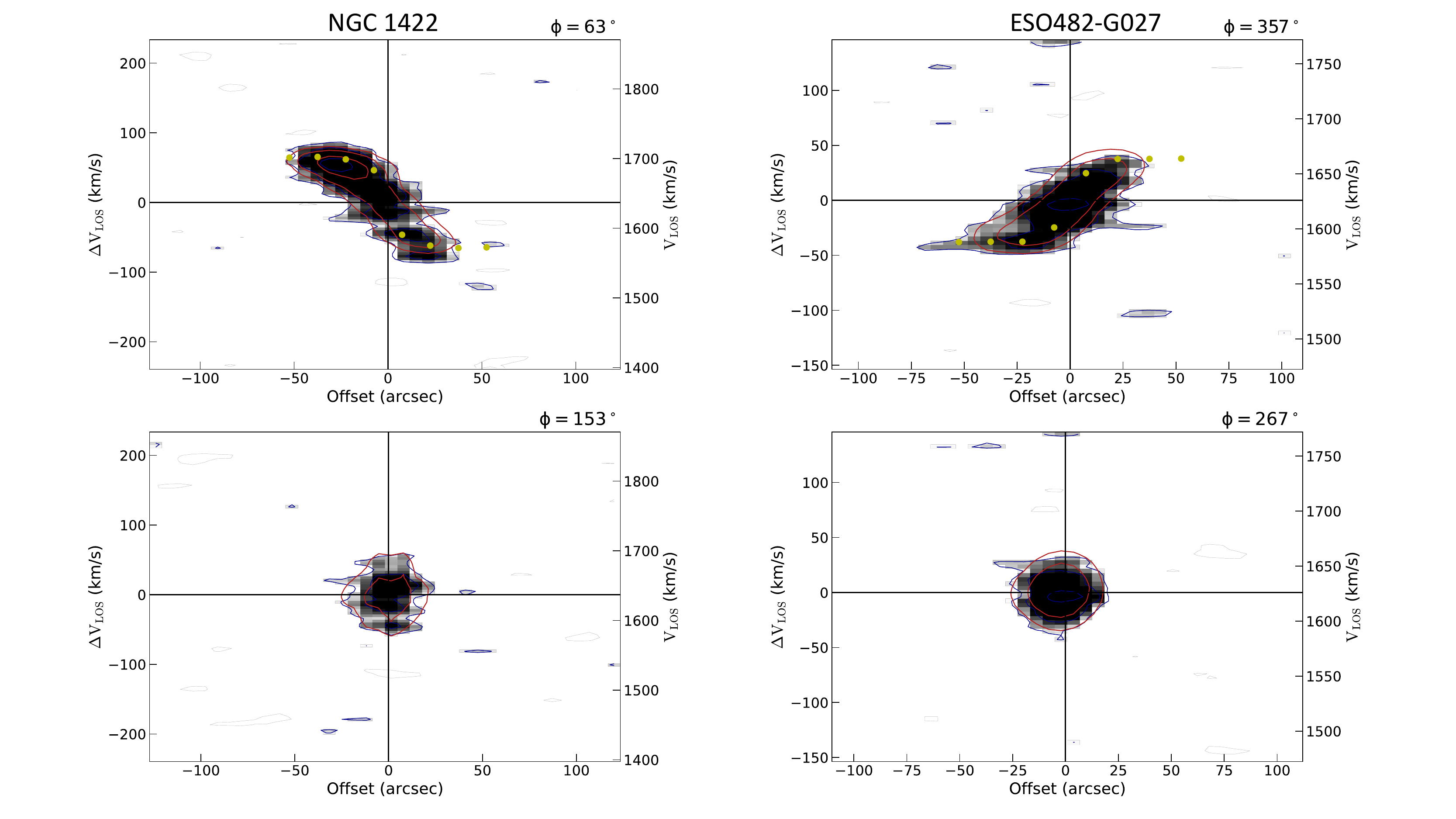}
    \caption{\textit{Left}: Top panel shows the position-velocity (PV) diagram along the major axis of NGC 1422 produced by \texttt{3DBarolo} after completing the fitting routine. The blue contours represent the data, and the red contours correspond to the final tilted-ring model, respectively. The yellow points correspond to the line-of-sight velocities $v_{\textrm{\tiny LOS}} = v_{\mathrm{rot}}\sin(i)$ derived from the fit. Bottom panel shows the PV diagram along the minor axis of NGC 1422. \textit{Right}:  PV diagrams along the major (top panel) and minor axis (bottom panel) of ESO482-G027, respectively. The contours and points are as defined before. We find that for both galaxies the model and the derived $v_{\textrm{\tiny LOS}}$ values are in good agreement with the observed data.} 
    \label{Fig.A2}
\end{figure*}

\section{Computing the ratio of specific AM of the \h1 gas and stars}
\label{appendix:j_fracVsMb}

\begin{figure}
    \centering
    \includegraphics[width=8cm,height=6.2cm]{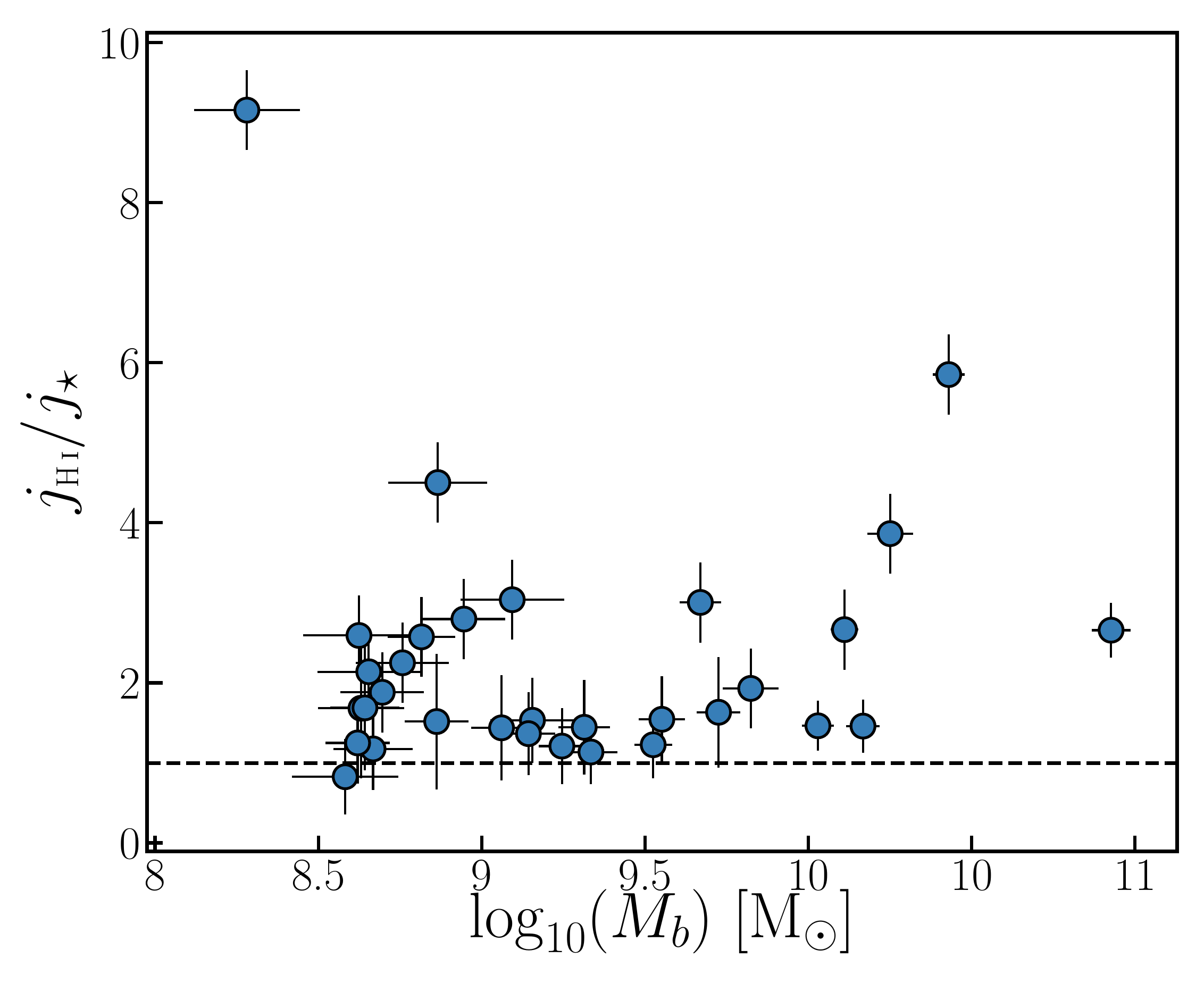}
    \caption{$j_{\textrm{\h1}}/j_{\star}$ ratio plotted against the total baryonic mass, $M_{\mathrm{b}}$. We see no correlation between the two parameters and find a mean value of $j_{\textrm{\h1}}/j_{\star} \sim 2$. The black dashed line represents $j_{\textrm{\h1}}/j_{\star} = 1$.} 
    \label{Fig.B1}
\end{figure}

We compute the total specific angular momentum of the \h1 gas and the stellar specific angular momentum following a method similar to computing $j_{\mathrm{b}}$ (see section~\ref{subsec:measuring_AM}), and as follows,
\begin{equation}
    j_{\textrm{\h1}} =\frac{\sum_{i} M_{\mathrm{\tiny \textrm{\h1}},i}v_{\mathrm{rot},i}r_{i}}{\sum_{i} M_{\mathrm{\tiny \textrm{\h1}},i}},
\end{equation}

\begin{equation}
    j_{\star} =\frac{\sum_{i} M_{\star,i}v_{\mathrm{rot},i}r_{i}}{\sum_{i} M_{{\star,i}}},
\end{equation}

\noindent where $M_{\textrm{\h1},i}$, $M_{\star,i}$, $v_{\mathrm{rot},i}$ and $r_{i}$ are the \h1 mass, stellar mass, rotation velocity and radius corresponding to the $i^{th}$ ring, respectively. Following this we compute the ratio $j_{\textrm{\h1}/j_{\star}}$ for the Eridanus sample. Fig.~\ref{Fig.B1} shows the $j_{\textrm{\h1}/j_{\star}}$ values plotted against the total baryonic mass ($M_{\mathrm{b}}$) of the galaxies.

\section{Comparing the various \h1 deficiency parameters}
\label{appendix:HI-deficiency}

\begin{figure}
    \centering
    \includegraphics[width=8cm,height=6.2cm]{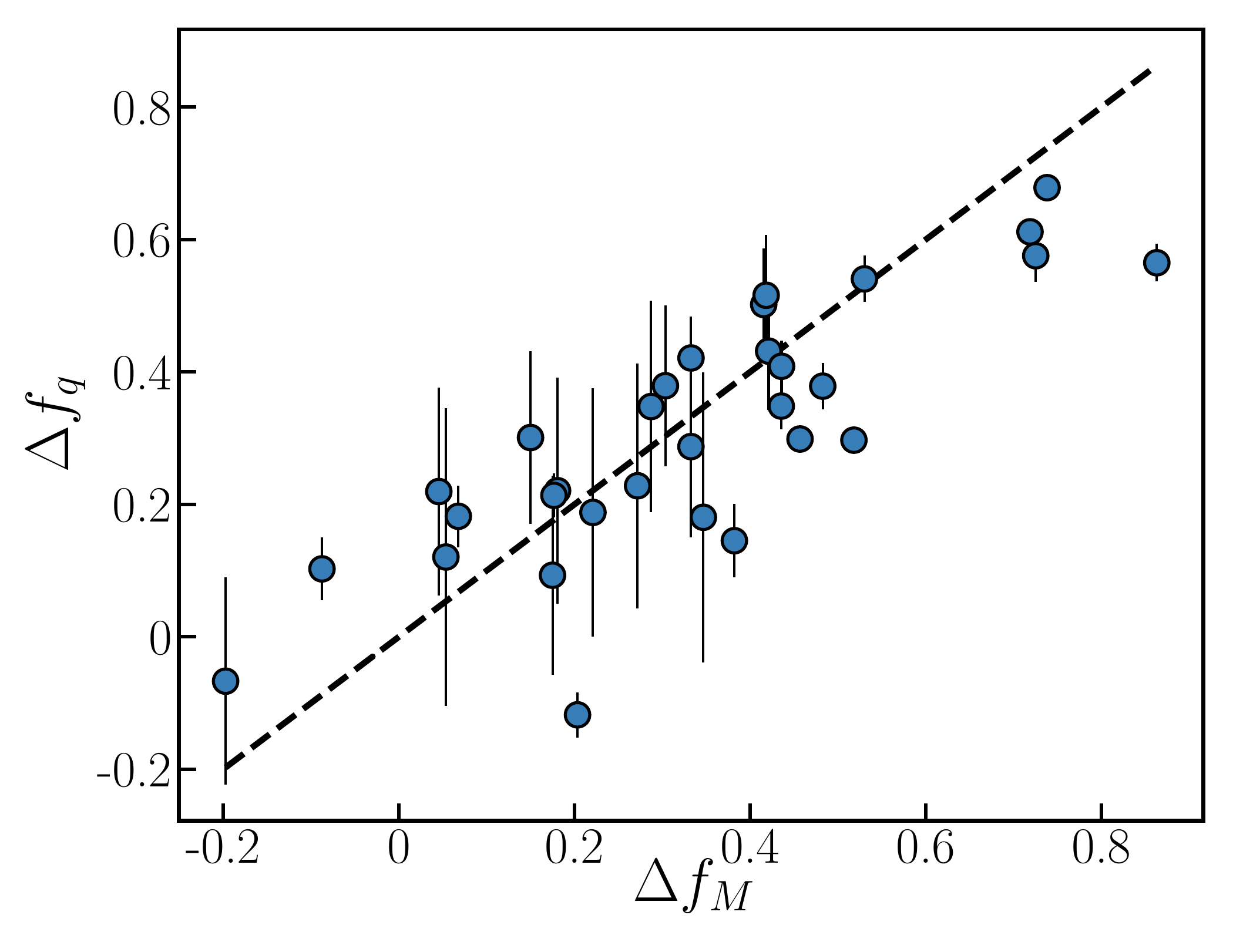}
    \caption{The $\Delta f_q$ values plotted against the \h1 deficiency parameter $\Delta f_M$ (see \citetalias{obreschkow16}).}
    \label{Fig.C1}
\end{figure}

\begin{figure}
    \centering
    \includegraphics[width=8cm,height=6.2cm]{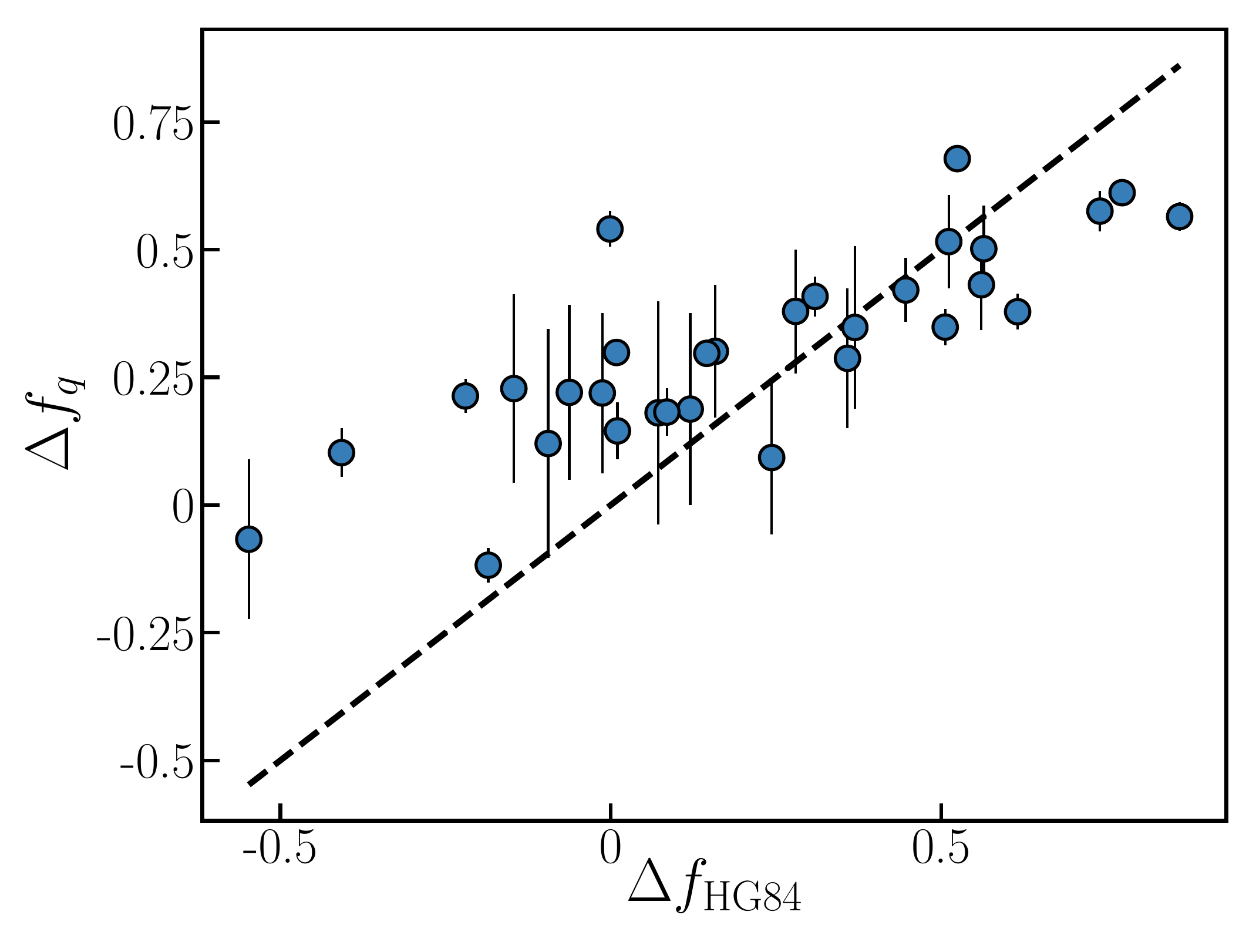}
    \caption{The $\Delta f_q$ values plotted against the \h1 deficiency parameter introduced by \citet{haynes84}.}
    \label{Fig.C2}
\end{figure}

\begin{figure}
    \centering
    \includegraphics[width=8cm,height=6.2cm]{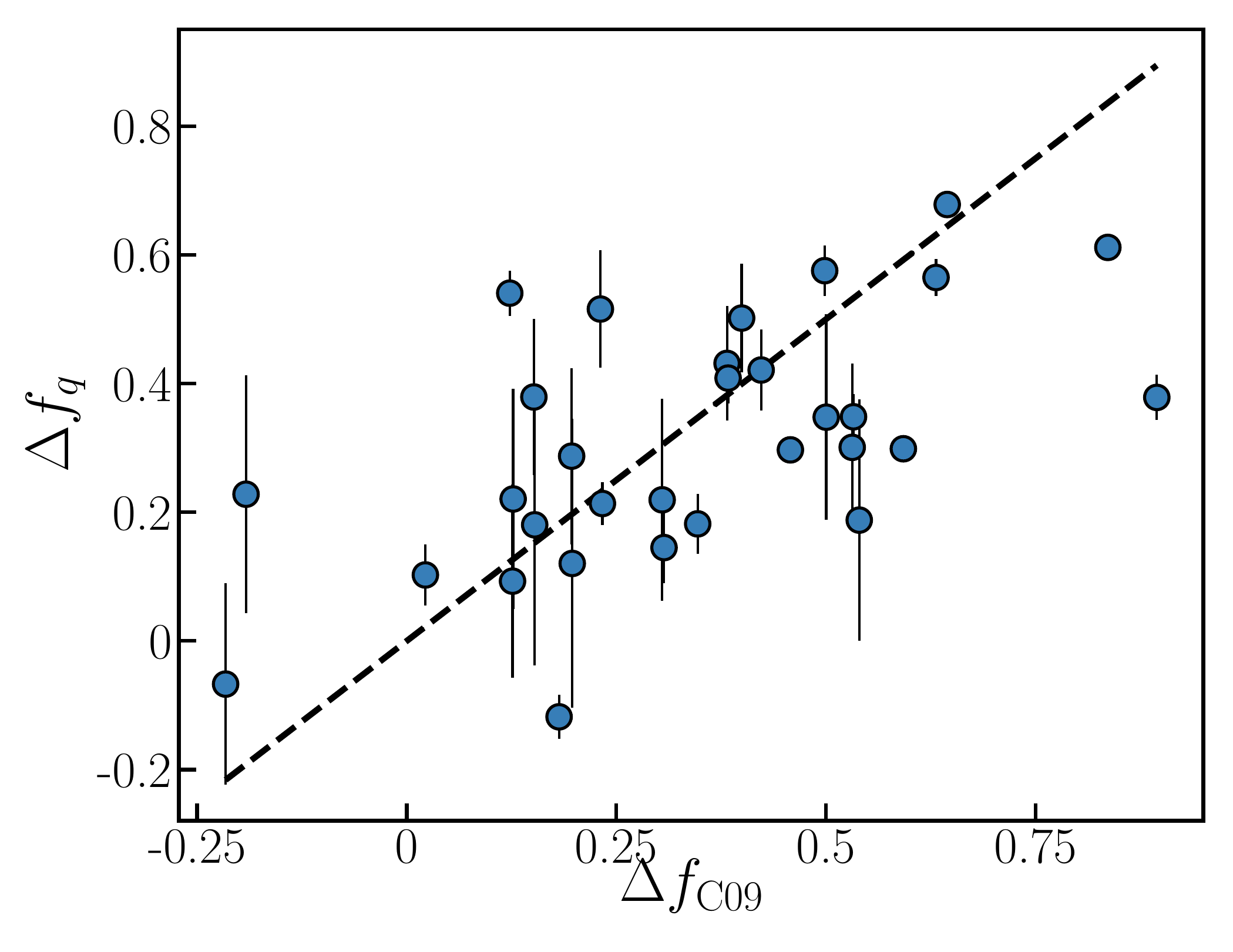}
    \caption{The $\Delta f_q$ values plotted against the \h1 deficiency parameter as defined in \citet{Chung09}.}
    \label{Fig.C3}
\end{figure}

\begin{figure}
    \centering
    \includegraphics[width=8cm,height=6.2cm]{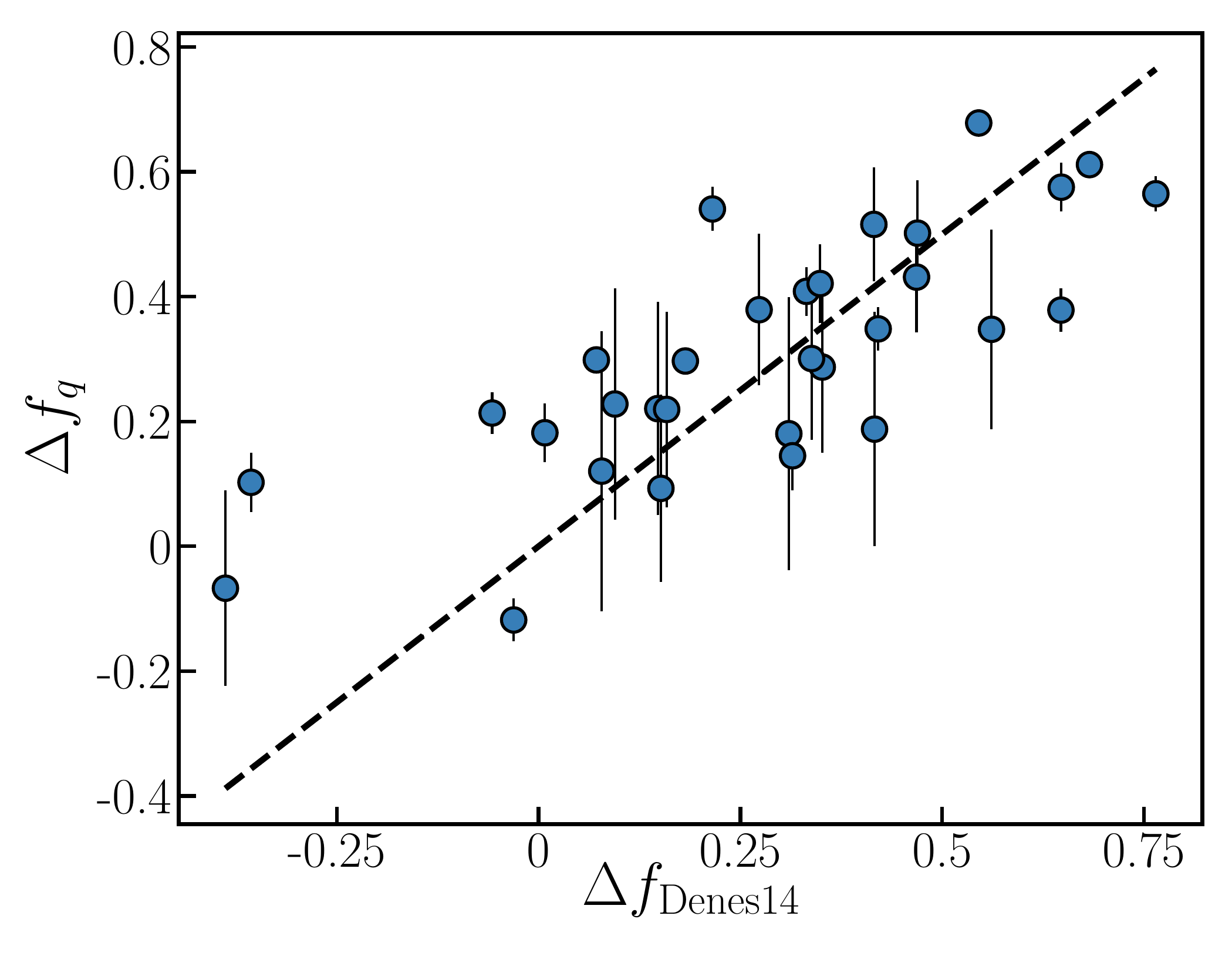}
    \caption{The $\Delta f_q$ values plotted against the \h1 deficiency parameter defined by \citet{denes14}.}
    \label{Fig.C4}
\end{figure}

Here we compare the $\Delta f_q$ values for the Eridanus galaxies with other widely used \h1 deficiency parameters in the literature. We first compare the $\Delta f_q$ values with the \h1 deficiency parameter ($\Delta f_M$), computed from the $f_{\mathrm{atm}} - M_{\mathrm{b}}$ scaling relation (see Fig.3 in \citetalias{obreschkow16}), using Eq.7 in \citet{Li20}. This is shown in Fig.~\ref{Fig.C1}, and we find a good one-to-one agreement between the two \h1 deficiency parameters. In addition, we also compare the $\Delta f_q$ values with more traditional \h1 deficiency parameters that are based on empirical scaling relations, such as the morphology dependent \h1 deficiency parameter first introduced by \citet{haynes84} (see Fig.~\ref{Fig.C2}), the morphology independent \h1 deficiency parameter calibrated for cluster-like galaxies by \citet{Chung09}, as shown in Fig.~\ref{Fig.C3} and finally with the recently calibrated \h1 deficiency parameter derived by \citet{denes14}, based on the B-band photometric diameter of galaxies (see Fig.~\ref{Fig.C4}). We see that the $\Delta f_q$ \h1 deficiency parameter overall shows a good one-to-one agreement with all the above \h1 deficiency parameters. This indicates the robustness of this physically motivated \h1 deficiency parameter, which is simply based on the deviation of the \h1 gas fraction of galaxies from the theoretical \h1 saturation, predicted by the analytical model in \citetalias{obreschkow16}. 

\section{Table -- Properties of the sample galaxies}
\label{appendix:properties_table}

\begin{landscape}
\begin{table}
\mysize
\caption{The sample properties. Cols (1) - (2): WALLABY designation and common name; Cols (3) - (4): RA and DEC corresponding to the optical center; Col (5): Systemic velocity; Col (6): Luminosity distance; Col (7): Median value of position angle; Col (8): Median value of the inclination angle; Cols (9),(10),(11): \h1, stellar and total baryonic mass respectively; Col (12): total baryonic specific AM; Col (13): atomic gas fraction; Col (14): disc stability parameter; Col (15): stability model based \h1 deficiency parameter}
\begin{tabular}{llccccccccccccc}
\hline \hline
WALLABY Name & Name	&	RA	&	Dec	&	$V_{\textrm{sys}}$	&	$D$	&	PA	&	$i$	&	$\log_{10}(M_{\textrm{\h1}})$	& $\log_{10}(M_{\star})$	&	$\log_{10}(M_{\mathrm{b}})$	&	$j_{\mathrm{b}}$	&	$f_{\mathrm{atm}}$	&	$q$	& $\Delta f_q$	\\
   &  & (J2000) & (J2000) & (\kms) & (Mpc) & ($\degr$) & ($\degr$) & (M$_{\odot}$) & (M$_{\odot}$) & (M$_{\odot}$) & (kpc~\kms) &  & & \\
   (1)   & (2) & (3) & (4) & (5) & (6) & (7) & (8) & (9) & (10) & (11) & (12) & (13) & (14) & (15) \\
   \hline
   \textbf{Eridanus supergroup} & & & & & & & & & & & & & & \\ 
WALLABYJ032425-213233	&	NGC1325	&	03h24m25.57s	&	-21d32m38.5s	&	1588	&	20	&	232	&	72	&	9.32	$\pm$	0.2	&	9.99	$\pm$	0.05	&	10.1	$\pm$	0.06	&	786	$\pm$	56	&	0.22	$\pm$	0.05	&	0.14	$\pm$	0.05	&	0.18	\\
WALLABYJ032455-214701	&	ESO548-G011	&	03h24m55.30s	&	-21d47m02.0s	&	1456	&	18	&	142	&	57	&	7.98	$\pm$	0.25	&	7.74	$\pm$	0.15	&	8.27	$\pm$	0.18	&	52	$\pm$	16	&	0.7	$\pm$	0.22	&	0.66	$\pm$	0.31	&	0.18	\\
WALLABYJ032735-211339	&	ESO548-G021	&	03h27m35.29s	&	-21d13m42.2s	&	1686	&	21	&	72	&	67	&	8.51	$\pm$	0.22	&	8.6	$\pm$	0.09	&	8.92	$\pm$	0.12	&	167	$\pm$	38	&	0.52	$\pm$	0.13	&	0.47	$\pm$	0.19	&	0.3	\\
WALLABYJ032831-222957	&	ESO481-G028	&	03h28m32.22s	&	-22d30m04.8s	&	1774	&	23	&	338	&	53	&	8.17	$\pm$	0.24	&	8.3	$\pm$	0.11	&	8.6	$\pm$	0.13	&	61	$\pm$	15	&	0.5	$\pm$	0.14	&	0.35	$\pm$	0.15	&	0.29	\\
WALLABYJ032937-232103	&	ESO481-G030	&	03h29m38.31s	&	-23d21m01.2s	&	1657	&	21	&	230	&	63	&	8.29	$\pm$	0.23	&	8.1	$\pm$	0.1	&	8.59	$\pm$	0.16	&	124	$\pm$	34	&	0.68	$\pm$	0.19	&	0.74	$\pm$	0.32	&	0.19	\\
WALLABYJ032941-221642	&	NGC1347	&	03h29m41.50s	&	-22d17m06.0s	&	1755	&	22	&	315	&	34	&	8.49	$\pm$	0.22	&	9.13	$\pm$	0.07	&	9.24	$\pm$	0.08	&	100	$\pm$	14	&	0.24	$\pm$	0.06	&	0.13	$\pm$	0.05	&	0.15	\\
WALLABYJ033047-210333	&	ESO548-G029	&	03h30m47.17s	&	-21d03m29.6s	&	1292	&	16	&	95	&	53	&	7.7	$\pm$	0.27	&	8.8	$\pm$	0.11	&	8.84	$\pm$	0.1	&	47	$\pm$	9	&	0.1	$\pm$	0.03	&	0.16	$\pm$	0.06	&	0.56	\\
WALLABYJ033257-210513	&	ESO548-G034	&	03h32m57.63s	&	-21d05m21.9s	&	1665	&	21	&	81	&	29	&	8.07	$\pm$	0.26	&	8.99	$\pm$	0.1	&	9.06	$\pm$	0.1	&	87	$\pm$	13	&	0.14	$\pm$	0.04	&	0.18	$\pm$	0.06	&	0.58	\\
WALLABYJ033302-240756	&	ESO482-G005	&	03h33m02.25s	&	-24d07m58.3s	&	1915	&	25	&	261	&	79	&	8.95	$\pm$	0.21	&	8.37	$\pm$	0.07	&	9.15	$\pm$	0.17	&	398	$\pm$	104	&	0.84	$\pm$	0.22	&	0.65	$\pm$	0.28	&	0.12	\\
WALLABYJ033326-234246	&	IC1952	&	03h33m26.67s	&	-23d42m46.0s	&	1810	&	23	&	322	&	74	&	8.8	$\pm$	0.21	&	9.65	$\pm$	0.07	&	9.72	$\pm$	0.07	&	277	$\pm$	30	&	0.16	$\pm$	0.04	&	0.12	$\pm$	0.04	&	0.35	\\
WALLABYJ033341-212844	&	IC1953	&	03h33m41.87s	&	-21d28m43.1s	&	1859	&	24	&	129	&	44	&	9.07	$\pm$	0.2	&	9.93	$\pm$	0.06	&	10.01	$\pm$	0.06	&	584	$\pm$	44	&	0.16	$\pm$	0.03	&	0.13	$\pm$	0.04	&	0.21	\\
WALLABYJ033347-192946	&	NGC1359	&	03h33m47.71s	&	-19d29m31.4s	&	1964	&	25	&	322	&	58	&	9.56	$\pm$	0.2	&	9.42	$\pm$	0.06	&	9.88	$\pm$	0.13	&	647	$\pm$	92	&	0.65	$\pm$	0.16	&	0.2	$\pm$	0.07	&	-0.07	\\
WALLABYJ033501-245556	&	NGC1367	&	03h35m01.34s	&	-24d55m59.6s	&	1458	&	18	&	136	&	47	&	9.64	$\pm$	0.2	&	10.3	$\pm$	0.06	&	10.41	$\pm$	0.07	&	1512	$\pm$	96	&	0.23	$\pm$	0.05	&	0.14	$\pm$	0.04	&	0.1	\\
WALLABYJ033527-211302	&	ESO548-G049	&	03h35m28.07s	&	-21d13m01.3s	&	1518	&	19	&	115	&	64	&	8.22	$\pm$	0.23	&	8.39	$\pm$	0.08	&	8.67	$\pm$	0.12	&	99	$\pm$	21	&	0.48	$\pm$	0.12	&	0.49	$\pm$	0.19	&	0.38	\\
WALLABYJ033537-211742	&	IC1962	&	03h35m37.49s	&	-21d17m38.6s	&	1802	&	23	&	358	&	74	&	8.86	$\pm$	0.21	&	8.74	$\pm$	0.07	&	9.18	$\pm$	0.13	&	255	$\pm$	55	&	0.64	$\pm$	0.16	&	0.39	$\pm$	0.15	&	0.22	\\
WALLABYJ033617-253615	&	ESO482-G011	&	03h36m17.39s	&	-25d36m17.0s	&	1590	&	20	&	68	&	69	&	8.19	$\pm$	0.26	&	8.66	$\pm$	0.1	&	8.82	$\pm$	0.11	&	89	$\pm$	17	&	0.32	$\pm$	0.09	&	0.31	$\pm$	0.12	&	0.43	\\
WALLABYJ033653-245445	&	ESO482-G013	&	03h36m53.87s	&	-24d54m45.7s	&	1842	&	24	&	61	&	61	&	8.5	$\pm$	0.22	&	8.51	$\pm$	0.14	&	8.88	$\pm$	0.14	&	92	$\pm$	26	&	0.57	$\pm$	0.15	&	0.29	$\pm$	0.12	&	0.09	\\
WALLABYJ033728-243010	&	NGC1385	&	03h37m28.85s	&	-24d30m01.1s	&	1497	&	19	&	181	&	50	&	9.24	$\pm$	0.2	&	10.08	$\pm$	0.06	&	10.16	$\pm$	0.06	&	341	$\pm$	27	&	0.16	$\pm$	0.03	&	0.05	$\pm$	0.02	&	-0.12	\\
WALLABYJ033752-190024	&	NGC1390	&	03h37m52.17s	&	-19d00m30.1s	&	1218	&	15	&	10	&	70	&	8.11	$\pm$	0.22	&	8.45	$\pm$	0.08	&	8.66	$\pm$	0.1	&	115	$\pm$	22	&	0.38	$\pm$	0.09	&	0.59	$\pm$	0.22	&	0.52	\\
WALLABYJ033854-262013	&	NGC1398	&	03h38m52.13s	&	-26d20m16.2s	&	1373	&	17	&	278	&	44	&	9.32	$\pm$	0.2	&	10.89	$\pm$	0.07	&	10.9	$\pm$	0.06	&	1488	$\pm$	103	&	0.04	$\pm$	0.01	&	0.04	$\pm$	0.01	&	0.3	\\
WALLABYJ033921-212450	&	LEDA13460	&	03h39m21.43s	&	-21d24m56.6s	&	1622	&	21	&	240	&	45	&	8	$\pm$	0.29	&	8.55	$\pm$	0.14	&	8.69	$\pm$	0.13	&	69	$\pm$	15	&	0.28	$\pm$	0.09	&	0.33	$\pm$	0.13	&	0.49	\\
WALLABYJ033941-235054	&	ESO482-G027	&	03h39m41.21s	&	-23d50m39.8s	&	1622	&	21	&	356	&	69	&	8.28	$\pm$	0.22	&	8.07	$\pm$	0.1	&	8.58	$\pm$	0.15	&	98	$\pm$	15	&	0.69	$\pm$	0.18	&	0.61	$\pm$	0.2	&	0.23	\\
WALLABYJ034002-192200	&	ESO548-G065	&	03h40m02.70s	&	-19d21m59.8s	&	1216	&	15	&	38	&	66	&	8.01	$\pm$	0.23	&	8.42	$\pm$	0.08	&	8.6	$\pm$	0.09	&	97	$\pm$	15	&	0.35	$\pm$	0.08	&	0.56	$\pm$	0.2	&	0.5	\\
WALLABYJ034040-221711	&	ESO548-G070	&	03h40m40.99s	&	-22d17m13.4s	&	1774	&	23	&	64	&	67	&	8.35	$\pm$	0.24	&	8.35	$\pm$	0.12	&	8.72	$\pm$	0.14	&	227	$\pm$	64	&	0.58	$\pm$	0.16	&	1	$\pm$	0.43	&	0.35	\\
WALLABYJ034056-223350	&	NGC1415	&	03h40m56.86s	&	-22d33m52.1s	&	1552	&	20	&	334	&	66	&	8.87	$\pm$	0.21	&	10.2	$\pm$	0.08	&	10.23	$\pm$	0.07	&	465	$\pm$	41	&	0.06	$\pm$	0.01	&	0.06	$\pm$	0.02	&	0.3	\\
WALLABYJ034057-214245	&	NGC1414	&	03h40m57.04s	&	-21d42m47.3s	&	1695	&	22	&	354	&	65	&	8.42	$\pm$	0.22	&	9	$\pm$	0.08	&	9.13	$\pm$	0.08	&	171	$\pm$	22	&	0.26	$\pm$	0.06	&	0.3	$\pm$	0.1	&	0.42	\\
WALLABYJ034114-235017	&	ESO482-G035	&	03h41m14.83s	&	-23d50m13.0s	&	1885	&	24	&	180	&	47	&	8.64	$\pm$	0.22	&	9.47	$\pm$	0.08	&	9.55	$\pm$	0.08	&	242	$\pm$	27	&	0.17	$\pm$	0.04	&	0.16	$\pm$	0.05	&	0.41	\\
WALLABYJ034131-214051	&	NGC1422	&	03h41m31.07s	&	-21d40m53.5s	&	1644	&	21	&	64	&	61	&	8.32	$\pm$	0.23	&	9.22	$\pm$	0.09	&	9.29	$\pm$	0.08	&	144	$\pm$	17	&	0.14	$\pm$	0.04	&	0.17	$\pm$	0.06	&	0.38	\\
WALLABYJ034219-224520	&	ESO482-G036	&	03h42m18.80s	&	-22d45m09.2s	&	1569	&	20	&	193	&	50	&	8.35	$\pm$	0.23	&	9.25	$\pm$	0.09	&	9.32	$\pm$	0.09	&	186	$\pm$	26	&	0.14	$\pm$	0.04	&	0.21	$\pm$	0.07	&	0.54	\\
WALLABYJ034337-211418	&	ESO549-G006	&	03h43m38.17s	&	-21d14m15.7s	&	1612	&	21	&	209	&	66	&	8.44	$\pm$	0.22	&	8.29	$\pm$	0.09	&	8.75	$\pm$	0.15	&	150	$\pm$	37	&	0.65	$\pm$	0.17	&	0.61	$\pm$	0.25	&	0.22	\\
WALLABYJ034456-234158	&	LEDA13743	&	03h44m56.61s	&	-23d41m58.9s	&	1819	&	24	&	213	&	45	&	8.03	$\pm$	0.28	&	8.43	$\pm$	0.13	&	8.62	$\pm$	0.13	&	46	$\pm$	10	&	0.35	$\pm$	0.11	&	0.26	$\pm$	0.1	&	0.23	\\
WALLABYJ034517-230001	&	NGC1438	&	03h45m17.23s	&	-23d00m08.9s	&	1546	&	20	&	225	&	67	&	8.34	$\pm$	0.23	&	9.62	$\pm$	0.07	&	9.65	$\pm$	0.07	&	264	$\pm$	26	&	0.07	$\pm$	0.02	&	0.14	$\pm$	0.04	&	0.68	\\
WALLABYJ034814-212824	&	ESO549-G018	&	03h48m14.08s	&	-21d28m28.1s	&	1586	&	20	&	211	&	54	&	8.25	$\pm$	0.24	&	9.47	$\pm$	0.06	&	9.51	$\pm$	0.06	&	192	$\pm$	24	&	0.07	$\pm$	0.02	&	0.14	$\pm$	0.05	&	0.61	\\
\hline
\textbf{Background galaxies} & & & & & & & & & & & & & & \\ 
WALLABYJ032434-215824	&	ESO548-G008	&	03h24m34.33s	&	-21d58m18.8s	&	4376	&	59	&	136	&	60	&	8.78	$\pm$	0.38	&	8.57	$\pm$	0.15	&	9.1	$\pm$	0.25	&	171	$\pm$	76	&	0.65	$\pm$	0.3	&	0.32	$\pm$	0.19	&	0.03	\\
WALLABYJ033138-250029	&	ESO482-G002	&	03h31m38.38s	&	-25d00m29.2s	&	6483	&	89	&	254	&	59	&	9.71	$\pm$	0.26	&	10.28	$\pm$	0.09	&	10.43	$\pm$	0.09	&	1150	$\pm$	172	&	0.25	$\pm$	0.07	&	0.1	$\pm$	0.03	&	-0.13	\\
WALLABYJ033147-211309	&	LEDA831121	&	03h31m48.07s	&	-21d13m08.2s	&	4244	&	57	&	56	&	43	&	8.81	$\pm$	0.33	&	8.69	$\pm$	0.15	&	9.16	$\pm$	0.2	&	209	$\pm$	73	&	0.6	$\pm$	0.23	&	0.34	$\pm$	0.17	&	0.09	\\
WALLABYJ033351-210313	&	LEDA13198	&	03h33m51.02s	&	-21d03m14.8s	&	6728	&	92	&	61	&	49	&	9.68	$\pm$	0.29	&	10.13	$\pm$	0.1	&	10.32	$\pm$	0.11	&	689	$\pm$	117	&	0.31	$\pm$	0.09	&	0.08	$\pm$	0.03	&	-0.34	\\
WALLABYJ033351-212725	&	ESO548-G040	&	03h33m51.32s	&	-21d27m19.7s	&	4064	&	55	&	309	&	59	&	9.46	$\pm$	0.16	&	9.74	$\pm$	0.07	&	10	$\pm$	0.07	&	820	$\pm$	105	&	0.39	$\pm$	0.07	&	0.19	$\pm$	0.06	&	0	\\
WALLABYJ033757-213938	&	ESO548-G056	&	03h37m57.77s	&	-21d39m42.7s	&	4150	&	56	&	137	&	61	&	8.95	$\pm$	0.32	&	9.41	$\pm$	0.07	&	9.6	$\pm$	0.11	&	405	$\pm$	65	&	0.3	$\pm$	0.1	&	0.24	$\pm$	0.08	&	0.22	\\
WALLABYJ033810-194412	&	ESO548-G057	&	03h38m10.33s	&	-19d44m11.0s	&	4166	&	56	&	42	&	72	&	9.11	$\pm$	0.33	&	8.79	$\pm$	0.16	&	9.4	$\pm$	0.23	&	380	$\pm$	158	&	0.7	$\pm$	0.28	&	0.35	$\pm$	0.2	&	0.05	\\
WALLABYJ033859-223502	&	ESO482-G024	&	03h38m59.73s	&	-22d35m02.4s	&	4337	&	59	&	335	&	53	&	9.28	$\pm$	0.22	&	8.91	$\pm$	0.07	&	9.56	$\pm$	0.17	&	554	$\pm$	48	&	0.72	$\pm$	0.22	&	0.36	$\pm$	0.11	&	0.04	\\
WALLABYJ034421-204443	&	LEDA135108	&	03h44m21.72s	&	-20d44m43.2s	&	8008	&	111	&	269	&	63	&	9.62	$\pm$	0.26	&	9.87	$\pm$	0.07	&	10.14	$\pm$	0.11	&	1254	$\pm$	214	&	0.41	$\pm$	0.12	&	0.21	$\pm$	0.08	&	0.03	\\
\end{tabular}
\label{tab:sample}
\end{table}
\end{landscape}

\section{\h1-deficiency parameter Vs $\Sigma_2$}
\label{appendix:Delta_f_qVsSigma2}

\begin{figure}
\centering
    \includegraphics[width=8cm,height=6cm]{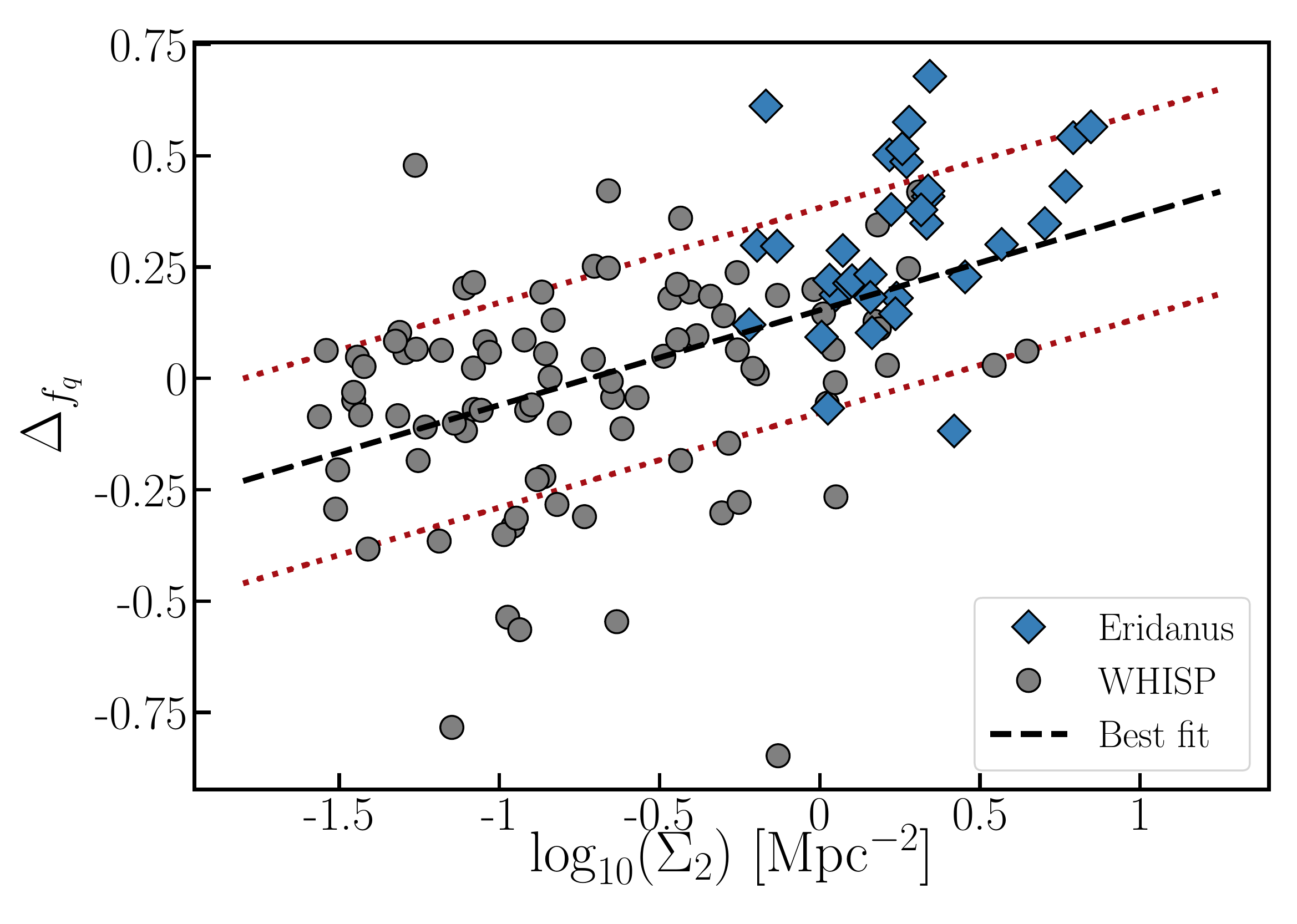}
    \caption{The $\Delta f_q$ \h1-deficiency parameter values for the Eridanus (blue diamonds) and the WHISP sample (grey circles) plotted against their $\log_{10}(\Sigma_2)$ local environment density estimates. The black dashed line is the best fit line to the data, and the red dotted lines represent the one sigma scatter about the distribution.} 
    \label{Fig.E1}
\end{figure}

As shown in Fig.~\ref{Fig.4} in section~\ref{Subsec:Env. effect}, galaxies in the Eridanus supergroup were not only observed to have higher $\Sigma_2$ values compared to galaxies in the WHISP sample, but it was also observed that those galaxies deviating the most from the \fatm relation are from denser environments. Fig.~\ref{Fig.E1} below highlights this trend, where the $\Delta f_q$ values of both the Eridanus and WHISP galaxies are plotted against their local environment density ($\Sigma_2$). We find that there is a general trend of increasing \h1-deficiency with increasing $\log_{10}(\Sigma_2)$ among both samples, clearly indicating that the \h1 gas in galaxies located in higher density environments have likely been affected by various environmental processes. We fit a linear regression of the form $\Delta f_q = \alpha \log_{10}(\Sigma_2) + b$ to the data and find a best fitting line with slope $\alpha=0.21$ and a scatter of 0.23 dex. The few galaxies in the WHISP sample that lie below the lower $1\sigma$ scatter threshold are those galaxies that are currently interacting with their companions and accreting gas, therefore their \h1 gas fractions are boosted (implying $\Delta f_q$ will be lower). These galaxies are discussed in more detail in \citetalias{Murugeshan20}.

\bsp	
\label{lastpage}
\end{document}